\documentclass[12pt,a4paper]{article}
\usepackage{psfig,amssymb}
\def\sp{{\mathbf s}_t}
\newcommand{\sm}{{\mathbf s_{\bar t}}}
\newcommand{\kt}{{\hat{\mathbf k}}_t}
\newcommand{\ph}{\hat{\mathbf p}}

\newcommand{\eins}{  1\!{\rm l}  }
\def\vec#1{\mathchoice{\mbox{\boldmath$\mathrm\displaystyle#1$}}
{\mbox{\boldmath$\mathrm\textstyle#1$}}
{\mbox{\boldmath$\mathrm\scriptstyle#1$}}
{\mbox{\boldmath$\mathrm\scriptscriptstyle#1$}}}
\newcommand{\bm}[1]{\mbox{\boldmath$#1$}}  
\renewcommand{\vec}{\bm}
\newcommand{\nn}{\nonumber}
\newcommand{\ds}{\displaystyle}
\newcommand{\ktb}{\mathbf{\hat{k}}_{\bar t}}
\newcommand{\lp}{\mathbf{\hat{q}}_+}
\newcommand{\nhat}{\mathbf{\hat{n}}}
\newcommand{\nv}{\mathbf{n}}
\newcommand{\lm}{\mathbf{\hat{q}}_-}
\newcommand{\qb}{\mathbf{\hat{q}}_b}
\newcommand{\qbb}{\mathbf{\hat{q}}_{\bar b}}
\newcommand{\qr}{\mathbf{\hat{q}}_r}
\newcommand{\qbr}{\mathbf{\hat{q}}_{\bar r}}
\newcommand{\qW}{\mathbf{\hat{q}}_{W^+}}
\newcommand{\qWb}{\mathbf{\hat{q}}_{W^-}}
\newcommand{\bea}{\begin{eqnarray}}
\newcommand{\eea}{\end{eqnarray}}
\newcommand{\be}{\begin{equation}}
\newcommand{\ee}{\end{equation}}
\newcommand{\<}{\langle\,}
\renewcommand{\>}{\,\rangle}
\renewcommand{\exp}[1]{{\prec\! {#1}\!\succ}}
\begin{document}
\begin{titlepage}
\begin{flushright}
CERN-TH/98-390 \\
PITHA 98/41
\end{flushright}
\vspace{0.8cm}
\begin{center}
\renewcommand{\thefootnote}{\fnsymbol{footnote}}
{\bf\LARGE Effects of Higgs sector CP violation \\
in top-quark pair production at the LHC\footnote{
Work supported by D.F.G. and by BMBF contract 05 7AC 9EP.}\\}
\setcounter{footnote}{0}
\renewcommand{\thefootnote}{\arabic{footnote}}
\vspace{2cm}
{\bf W.\ Bernreuther$^{a,b}$, A. Brandenburg$^a$
and M.\  Flesch$^a$}
\par\vspace{1cm}
$^a$ Institut f.\ Theoretische Physik, RWTH Aachen, D-52056 Aachen, Germany\\
$^b$ Theory Division, CERN, CH-1211 Geneva 23, Switzerland
\par\vspace{3cm}
%                                                
%  Abstract                                      
%                                                
{\bf Abstract:}\\
\parbox[t]{\textwidth}
{A striking manifestation of CP violation in the electroweak symmetry
breaking sector would be the existence of neutral Higgs boson(s) with 
undefined CP parity. We analyse  signatures of such  a boson, with a 
mass of about 300 GeV or larger, produced in high energy proton-proton 
collisions at LHC energies in its top-quark antitop-quark decay channel. 
The large irreducible $t\bar t$ background is taken into account. 
We propose, both for the dilepton and the lepton + jets decay
channels of $t\bar t$, several correlations and asymmetries with which
(Higgs sector) CP violation can be traced.
We show that for judiciously chosen cuts on the
$t\bar t$ invariant mass these CP observables yield,
for an LHC integrated luminosity of 100 $\rm{fb}^{-1}$, statistically 
significant signals
for a range of  Higgs boson masses and  Yukawa couplings. }
\end{center}
\vspace{2cm}
PACS number(s): 11.30.Er, 12.60.Fr, 14.65.Ha, 14.80.Cp
\end{titlepage}
\section{Introduction}
The clarification of the mechanism of 
electroweak gauge-symmetry breaking will be among the 
most important physics issues  at the CERN Large Hadron Collider (LHC).
According to the concept used in  the Standard Model (SM) and in 
many of its extensions,
this amounts to searching for Higgs bosons.
While in the standard electroweak theory (SM) only one
neutral Higgs boson is associated with electroweak symmetry breaking, 
 many extensions of the SM entail a more
complex scalar sector. Apart from predicting  a number of  Higgs 
particles, an extended scalar sector  can  also have  a bearing  
on another phenomenon of  unclarified origin,
 namely CP non-conservation. As is well-known  CP
can be violated by an extended Higgs sector \cite{Lee}. As far as the neutral 
Higgs bosons are concerned, this would manifest itself in neutral
spin-zero states of undefined CP parity, i.e. particles that have both
 scalar and pseudoscalar Yukawa couplings to quarks and leptons \cite{DepMa}. 
This possibility arises naturally already in two-Higgs doublet extensions
of the SM \cite{Lee,Branco,Wolf,Wein}. As to  supersymmetric extensions
of the SM, mixing of the $\rm{CP}=+1$ and  
$\rm{CP}=-1$ neutral Higgs-boson states, leading to
CP-impure mass eigenstates,  can also occur at tree level
in its next-to-minimal extension (NMSSM) \cite{Matsuda}, while in the minimal
extension (MSSM) mixing is induced radiatively and  can be
quite substantial \cite{Babu,Pil}, depending
on the parameters of the model.
\par
If Higgs  boson(s)  will be discovered, the most direct way 
to study their CP properties at the LHC will be the investigation of their
Yukawa interactions with top quarks. For a neutral Higgs boson $\varphi$ of
arbitrary CP parity and mass $m_{\varphi} > 2m_t$,
 CP violation occurs already at 
the Born level in its decay\footnote{For the channels $\varphi\to W^+W^-, ZZ$
CP violation can take place only at the one-loop level (see, e.g., 
\cite{Chang,Osland}) and is therefore expected to lead to smaller effects.}
$\varphi\to t\bar{t}$ and shows up in spin-spin correlations 
\cite{BeBra1} -- \cite{corr}. However, the effect is diluted by interference
with the $t\bar{t}$ background \cite{BeBra1,BeBra2}.
Effects of CP-violating $\varphi$ boson exchange 
in $p p\to t\bar{t}X$ were  analysed in \cite{Peskin,BeBra1,BeBra2,Zhou}.
Concerning the study of Yukawa couplings of
 a light neutral Higgs boson $\varphi$ with a mass of about 100 GeV at the LHC,
associated $t\bar{t}\varphi$ production \cite{He} and the decay
mode $\varphi \to \tau^+\tau^-$ \cite{BeBra1,corr}
offer further possibilities.
CP effects in $t\bar t$ production and decay
at  hadron colliders were also investigated in terms of form factor
parametrizations \cite{KLY} -- \cite{Lampe}, within the MSSM
\cite{Schmidt,Zhou},
and for single top-quark  production  within two-Higgs doublet models and 
the MSSM \cite{Atwood,Bar}. 
\par
In this article  we investigate the signatures of a  heavy Higgs boson of 
arbitrary CP parity
in the $p p\to t\bar{t} X$  channel at the LHC, taking the irreducible
$t\bar t$ background into account. We extend previous analyses and
obtain several new results.
We propose, for the dilepton and the lepton + jets decay channels of
$t\bar{t}$, a number of CP observables and associated asymmetries that
should be rather robust with respect to measurement uncertainties. 
We analyse, within two-Higgs doublet and supersymmetric extensions of the SM,
the sensitivity
of these observables to CP-violating contributions 
from the $t\bar t$ production and
top-quark decay amplitudes. 
By using appropriate bins in the $t\bar{t}$ invariant-mass distribution
we show that, for a range of Higgs boson masses and Yukawa couplings, 
these
observables and asymmetries exhibit CP effects at the percent level.
Thus these quantities should be good tools for tracing Higgs sector CP 
violation in future LHC experiments. 
\par                                             
Our paper is organized as follows. In section 2 we recapitulate the 
salient features of neutral Higgs sector CP violation within two-Higgs
doublet and  supersymmetric extensions of the SM,
which are relevant to our analysis, and we outline the strategy of how to 
trace  possible CP effects caused by a heavy Higgs boson $s$-channel
resonance in $pp\to t\bar{t} X$.  In section 3 we discuss the general
structure of the squared matrix elements for the parton reactions
 $gg, q\bar{q}\to t\bar{t}\to W^+ b W^-\bar{b}\to  6f$ in the on-shell 
approximation 
for $t$ and $\bar t$ quarks using the spin-density matrix formalism. 
We exhibit the two types of CP-odd spin-momentum correlations which are 
induced at the level of the $t\bar t$ states and argue that CP
violation effects in $t\to W b$ are small  compared with (quasi)resonant
$\varphi$ exchange in  $t\bar t$
production. In section 4 we introduce a number of
CP observables and  asymmetries with which CP violation effects
can be traced in $pp\to t\bar{t} X$ in the dilepton and the lepton + jets
channels, and we derive  relations between expectation values of
observables and corresponding asymmetries.
We show that those of our observables that are T-odd are predominantly
sensitive to
CP-violating terms in the $t\bar t$ production amplitude, irrespective
of whether $\varphi$ is heavy or light. 
In section 5 we compute and plot, for the c.m. energy $\sqrt{s}=14$ TeV
and for a set  of 
Higgs boson masses and Yukawa couplings, the expectation values of 
two CP observables
as a function of the $t\bar t$ invariant mass $M_{t\bar t}$, 
taking resonant and non-resonant
Higgs boson exchange and the irreducible $t\bar t$ background into
account. These plots serve to  select appropriate $M_{t\bar t}$ bins
in the computation of the expectation values and asymmetries of the CP
observables of section 4. 
These calculations are then performed  
for the dilepton and lepton + jets channels, using  phase-space 
cuts on the transverse momenta of the final state leptons and partons. 
Finally we estimate, for an LHC integrated luminosity of 
100 ${\rm fb}^{-1}$, the statistical sensitivity of these observables to 
Higgs sector CP violation, i.e. the sensitivity of these observables to detect
a non-zero product of scalar and pseudoscalar Yukawa couplings.
We conclude in section 6. In the appendix we give a compact formula for 
computing the expectation values of the CP observables.
\section{Higgs sector CP violation}
A number of extensions of the Standard Model allow for the possibility of 
CP violation in the Higgs sector, the simplest ones being extensions by an 
additional Higgs doublet \cite{Lee,Branco,Wolf,Wein}. For definiteness we 
consider two-Higgs doublet extensions with natural flavour conservation at 
the tree level and explicit CP violation, both by  complex Yukawa coupling 
matrices, which lead to  a Kobayashi-Maskawa (KM) phase \cite{KM}, and by 
the tree-level Higgs potential $V(\Phi_1,\Phi_2)$ (see, e.g., \cite{Wein}).
As a consequence of the latter the three physical neutral scalar mass 
eigenstates $\varphi_{1,2,3}$ of these models are not CP eigenstates, i.e.
they couple to both scalar and pseudoscalar quark and lepton currents.
Using the same symbols  for the corresponding scalar fields we have
\begin{equation}
{\cal L}_{Y}=-(\sqrt{2}G_F)^{1/2} \sum_{j,f} m_f(a_{jf} \bar{f}f +
\tilde{a}_{jf} \bar{f}i\gamma_5 f)\,\varphi_j \,\, , 
\label{lagr}
\end{equation}
where $G_F$ is Fermi's constant, $f$ denotes a quark or lepton field  and 
$m_f$ its associated mass, and $a_{jf}, \tilde{a}_{jf}$ are the reduced 
scalar and pseudoscalar Yukawa couplings, respectively, which depend on the 
parameters of the scalar potential and on the type of model. For example, 
in the model where the Higgs doublet $\Phi_2 (\Phi_1)$ gives mass to the 
$u (d)$-type quarks (called model II in the literature) these couplings 
read \cite{BSP}
\bea
\label{coupl}
a_{ju} = d_{2j}/\sin\beta \,\, , \quad \tilde{a}_{ju} = -d_{3j}\cot\beta \,\, ,  \nn \\
a_{jd} = d_{1j}/\cos\beta \,\, , \quad \tilde{a}_{jd} = -d_{3j}\tan\beta \,\, ,
\eea
where $u,d$ labels the charge $2/3$ and $-1/3$ quarks, respectively, 
 $\tan\beta = v_2/v_1$ is the ratio of (the moduli of) vacuum expectation 
values of the two Higgs doublets, and $(d_{ij})$ is a $3\times 3$ orthogonal 
matrix that describes the mixing of the neutral scalar states.
In the model where one Higgs doublet gives mass to both the $u$- and $d$-type
quarks (called model I) one has $a_{jd}=a_{ju}$ and 
 $\tilde{a}_{jd}=\tilde{a}_{ju}$. If this doublet is $\Phi_2$
then $a_{ju},\tilde{a}_{ju}$ are the same as in (\ref{coupl}).
 (In the SM $a=1$, ${\tilde a}=0$.) 
At  the Born level only the $\rm{CP}=+1$ component of $\varphi$ couples to
$W^+W^-$ and to $ZZ$.
The couplings are given by the respective SM couplings times
the factor $g_{VV}=(d_{11}\cos\beta + d_{21}\sin\beta), V=W,Z.$
\par
In addition, the particle spectrum contains a charged Higgs boson $H^\pm$.
In models with natural flavour conservation at tree level, $H^\pm$ 
exchange transports
the KM phase and does not lead to  significant CP effects of the type discussed
below. Experimental data on $b\to s +\gamma$ yield, for type II models, a lower 
bound on the mass of $H^\pm$, which is $m_{H^+} > 200$ GeV  for $\tan\beta \leq 1$
\cite{BorzG}.
\par
The effects analysed below are  significant only if the reduced 
Yukawa couplings of $\varphi$ to the $t$ quark are not very much suppressed
as compared with those of the SM -- if suppressed at all. Within two-Higgs doublet 
extensions  one should have $\tan\beta$ of order 1  or smaller. On the other hand
the measured strength of neutral $K$ and $B$ meson mixing provides a lower 
bound on $\tan\beta$. From studies of the meson mixing amplitudes in these 
models \cite{Buras} one infers that $\tan\beta \gtrsim 0.3$. An important 
constraint on non-standard (Higgs sector) CP violation is provided by the 
experimental upper bounds of the electric dipole moments (EDMs) of the 
neutron \cite{neutr} and of the electron \cite{elect}. A  reanalysis 
of these moments within  two-Higgs doublet models was made in \cite{Hayashi},
assuming so-called maximal CP violation in the neutral Higgs sector, which in 
our parametrization may be defined by putting 
$|d_{11}|=|d_{21}|=|d_{31}|=1/\sqrt{3}$. From this study one deduces that
$|a_{jt}\tilde{a}_{jt}|\lesssim 2$ is compatible with these 
low-energy constraints. It should be noted that this 
bound is subject to  considerable theoretical uncertainties in the estimates 
of hadronic matrix elements that are needed in the computation of the neutron 
EDM \cite{Chemtob,CPrev}.
\par
In the next-to-minimal supersymmetric extension of the SM with two
$SU(2)$ Higgs doublets $\Phi_{1,2}$ and one gauge singlet $N$, the Higgs 
potential $V(\Phi_1,\Phi_2,N)$ can be CP-violating at tree level, leading 
again to neutral scalar mass eigenstates with undefined CP parity 
\cite{Matsuda}. As far as the minimal supersymmetric extension of 
the SM with soft supersymmetry breaking terms is concerned, it is well-known 
that the tree-level Higgs potential $V(\Phi_1,\Phi_2)$ is CP-invariant. 
Nevertheless, CP-violating phases in the soft SUSY-breaking terms can induce 
mixing of the $\rm{CP}=+1$ neutral scalar Higgs boson states $h,H$ with the
$\rm{CP}=-1$ state $A$ at the one-loop level. It was recently pointed out 
\cite{Pil} that in a large region of the parameter space of the MSSM,  
where the heavy states $H$ and $A$ are almost mass-degenerate at tree level, 
sizeable radiative mixing of $H$ and $A$ can be induced by CP-violating 
Yukawa couplings involving scalar quarks of the third generation.
\par
In the following we study, in a quite model-independent fashion,  
the effects on $t\bar t$ production at the LHC of a heavy  Higgs boson 
resonance $\varphi$  with mass 
 $m_{\varphi} > 300$ GeV, which is not a CP eigenstate. 
For definiteness we choose 
in section 5  a range of 
reduced scalar and pseudoscalar Yukawa couplings to $t$ quarks, denoted 
by $a,\tilde{a}$ in the following, between $0.3\leq |a|,|\tilde{a}|\leq 1$,
and a reduced coupling $|g_{VV}|\le 0.4$ to $W^+W^-$ and $ZZ$.
A straightforward exercise yields  that the chosen range 
of couplings is compatible with the empirical bound $\tan\beta \gtrsim 0.3$.
Larger values of $|a\tilde{a}|$ would, of course, increase the effects
computed in section 5. Our choice of values for $g_{VV}$ is somewhat arbitrary.
However, because the CP effects below decrease for $|g_{VV}|\to 1$,
we are interested in $\varphi$-boson couplings to $W^+W^-$ and $ZZ$
bosons that are smaller than the corresponding SM couplings.  
For instance this is the
case for $\varphi$ states with a substantial pseudoscalar component.  
\par
If $m_{\varphi} > 2 m_t$, the total $\varphi$ decay width is then given, to 
good approximation, by the sum of the widths of $\varphi$ decay into 
$t\bar{t}$, $W^+W^-$, and $ZZ$, which we compute in terms of $a,\tilde a$, 
and $g_{VV}$ (see Eq. (\ref{width})). The parameter
\be
\gamma_{CP} \equiv -a\tilde a
\ee
serves as a measure of Higgs sector CP violation.
\par
Finally a few words on why it is justified to  take into account below 
the exchange of only one $\varphi_j$ boson. One should recall that a 
necessary condition  of observable CP violation in the neutral Higgs 
sector is a non-degenerate mass spectrum of the neutral states. 
Even if the products 
of couplings  $a_{jt}\tilde{a}_{jt}$ are of the same order of magnitude
the CP effects  caused
by  $\varphi_j$  bosons with  mass markedly below $2m_t$ 
are significantly smaller \cite{Peskin,BeBra2} than the effects due to
heavier bosons which appear as
resonances in   $ g g\to t \bar t$.  
If one evaluates CP observables $\cal O$ on the whole $t\bar t$ sample,
 all $\varphi_j$ exchanges must be taken into account. Even if
some or all of the products 
$|a_{jt}\tilde{a}_{jt}|$ are of order 1,
considerable  cancellations among the contributions to $\<{\cal O}\>$
may nevertheless occur. This is  because of  the orthogonality of the mixing 
matrix $(d_{ij})$, $\sum_j d_{1j}d_{3j} = 0$.
However, what we have in mind are CP studies {\it after} one heavy Higgs 
boson (or several) would have been discovered, and seen as a resonance in 
the invariant $t\bar t$ mass distribution $d\sigma/dM_{t\bar t}$. 
It was shown in \cite{BFH}
that there is a statistically significant signal in this spectrum for
the above range of couplings. Then the CP observables proposed in section 4
are to be measured on data samples which will be selected by appropriately
chosen cuts on $M_{t\bar t}$ in the vicinity of $m_{\varphi}$. In this 
way the CP properties of this boson can be investigated experimentally.
\section{Matrix elements}
Next we investigate $t\bar t$ production in high-energy $p p$ collisions
by the main parton reactions $gg, q\bar{q}\to t\bar{t}\to 6f$. 
At the level of these reactions  the exchange of a Higgs boson $\varphi$ 
with couplings (\ref{lagr}) induces CP-odd $t$ and $\bar t$ spin-momentum 
correlations, which,  through $t$ and $\bar t$ decay, 
lead to characteristic angular correlations and asymmetries among the decay 
products of these quarks.  
\par
At high energy hadron colliders, such as the LHC, a  $\varphi$ boson with 
mass not more than about $600$ GeV will be produced dominantly by gluon-gluon 
fusion through a virtual top-quark loop. This holds true also for the 
multi-Higgs doublet extensions of the SM in the parameter space of interest 
to us, namely $\tan\beta$ of order 1. Then $gg\to \varphi$ 
production through $b$-quark loops can be safely neglected. Likewise, when 
considering (N)MSSM models, squark loops can be neglected in this production 
mechanism if squark masses are larger than $400$ GeV \cite{SSH}, which we shall 
assume. The next-to-leading QCD corrections for scalar and pseudoscalar 
Higgs-boson  production  were computed in \cite{SH} -- \cite{SH3} 
and were found to 
increase the cross section significantly. (For the  SM electroweak 
corrections to $t\bar t$ production, including $s$-channel Higgs boson
exchange, see \cite{been}.)
Here we study the signal 
 and the background   
amplitudes only to lowest order in the QCD coupling. (Computing the QCD corrections
to these amplitudes for $polarized$ $t$ and $\bar t$ quarks is a task
beyond  the scope of this paper.) This can be justified as  we do not
investigate cross sections but normalized distributions. More specifically,
non-zero expectation values of the  observables 
which we study in sections 4 and 5 cannot be generated 
by QCD corrections --  but by $\gamma_{CP} \neq 0$. 
\par
The $\varphi\to t\bar t$ decay channel is affected by the large non-resonant 
$t\bar t$ background. The amplitudes $gg\to \varphi \to t\bar t$ and 
$gg\to t\bar t$ interfere and produce at the parton level a characteristic 
peak-dip structure in a number of observables \cite{Gaemers,BeBra1,BeBra2,Dicus,BFH}
 if the $t \bar t$ invariant mass lies in the vicinity of the Higgs boson mass. 
In \cite{BFH} it was found that a heavy $\varphi$ boson of arbitrary CP nature 
yields -- for a range of  reduced Yukawa and vector boson couplings 
$a,{\tilde a}$ and  $g_{VV}$ -- a statistically significant signal in the 
 $t\bar t$ invariant mass distribution. Here we are interested in $\varphi$-induced 
$t$ and $\bar t$ spin polarization and spin-correlation effects which reveal 
the nature of the Yukawa couplings of $\varphi$.  
For this purpose we consider  the  $t\bar t$ production density matrices for 
the parton processes $\lambda\bar\lambda\rightarrow t\bar t$ 
($\lambda=q,g$). They are defined by
\begin{equation}
R^{(\lambda)}_{\alpha\alpha' ,\beta\beta'}= \frac{1}{n_{(\lambda)}}
\sum_{{{\rm\scriptscriptstyle colours} \atop 
{\rm\scriptscriptstyle initial}\;
{\rm\scriptscriptstyle spins} }}
\langle t_\alpha\bar t_\beta |{\cal T}|
\,\lambda\bar\lambda\,\rangle\;
\langle \,\lambda\bar\lambda\,|{\cal T}^\dagger|
t_{\alpha'}\bar t_{\beta'}\rangle \,\, ,
\end{equation}
where the factor $n_{(\lambda)}$ averages over spin and colour
of the initial state partons; $n_{(q)}=(2N_C)^2=36$,
$n_{(g)}=(2(N^2_C-1))^2=256$.
The matrices $R^{(\lambda)}$ are
of the form
\par
\hfill\parbox{13.4cm}{
\begin{eqnarray*}
R^{(\lambda)}_{\alpha\alpha',\beta\beta'}&=&
A^{(\lambda)}\delta_{\alpha\alpha'}\delta_{\beta\beta'}
+B^{(\lambda)}_{+i} \hat{k}_i(\sigma^i)_{\alpha\alpha'}
\delta_{\beta\beta'} + B^{(\lambda)}_{-i} \delta_{\alpha\alpha'}
(\sigma^i)_{\beta\beta'} \\
&&+C^{(\lambda)}_{ij}(\sigma^i)_{\alpha\alpha'}
(\sigma^j)_{\beta\beta'}  \,\, ,
\label{Rlam}
\end{eqnarray*} }
\hfill\parbox{0.8cm}{\begin{eqnarray}  \end{eqnarray} }
\par\noindent
where $\sigma^i$ are the Pauli matrices. Using rotational invariance the 
vectors $B^{(\lambda)}_{\pm i}$ and tensors $C^{(\lambda)}_{ij}$ can be 
further decomposed. A general discussion of the symmetry properties of 
these matrices and their decomposition in the $t$ and $\bar t$ spin spaces
is given in \cite{BeBra1,BeBra2}.
\par
As we are concerned with CP-violating Higgs-boson effects, we take 
into account the QCD Born amplitudes for $q {\bar q}, gg \to t\bar{t}$ 
and all $\varphi$ exchange terms proportional to $\alpha_s\gamma_{CP}$,
that is,  the CP-violating pieces of the one-loop self-energy, vertex, and box 
diagram  contributions. If $m_{\varphi}> 2m_t$
the $s$-channel $\varphi$-exchange diagram
 $gg \to\varphi \to  t\bar{t}$  is by far the most important $\varphi$ 
contribution. The spin density matrices $R^{(\lambda)}$ 
were computed in \cite{BeBra1,BeBra2} and these calculations were recently 
confirmed in \cite{Zhou}.
\par
Heavy $\varphi$ exchange leaves, for $\gamma_{CP}$ of order 1,
also a markedly larger signal in  our CP observables (given below and in 
the next section) 
 than CP-violating gluino exchange in $q\bar{q}, gg\to t\bar t$,
which was computed in \cite{Schmidt,Zhou}. Therefore we omit such
 effects below.
\par
At the level of the $t\bar t$ states the CP-violating interaction (\ref{lagr})
generates two types of CP-odd spin-momentum correlation terms in (\ref{Rlam}),
 namely
\begin{equation}
\kt\cdot(\sp - \sm) f_e(z) \,\, ,
\label{kabs}
\end{equation}
\begin{equation}
\ph\cdot(\sp - \sm) f_o(z) \,\, ,
\label{pabs}
\end{equation}
and 
\begin{equation}
\kt\cdot(\sp \times  \sm) h_e(z) \,\, ,
\label{kdisp}
\end{equation}
\begin{equation}
\ph\cdot(\sp \times  \sm) h_o(z) \,\, .
\label{pdisp}
\end{equation}
Here $\sp,\sm$ are the spin operators of $t$ and $\bar t$, and $\kt$ and 
$\hat{\bf p}$ are the unit vectors of the momenta of the top quark and of the 
initial parton $\lambda$, respectively, defined in the parton c.m.\ system. 
Here $\lambda=g,q,\bar{q}$ is the parton in the proton that moves by definition 
along the $+z$-direction. (Neglecting transverse parton momenta implies that 
$\hat{\bf p}$ is equal to the direction of the proton beam in the laboratory frame.)
The functions $f_e, h_e$ and $f_o, h_o$ denote even and odd functions of the 
scattering angle $z =\cos\theta = \ph\cdot\kt$. 
The expressions (\ref{kabs}) -- (\ref{pdisp}) constitute a complete
set of CP observables in the case at hand. Here the expectation value
of an observable $\cal O$ for the respective parton reaction 
is defined as $\<{\cal O}\>_{\lambda}=\int_{-1}^1 dz
{\rm Tr}(R^{(\lambda)}{\cal O})/N_{\lambda}$, where $N_{\lambda}$ is a 
normalization factor. 
The observables (\ref{kabs}), (\ref{pabs}) are CP-odd but 
T-even, i.e. do not change sign under a naive T transformation 
(reversal of momenta and spins, 
but no interchange of initial and final states), whereas  (\ref{kdisp}), (\ref{pdisp}) 
are CP-odd and T-odd. This implies that non-zero expectation values of (\ref{kabs}), 
(\ref{pabs}) require $\gamma_{CP} \neq 0$ and a non-zero absorptive part of the 
respective scattering amplitude, whereas  (\ref{kdisp}), (\ref{pdisp}) are 
``dispersive'' CP observables. (The CP asymmetry 
 $\Delta N_{LR} = [N(t_L{\bar t}_L) - N(t_R{\bar t}_R)]/({\rm all}\, t{\bar t})$
considered in \cite{Peskin} corresponds to the basic longitudinal polarization
asymmetry $\<\kt\cdot(\sm - \sp)\>$.) 
\par 
Furthermore it is worth  pointing out the following. One might naively think 
that the CP-odd and T-odd observable
\be
{\cal P}_{CP} \, = \, \nhat\cdot(\sp - \sm)\, w(z) \,\, ,
\label{ptrans}
\ee
where $\nhat$ is the unit vector corresponding to $\nv={\bf p}\times{\bf k}_t$
and $w$ is some  function of $z$, would also be of relevance here.
(For  $g g \to t\bar{t}$ Bose symmetry requires $w(z)$ to be an 
odd function.) It amounts to searching
for a difference in the normal components of the $t$
and $\bar t$ spin-polarizations.  However,
 $\<{\cal P}_{CP}\>_{\lambda} \neq 0$ requires
C- and CP-violating interactions -- but QCD and the Yukawa interaction
(\ref{lagr}) are C-invariant and thus do not generate terms of the
form (\ref{ptrans}) in the above density matrices. 
The quantity ${\cal P}_{CP}$ and the corresponding
observables involving the charged lepton momenta 
from semileptonic $t$ or $\bar t$ decay instead of the spin
vectors $\sp$ and $\sm$
(see section 4 for the analogous transcriptions of
(\ref{kabs}) and (\ref{kdisp})) may
nevertheless be used experimentally to 
check for C- and CP-violating interactions in $t\bar t$ production.
Exchange of $W$ and/or $Z$ bosons in conjunction with $\varphi$ exchange
leads to such effects, but they are very small.
(We note in passing
that for $e^+e^- \to (\gamma^*, Z^*) \to t\bar{t}$ the difference of the
normal polarizations of $t$ and $\bar t$ is a relevant CP observable
for tracing Higgs-sector CP violation \cite{BO}.) On the other hand,
absorptive parts due to QCD in the scattering amplitudes of the above
parton reactions generate equal normal polarizations of $t$ and $\bar t$,
i.e. lead to T-odd terms of the form
\be
 \nhat\cdot(\sp + \sm)\, w(z)
\label{pQCD}
\ee
in the respective density matrices. The QCD-induced normal polarizations were
computed in \cite{BBU,Goldstein}. It is also worth recalling the observation 
\cite{BeBra1} that interactions being   P- and CP-invariant cannot induce 
T-odd spin-spin correlations. 
\par
We have found that, for LHC energies, the spin-polarisation and spin-correlation
observables  (\ref{kabs}),(\ref{kdisp}) which involve the helicity
axis $\kt$ are more sensitive to a non-zero product $\gamma_{CP}$ of reduced Yukawa
couplings than the corresponding observables  (\ref{pabs}),(\ref{pdisp})
which involve the  beam axis $\ph$. Therefore we shall consider only
the former set of observables below. Moreover it was shown in \cite{corr,BFH}
that the CP-even spin-spin correlation observable $\sp\cdot\sm$ is also 
sensitive to the Yukawa couplings $a,\tilde{a}$
 and should therefore also be taken into 
account in these kind of investigations. 
\par
The $t$ and $\bar t$ quarks auto-analyse their spins by their parity-violating 
weak decays. We shall assume that $t\to Wb$ is the dominant decay mode, as 
predicted by the SM. It is well-known that in the SM the most efficient analyser 
of the $t$ spin is the charged lepton from subsequent $W$ decay.
Its spin analyser quality is more than twice as high as the $W$ or 
$b$ quark direction of flight.
\par
 We use the narrow width approximation for $t\bar t$ production and 
decay, which is justified because of $\Gamma_t/m_t,\, \Gamma_W/m_W \ll 1$ 
and because we are concerned only with  normalized distributions and 
expectation values of observables. Moreover we take the leptons 
and the light quarks, including the $b$ quark, to be massless. 
 If  the charged lepton from  semileptonic $t$ decay, $t\to \ell^+\nu_{\ell}b$,
acts as $t$-spin analyser then, integrating out the $b,\nu_{\ell}$ momenta,
the $t$ decay density matrix $\rho_{\alpha'\alpha}$ in the $t$ rest frame 
is of the form 
 $\rho_{\alpha'\alpha} = f(E_\ell) (\eins + \vec{\sigma}\cdot\lp)_{\alpha'\alpha}$,
where $\alpha',\alpha$ are $t$ spin indices, $\lp$
is the $\ell^+$ direction of flight, and $f$ is a function of the lepton energy
$E_\ell$ (see, e.g., \cite{BNOS,MB}). 
This expression, which holds to lowest order in the SM, is 
respected by QCD corrections to a high degree of accuracy \cite{Kuhn}.
(The $t$, respectively $\bar{t}$ rest frame 
is defined by performing a rotation-free Lorentz boost from the parton  
c.m.\ system. The parton c.m. frame can be obtained from the laboratory
frame by a rotation-free boost along the beam axis that depends on the
momentum fractions $x_{1,2}$ of the partons.
The $t$ and $\bar{t}$ rest frames defined in this way differ
by a Wigner rotation from the respective rest frames obtained 
by a direct rotation-free boost from the laboratory frame.)
\par
One may also choose  the $b$ quark -- or  the $W$ boson -- to be 
the spin-analyser
of the $t$ quark. This is an  obvious choice for 
 non-leptonic $t$ decay and we 
shall employ
it below\footnote{In principle one could do better by identifying the 
$u$ or $d$-type
quark from $W\to q{\bar q'}$ decay either by tagging methods or from decay 
distribution characteristics.}. In this case the $t$ decay density matrix is
given in the $t$ rest frame by $\rho_{\alpha'\alpha}
 = (\eins -\kappa\,\vec{\sigma} \cdot \qb)_{\alpha'\alpha}$,
 where the $t$ spin-analyser quality factor $\kappa =(1-2\mu)/
(1+2\mu)$ with $\mu=m_W^2/m_t^2.$
If the reconstructed $W$ direction of flight is used instead of $\qb$, the 
corresponding $t$ decay  density matrix is obtained from this expression 
by substituting $\qb \to -\qW$. Numerically, $\kappa \approx 0.41$, which shows 
that in these cases the $t$ spin analysing power is lower by more than a 
factor of two as compared to the charged lepton. The density matrices for 
$\bar t$ decay are obtained by substituting $\lp \to - \lm,
 \qb \to -\qbb, \qW \to -\qWb$ in the above expressions.
\par
Suffice it to mention that SM CP violation in $t\to W b$ is tiny \cite{MB}, as in 
$t\bar{t}$ production. 
There is no effect to one-loop 
approximation in the amplitudes $q{\bar q}, gg\to t\bar t$.
What about CP effects in $t\to W b$ within the models discussed 
in the previous section? Although 
neutral $\varphi$ exchange induces a CP-violating 
form factor\footnote{For a discussion of the symmetry properties of the form 
factors that can appear in this decay mode, see \cite{BNOS,MB}.}  
in the $t\to Wb$ vertex at the one-loop order,
 the resulting  CP effect in top-quark decay 
is  only a few per mille \cite{GG} and thus markedly
smaller than the effects in $t\bar t$ production (see section 5).
 Within the (N)MSSM the form factors that are 
induced by CP-violating gluino \cite{BO}, neutralino, and chargino 
exchanges \cite{Bar} 
also lead to CP effects at the per mille level only. We shall omit these
effects below. (A more specific 
discussion is
given in section 4.)
\par
Besides $t\to W b$ also other decay modes/mechanisms may
be relevant to top quark decay; for instance $t\to H^+b$, respectively
 $ t\to b\tau\nu_{\tau}, bq\bar{q}'$ mediated by virtual $H^+$ exchange.
In non-supersymmetric $n$-Higgs doublet models ($n\ge 2$) of type II, the lower bound
 on the mass of $H^+$ quoted above implies that 
in  these models the $H^+$ mediated part of  the decay amplitude 
${\cal T}(t\to f)$ will be small with respect to the dominant part from 
(on-shell) $W^+$ exchange. (Here $f=b\ell\nu_{\ell},bq\bar{q}'$.)
However, charged Higgs bosons in type I or in SUSY models are not restricted
by these data \cite{BorzG,Sola}. Also a direct search made by the
CDF collaboration at the Tevatron
\cite{CDF} does not yet severely constrain the  decay mode $t\to H^+b$.
\par
In any case, our analysis below is set up in a modular way and can always be 
straightforwardly extended if  significant top decay modes and/or decay 
mechanisms other than $t\to Wb$ should be discovered. Moreover, in the 
next section we define observables that are mainly sensitive to
 CP violation in  
the $t\bar t$ production amplitude.
\par
In summary, the squared matrix elements of the reactions $\lambda\bar{\lambda}
\to t\bar t\to f_1 f_2$, which we use below,  are  of the form
\begin{equation}
{\rm Tr}\;[\rho^{(f_1)} R^{(\lambda)}\rho^{(f_2)}]\equiv\rho^{(f_1)}_{\alpha'\alpha}
R^{(\lambda)}_{\alpha\alpha',\beta\beta'}{\rho}^{(f_2)}_{\beta'\beta} \,\, ,
\label{trace}
\end{equation}
with production and decay density matrices as discussed above. At least as far 
as Higgs bosons with mass $m_{\varphi} \gtrsim$ 300 GeV and reduced Yukawa
couplings $a,\tilde a$ of order 1 in 
simple two-Higgs doublet models and the (N)MSSM are  concerned, the most 
significant CP-violating contribution to this expression  
 comes from  (quasi-) resonant $\varphi$ $s$-channel exchange  
 and resides in $R^{(g)}$.  
\section{Observables}
In the following  we consider two types of $t\bar t$ decay 
channels: first the ``dilepton + jets''  channels,  where both $t$ and 
$\bar t$ decay semileptonically,
\begin{equation}
t+\bar t \to W^+  b + W^- {\bar b} \to \ell^+
\nu_{\ell} b +  \ell'^- \bar{\nu}_{\ell'}\bar b \,\,.
\label{dilepton}
\end{equation}
As mentioned above, the directions of flight of the charged leptons 
are the best analysers of the $t, \bar t$ spins.
\par
Secondly we study  the ``lepton + jets'' channels, where the 
$t$ quark decays semileptonically and the $\bar t$ quark non-leptonically and
 vice versa.  These channels also have  a  good signature for top-quark 
identification and they are suitable for determining the $t \bar t$ 
invariant-mass spectrum. Events where the top quark decays semileptonically 
and the top antiquark decays hadronically will be called sample ${\cal A}$:
\be
{\cal A}:\,\left\{\,\,
\begin{array}{l}
t\to W^+ b\to \ell^+ \nu_{\ell}b \,\, , \\ 
\bar{t}\to W^- \bar{b}\to q \bar{q}'\bar{b} \,\, ,
\end{array}\right.
\label{sampleA}
\ee 
while  ${\bar{\cal A}}$ will denote the sample that consists of the charge-conjugated 
decay channels of the $t\bar{t}$ pairs:
\be
{\bar{\cal A}}:\,\left\{\,\,
\begin{array}{l}
t\to  W^+ b\to \bar{q} q' b \,, \\ 
\bar{t}\to  W^- \bar{b}\to \ell^- \bar{\nu}_{\ell} \bar{b} \,\, .
\end{array}\right.
\label{sampleAb}
\ee
In (\ref{dilepton}) -- (\ref{sampleAb})
 we take into account only semileptonic top decays into  either an 
electron or a muon. Then the SM predicts, to a good approximation,
a fraction of $24/81$ of all $t\bar t$ events to decay into the 
single lepton channels and $4/81$ into the dilepton channels.
\par
For the dilepton channel  the cross section measure  reads in the narrow 
width approximation (\ref{trace}):
\begin{eqnarray}
\label{dsigma}
%\int d\sigma \!\!\!\!\! & ( &\!\!\!\!\!\!pp \to t\bar t + X
\lefteqn{\int d\sigma (pp \to t{\bar t} X
\to {\ell^+} \nu_{\ell}  b + {\ell'^-} \bar{\nu}_{\ell'} {\bar b}+X) = }  \nonumber \\
&& {\cal N}\sum_{\lambda=q,\bar q,g} 
 \int_0^1 \!dx_1 \int_0^1 \!dx_2 \; N_\lambda(x_1)
   N_{\bar \lambda}(x_2) \Theta(\hat{s}-4m_t^2) \nonumber \\
&& \times \;
   \frac{\beta}{(4\pi)^2\hat s}\int d\Omega_{\kt}\;
   \frac{1}{\eta}\int_\mu^1 dy_+ y_+(1-y_+)
   \frac{1}{\eta}\int_\mu^1 dy_- y_-(1-y_-) \nonumber \\
&& \times 
   \int\frac{d\Omega_{\lp}}{4\pi}
   \int\frac{d\Omega_{\lm}}{4\pi}
   \Bigg\{ A^{(\lambda)}
   + B^{(\lambda)}_{+i}{\hat q}_{+i}
    - B^{(\lambda)}_{-i}{\hat q}_{-i} 
          - C^{(\lambda)}_{ij}{\hat q}_{+i}{\hat q}_{-j}\Bigg\}  \\
&& \times 
   \int_0^\infty dE_b\;
   \delta\left(E_b-\frac{m_t^2\!-\!m_W^2}{2m_t}\right) 
   \int\frac{d\Omega_{\qb}}{2\pi}\;
   \delta\left(\lp \!\!\!\cdot\!\qb -
   \frac{2\mu -y_+(1+\mu)}{y_+(1-\mu)}\right) \nonumber \\
&& \times \int_0^\infty dE_{\bar b}\;\delta\left(E_{\bar b}-
   \frac{m_t^2\!-\!m_W^2}{2m_t}\right) 
   \int\frac{d\Omega_{\qbb}}{2\pi}\;
   \delta\left(\lm \!\!\!\cdot\!\qbb -
   \frac{2\mu -y_-(1+\mu)}{y_-(1-\mu)}\right) \,\,. \nn
\end{eqnarray}
In the first line on the r.h.s. ${\cal N}=BR(t\to b\ell^+\nu_{\ell})
BR(\bar{t}\to\bar{b}\ell^-\bar{\nu}_{\ell})$
is the product of the semileptonic $t$ and $\bar t$ branching 
ratios, and $N_\lambda(x_1)$, $N_{\bar \lambda}(x_2)$ are the
parton distribution functions. The next two lines in 
Eq.\ (\ref{dsigma}) represent (up to the factor ${\cal N}$) 
the cross section $\int d\hat{\sigma}^{(\lambda)}$ for the partonic 
subprocesses 
$\lambda \bar{\lambda}\to t\bar{t}\to {\ell^+}
\nu_{\ell}b\,\, {\ell'^-}\bar{\nu}_{\ell'}{\bar b} $. 
The lepton momentum directions $\hat{\bf q}_\pm$
and the normalized lepton energies $y_\pm=2E_\pm/m_t$ are
defined in the $t$ and  $\bar t$ rest systems,
respectively. The minimal value of the normalized lepton 
energies is $\mu=m_W^2/m_t^2$.
Furthermore $\beta=(1-4m_t^2/\hat s)^{1/2}$ with the parton 
c.m.\ energy $\hat{s}=x_1x_2s$, and $\eta=(1-\mu)^2(1+2\mu)/6$.
The coefficients $A^{(\lambda)}$, $B^{(\lambda)}_{\pm i}$, and 
 $C^{(\lambda)}_{ij}$ are given\footnote{There
are some  misprints in the appendix of  \cite{BeBra2}:
in the formula for $\Gamma_Z$ in Eq. (A10)  
the denominator should read 32$\pi$ and the signs of $c_{g1}^{1(f)}, 
c_{g2}^{1(f)}$ in Eqs. (A21) and (A24) are to be changed.}
 in \cite{BeBra2}. 
\par
Eq.\ (\ref{dsigma}) has to be modified in an obvious way if  
phase-space cuts are imposed. The cuts must be CP-invariant
in order to avoid any bias in the evaluation of the expectation
values of the observables given below.
Without such cuts only $A^{(\lambda)}$ contributes to 
the rate, while the  coefficients $B^{(\lambda)}_{\pm i}$ and $C^{(\lambda)}_{ij}$ 
contain all the information about the  $t, \bar t$ spin-polarizations
and spin-spin correlations, respectively.
\par
Expectation values of observables ${\cal O}$ are defined as usual by
\be
\<{\cal O}\> \,  = \,  \frac{\int\! d\sigma \,{\cal O}}{\int\! d\sigma} \,\,.
\label{calo}
\ee
For the lepton + jets channels (\ref{sampleA}), (\ref{sampleAb}) a formula 
analogous to (\ref{dsigma}) holds. In this case the appropriate decay density 
matrices must be used in (\ref{trace}).
We note in passing that the cross section measure for $p\bar p$ collisions is 
simply obtained from Eq.\ (\ref{dsigma}) by substituting the parton distribution 
function $N_{\bar \lambda}(x_2)$ for $\bar{N}_{\bar\lambda}(x_2)$.
\par
Let us now define appropriate angular correlations and asymmetries
with which Higgs sector CP violation can
be traced in   the dilepton and single-lepton channels. Suffice
it to say that these correlations and asymmetries can of course
be used as tools
to search for CP violation in future data independent of any model.   
If both $t$ and $\bar t$ quark decay into semileptonic final states 
we consider the two observables
\bea
Q_{1} & = & \kt\cdot\lp - \ktb\cdot\lm \,\, ,
\label{qabs}
\eea
and
\bea
Q_{2} & = & (\kt - \ktb)\cdot(\lm\times\lp)/2 \,\, ,
\label{qdisp}
\eea
where $\kt,\, \ktb$ are the $t,\, \bar{t}$ momentum directions in the 
parton c.m.s.
and $\lp$,$\lm$ are the $\ell^+$,\, $\ell^-$ momentum directions
 in the $t$ and  $\bar{t}$ quark rest frames, respectively.
The channels $\ell^+,\, \ell'^-$ with $\ell, \ell' = e, \mu$ are summed over.
Obviously, (\ref{qabs}) and (\ref{qdisp}) are the transcriptions
of (\ref{kabs}) and (\ref{kdisp}) to the level of the final states,
i.e. (\ref{qabs}) and (\ref{qdisp}) serve as absorptive and dispersive
CP observables, respectively, taking account of the fact that the 
charged lepton is the most efficient analyser of the top-quark spin.
\par
The reconstruction of the $t$ and $\bar t$ directions of flight 
(up to ambiguities) in these channels is possible by solving kinematic
 constraints \cite{Aeppli}. 
Nevertheless, the measurement of these 
correlations  will  be a challenging task. 
Therefore, we define also corresponding asymmetries which should  be 
experimentally more robust than (\ref{qabs}), (\ref{qdisp}), because  only 
the signs of $Q_1,\, Q_2$ have to be measured, as follows
\bea
A(Q_{1}) & = & \frac{N_{\ell\ell}(Q_{1}>0) - N_{\ell\ell}(Q_{1}<0)}
                    {N_{\ell\ell}} \,\,,\nonumber \\
A(Q_{2}) & = & \frac{N_{\ell\ell}(Q_{2}>0) - N_{\ell\ell}(Q_{2}<0)}
                    {N_{\ell\ell}} \,\, .
\label{asq}
\eea
$N_{\ell\ell}$ is the number of $t\bar t$ events decaying into the 
dilepton + jets channels. 
If no phase-space cuts -- besides possible cuts on 
the $t\bar t$ invariant mass -- are imposed the following relations  can be derived 
between the asymmetries and  the expectation values of the corresponding
observables:
\bea
A(Q_{1}) & = & \quad \,\,\< Q_{1} \>_{\ell\ell}        \,\, , \nn \\
A(Q_{2}) & = & \frac{9\pi}{16} \< Q_{2} \>_{\ell\ell}  \,\, , 
\label{resa}
\eea
where the index $\ell\ell$ refers to the dilepton + jets sample.
For the lepton + jets channels we propose the following observables:
for  sample ${\cal A}$ we define 
\bea
O_{1} & = & \kt\cdot\lp             \,\, , \\
O_{2} & = & \kt\cdot(\lp\times\qbb) \,\, ,
\label{o12}
\eea
where $\qbb$ is the momentum direction of the
$\bar{b}$ quark jet in the $\bar{t}$ quark 
rest frame, while for the  sample $\bar{{\cal A}}$ we use  the charge-conjugated 
observables
\bea
\bar{O}_{1} & = & \ktb\cdot\lm            \,\, ,\\
\bar{O}_{2} & = & \ktb\cdot(\lm\times\qb) \,\, ,
\label{obar12}
\eea
with $\qb$ denoting the momentum direction of the  $b$ quark jet
 in the $t$ quark rest frame. 
In these channels the $t$ and  $\bar t$ momenta can be reconstructed
up to a twofold ambiguity \cite{Lad}.
Taking both samples one can define the quantities
\bea
{\cal E}_{1} & = & \<O_1\>_{\cal A} - \<\bar{O}_{1}\>_{\bar{{\cal A}}}\,\, ,
\label{cale1}
\eea 
\bea
{\cal E}_{2} & = & \<O_2\>_{\cal A} + \<\bar{O}_{2}\>_{\bar{{\cal A}}}\,\, .
\label{cale2}
\eea
Eq. (\ref{cale1}) is a transcription of the absorptive CP-odd
observable (\ref{kabs}). The dispersive spin-spin correlation
 observable (\ref{kdisp}) implies that in the non-leptonic
decay mode of the top quark either the $W$ boson or the $b$ quark jet
must act as $t$-spin analyser. In either case this costs
 a dilution factor $\kappa \approx 0.41$
(see section 3) -- although one gains in statistics as compared with the dilepton
channels.
For definiteness we have chosen  observables that involve the 
$b$ and $\bar b$ jet direction. If reconstruction of
the $W$ direction of flight in non-leptonic top quark decays should turn out
to be more efficient experimentally, then one should use observables with 
 $\qbb \to -\qWb$ and  $\qb  \to -\qW$  in (\ref{o12}),
(\ref{obar12}). The results given below apply to both cases.
\par
In addition we  define  corresponding asymmetries  as follows
\bea
A({\cal E}_{1}) & = & 
\frac{N_{\cal A}(O_{1}>0)-
N_{\cal A}(O_{1}<0)}{N_{\cal A}}
-
\frac{N_{\bar{\cal A}}(\bar{O}_{1}>0)-
N_{\bar{\cal A}}(\bar{O}_{1}<0)}{N_{\bar{\cal A}}} \,\,, \nonumber \\
A({\cal E}_{2}) & = & 
\frac{N_{\cal A}(O_{2}>0)-
N_{\cal A}(O_{2}<0)}{N_{\cal A}}
+
\frac{N_{\bar{\cal A}}(\bar{O}_{2}>0)-
N_{\bar{\cal A}}(\bar{O}_{2}<0)}{N_{\bar{\cal A}}} \,\,.
\label{ase}
\eea
Here $N_{\cal A}$ $(N_{\bar{\cal A}})$ is the number of $t\bar{t}$ events 
in sample ${\cal A}$ $(\bar{\cal A})$.
If no cuts -- besides  possible cuts on 
the $t\bar t$ invariant mass --  are imposed we derive  the
following  relations:
\bea
{\cal E}_{1}  & = & \quad \,\,  \<Q_{1}  \>_{\ell\ell} \,\, , \nn \\
{\cal E}_{2}  & = &  2\kappa\,  \<Q_{2}  \>_{\ell\ell} \,\, ,
\label{rele}
\eea
and
\bea
A({\cal E}_{1}) & = & \,\,\frac{3}{2} \,\,\, {\cal E}_{1} \,\,,  \nn\\
A({\cal E}_{2}) & = & \frac{9\pi}{16} \,     {\cal E}_{2} \,\,.
\label{rese}
\eea
Although an a priori classification of observables with respect to
CP cannot be made in the case at hand because the initial $p p$ state
is not a CP eigenstate, our observables and asymmetries
are, nevertheless, good indicators of CP violation for the reactions
$p p\to t\bar{t} X \to f_1 f_2 X$ above. The analysis of
possible contaminations from CP-invariant interactions which was made in
\cite{BeBra2} for CP observables that involve momenta
 defined in the laboratory frame can be applied in an analogous fashion
also to the above observables. Here  we  find that, within the parton model,
contributions from CP-invariant interactions 
to the reactions
$gg\to t\bar{t}\to f_1 f_2$ and $q\bar{q}\to t\bar{t}\to f_1 f_2$, 
 i.e. the CP-invariant part of the differential distributions
${\rm Tr}\;[\rho^{(f_1)} R^{(\lambda)}\rho^{(f_2)}]$,
cannot generate non-zero expectation values of our CP observables.
This statement holds to all orders in perturbation theory and 
can be shown in a straightforward fashion by writing down the
expectation values of $Q_1$,  $Q_2$, using the phase-space measure 
(\ref{dsigma}) and using the transformation properties with respect 
to CP of the coefficients of the production and decay density matrices
\cite{BeBra2,BNOS}. An analogous exercise can be performed for
${\cal E}_1$, ${\cal E}_2$. Kinematic cuts must be CP-invariant.
\par
Moreover, we remark that the T-odd observable (\ref{qdisp}) and the
quantity  ${\cal E}_2$ are predominantly sensitive to  CP violation 
in the $t\bar t$
production amplitude. Possible CP-violating form factors in the
amplitudes of the above $t$ and $\bar t$ decay channels do not contribute
to leading order in the couplings. In order to show this we first
note  that the above CP observables involve only the momentum of
one final-state particle $r$ ($r=\ell,b,W$) 
from $t$ and/or $\bar{t}$ decay. Then
the corresponding decay density matrices are of the form
\bea
\label{onep}
 \rho^{(f)}  & = & \quad  (\alpha_+\eins + \beta_+\vec{\sigma} \cdot \qr)
 \,\, , \nn \\ 
\rho^{(\bar{f})}
& = & \quad (\alpha_-\eins - \beta_- \vec{\sigma} \cdot \qbr) \,\, .
\eea
CP invariance implies that $\alpha_+ = \alpha_-$, $\beta_+ = \beta_-$.
Let us neglect, for a moment, contributions to (\ref{onep}) from
absorptive parts of the decay amplitudes. Then CPT invariance enforces the
same conditions on the coefficients $\alpha_\pm,\beta_\pm$ \cite{BNOS}.
 In other words,
there are no contributions to (\ref{onep})  from CP-violating dispersive terms 
in the decay amplitude if the interactions are CPT invariant. 
There can
be contributions to  (\ref{onep}) from CP-violating absorptive parts
to one-loop order. However, for CP-violating neutral $\varphi$ exchange they
are absent in the limit of vanishing $b$ quark mass. In supersymmetric 
models CP-violating gluino,  neutralino,
and chargino exchange leads to such absorptive parts in the $t\to W b$ amplitude
if $m_t > m_{\chi}
+ m_{\tilde q}$, where $m_{\chi}, m_{\tilde q}$ denote the masses of
a gluino or neutralino and scalar top quark, 
or those of a chargino and scalar
$b$ quark, respectively \cite{BO,Bar}.
\par
In the following $D=\rho^{(f)}\otimes\rho^{(\bar{f})}$ and the labels
$disp$ $(abs)$ and $I$ $(CP)$ denote the dispersive (absorptive) parts of the
CP-invariant (-violating) terms in the production and decay density
matrices, respectively. The T-odd observables $\<Q_2\>$, $A(Q_2)$,
${\cal E}_2$, and $A({\cal E}_2)$ are generated by the terms 
\be
\label{disp}
{\rm Tr} [R^{disp}_{CP}D^{disp}_I + R^{disp}_{I}D^{disp}_{CP} + 
R^{abs}_{CP}D^{abs}_I + R^{abs}_{I}D^{abs}_{CP}]
\ee
in the squared matrix elements (\ref{trace})
while the  T-even observables $\<Q_1\>$, $A(Q_1)$,
${\cal E}_1$, and $A({\cal E}_1)$ pick up the terms
\be
\label{abs}
{\rm Tr} [R^{abs}_{CP}D^{disp}_I + R^{abs}_{I}D^{disp}_{CP} + 
R^{disp}_{CP}D^{abs}_I + R^{disp}_{I}D^{abs}_{CP}] \,.
\ee
It remains to count powers of couplings in these expressions.  We distinguish
between two situations:\\
(a) For a light $\varphi$ boson
the terms $ R^{disp}_{I}, R^{abs}_{I}, 
R^{disp}_{CP}$, and $R^{abs}_{CP}$ are in our
normalization of the order $g_s^4, g_s^6/16\pi, g_s^4\lambda_{CP}/16\pi^2,$
 and $g_s^4\lambda_{CP}/16\pi$, respectively. Here $g_s$ denotes the
QCD coupling and $\lambda_{CP}$ is either  $\gamma_{CP}$ in the case of CP-violating
$\varphi$ exchange or ${\rm Im}(g_1g_2^*)$, where $g_1, g_2$
denote  couplings  associated with CP-violating  gluino, neutralino,
or chargino exchange.
 The terms
$D^{disp}_I$ and $D^{abs}_I$  are of order 1 and $g_s^2/16\pi$, respectively.
Further, $D^{disp}_{CP}=0$. As to  $D^{abs}_{CP}$, it is negligible
in the two-Higgs doublet models of section 2 and is of the 
order ${\rm Im}(g_1g_2^*)/16\pi$ in supersymmetric models.
Hence for T-odd observables the first term in (\ref{disp})
is the dominant one. Our  argumentation applies to any
other theory for which perturbation theory is applicable. This proves the above statement.
For  T-even observables the first and the last term in (\ref{abs})
are in general -- apart from the two-Higgs doublet models
in section 2 -- of the same order of magnitude. \\
(b) For a heavy $\varphi$ boson with mass 
$m_{\varphi} > 2m_t$, where the dominant contribution
to $R_{CP}$ in the resonance region $|\hat{s}- m_{\varphi}^2| \lesssim
m_\varphi\Gamma_{\varphi}$ comes from $gg\to\varphi\to t\bar{t}$, the situation
is the following.
Using the formula for $\Gamma_{\varphi}=\Gamma_{W}+\Gamma_{Z}+
\Gamma_{t}$ (see Eq.\ (\ref{width}) of the appendix) and assuming that the reduced
Yukawa couplings $a,\tilde a$ are of order 1, the inspection of the
$s$-channel contribution yields that now 
$R^{disp}_{CP}$ and $R^{abs}_{CP}$ are of order $g_s^4$ and $g_s^4/\pi$,
respectively, while
$R^{disp}_{I}, R^{abs}_{I}$ are of the same order as above. (Actually,
CP-invariant $s$-channel $\varphi$ exchange is  also  relevant to  these
terms.) In this case, both in (\ref{disp}) and
in (\ref{abs}), the first terms are the dominant ones. Thus for
a heavy $\varphi$ boson all the CP observables defined above are
predominantly sensitive to CP violation in $gg\to\varphi\to t\bar{t}$.
\par
It is clear that, especially in the case of resonant $\varphi$ production,
the sensitivity of the above observables to a non-zero product  
$\gamma_{CP}$ of Yukawa couplings can be enhanced considerably by judicious 
choices of cuts on the invariant $t\bar t$ 
mass $M_{t\bar t}=\sqrt{(k_t+k_{\bar t})^2}$. In fact such a cut is 
mandatory in view of the remarks made at the end of section 2. The spectrum
$d\sigma/dM_{t\bar t}$ is  obtained by multiplying the cross sections 
$\int d\hat{\sigma}^{(\lambda)}$ for the parton subprocesses
at ${\hat s}=M^2_{t\bar t}$ with the so-called luminosity functions
\be
L^{(\lambda)}(\tau)= {2\tau}
\int_\tau^{1/\tau}\frac{d\zeta}{\zeta}
N_{\lambda}(\tau\zeta)
N_{\bar \lambda}(\frac{\tau}{\zeta}) \,\,,
\label{lumi}
\ee
where $\tau=M_{t\bar t}/\sqrt{s}$. That is, one has
\be
\frac{d\sigma}{dM_{t\bar{t}}}=\frac{1}{\sqrt{s}}\sum\limits_{\lambda}
L^{(\lambda)}(M_{t\bar{t}}/\sqrt{s}) \, 
\int \!\! d\hat{\sigma}^{(\lambda)}(M_{t\bar{t}}^2) \,\,.
\label{dsigM}
\ee
Let us  define a ``differential expectation value'' $\exp{\cal O}$  
for a given invariant $t\bar t$ mass by 
\be
\exp{\cal O}(M_{t\bar{t}})\equiv \frac{\<{\cal O}\delta(\sqrt{\hat{s}}
-M_{t\bar{t}})\>} {\<\delta(\sqrt{\hat{s}}-M_{t\bar{t}})\>}
=\frac{\ds \sum\limits_{\lambda}L^{(\lambda)}(M_{t\bar{t}}/\sqrt{s})
\,\int\!\! d\hat{\sigma}^{(\lambda)}(M_{t\bar{t}}^2)\,{\cal O}}
{\ds \sum\limits_{\lambda}L^{(\lambda)}(M_{t\bar{t}}/\sqrt{s})
\int\!\! d\hat{\sigma}^{(\lambda)}(M_{t\bar{t}}^2)}
\,\,.
\label{diffexp}
\ee
These quantities, which we shall compute in the next section
for the observables  $Q_{1,2}$,  turn out to be very good indicators
of  how to choose appropriate $M_{t\bar t}$ mass bins for the evaluation
of the expectation values and asymmetries of the above observables.
\section{Results}
Throughout this section  we put 
$m_t=175$ GeV and $m_W=80.4$ GeV. In the computations 
we took the parton distribution functions (PDF) from 
 \cite{GRV}, evaluated at the factorization scale
$\Lambda=2m_t$, and convinced ourselves that our results do not 
change significantly if we vary $\Lambda$ 
or work with other PDF sets \cite{MRSCTEQ}.
Furthermore
 all sensitivity estimates are based on an LHC integrated luminosity
of 100 ${\rm fb}^{-1}$ at $\sqrt{s}=14$ TeV.
\par
Let us first assess the relative size of the  $gg\to\varphi\to t\bar t$ (which we call
resonant) and the remaining 
$\varphi$ contributions (which we call non-resonant)
to $t\bar t$ production. As already emphasized, all the interferences with
the non-resonant QCD background are included.
 For this purpose we have plotted in Figs. 1 and 2 
the resonant and all $\varphi$ contributions
to the differential expectation values (\ref{diffexp}) of $Q_1$ and $Q_2$ as a function
of the $t\bar t$ invariant mass $M_{t\bar t}$
for four different Higgs boson masses. 
 Figs. 1 and 2 illustrate the fact that for
$m_{\varphi}> 2 m_t$ the non-resonant contributions are
negligible with respect to  the $s$-channel $\varphi$ contribution. Therefore
we take the non-resonant terms into account only 
when evaluating observables for $m_{\varphi}\leq 2 m_t$. 
\par
In Figs. 3 -- 7 we show again the values of $\exp{Q_1}$ and $\exp{Q_2}$
as a function of $M_{t\bar t}$; this time for Higgs boson masses 
$m_{\varphi} \geq 2 m_t$ and for different values of the 
couplings $g_{VV},a,$ and $\tilde a$. 
Only the resonant $\varphi$ contributions are included.
The figures show the characteristic
peak-dip structure in the resonance region. The height and the width of the
peaks depend on the Higgs boson decay width $\Gamma_{\varphi}$ and on its Yukawa couplings
to top quarks. As usual the peaks get broader and less pronounced with increasing
$\Gamma_{\varphi}$. As expected, the height of a peak
is, for a given $\varphi$ mass and set of couplings,
larger for  $Q_2$ than for 
$Q_1$. Note that in the resonance region the sensitivity of these observables
to $\gamma_{CP}$ is high.
\par
These plots can be used to choose appropriate
$M_{t\bar t}$ mass bins
for the evaluation of the CP observables and asymmetries on the dilepton
and the lepton + jets samples.
The peak-dip structure of the signals in Figs. 3 -- 7
suggests to select, for a ``known''  Higgs boson mass, $M_{t\bar t}$
mass bins below 
(if kinematically possible) and above $m_{\varphi}$. Our choices are given
in Table 1 and will be used below.
\par
In addition  we shall 
apply further cuts, namely
\begin{equation}
|y(t)|\le 3\;,\quad\quad p_T\ge 20\,{\rm GeV}
\label{cuts}
\end{equation}
for the rapidities of the $t$ and $\bar t$ quarks  and for
the transverse
momenta in the laboratory frame
of the final-state  charged leptons and quarks in the dilepton and 
the lepton + jets samples.  
\par 
Implementing (\ref{cuts}) and the $M_{t\bar t}$ intervals of
Table 1 in the cross section measure (\ref{dsigma}) and in the analogous
measure for the lepton + jets channels, we have 
numerically computed $\<Q_1\>$, $\<Q_2\>$,
${\cal E}_1$, and ${\cal E}_2$.  In the same way we have computed the 
1 s.d. statistical errors $\delta {\cal O}\!=\![(\< {\cal O}^2 
\>\!-\!\< {\cal O} \>^2 )/N]^{1/2}$ of the CP observables  ${\cal O}$. 
Here $N$ is the number of events in the respective sample and 
$M_{t\bar t}$ interval
that pass the above cuts. 
We considered Higgs boson masses and reduced couplings 
in the ranges 320 GeV $\leq m_{\varphi} \leq$ 500 GeV, 0.3 
$\leq |a|,|{\tilde a}| \leq$ 1, $|g_{VV}| \leq$ 0.4. Without
loss of generality the signs of $a, \tilde{a}$
were chosen such that $\gamma_{CP} > 0$. 
\par
The results are given in Tables 2 -- 9. 
The non-resonant contributions have been included only for $m_\varphi=320$ GeV 
and $m_\varphi=350$ GeV. For each value 
of $m_{\varphi}, g_{VV}, a, \tilde a$ the number in the first column of the 
respective sub-table contains the expectation value
in percent and the second number is the statistical significance (in s.d.)
of the CP effect for  100 ${\rm fb}^{-1}$ of integrated LHC luminosity.
Tables 2 -- 5 show that the dispersive observable $Q_2$ has a higher sensitivity
to Higgs sector CP violation than the absorptive observable $Q_1$. 
The sensitivities of  ${\cal E}_1$
and ${\cal E}_2$ are, by and large, of the same order. (Recall that the
quality of ${\cal E}_2$ is diminished with respect to $\<Q_2\>$ because of the
limited $t$-spin analysing power of the $b$ quark jet.) 
\par
These investigations can be extended to the channels where both $t$ and $\bar t$
decay non-leptonically. In this case one may use observables which result 
from replacing $\lp,\lm \to \qb,\qbb$ in (\ref{qabs}), (\ref{qdisp}).
\par
In this context another relevant  observable is the $t\bar t$ spin-spin correlation
$\sp\cdot\sm$, which translates into 
\bea
Q_{3} & = & \lp\cdot\lm \,\, ,
\label{q3}
\eea
for the dilepton sample (\ref{dilepton}), where $\lp, \lm$ are the $\ell^+,\ell^-$
momenta as defined above. Although  $\sp\cdot\sm$ and (\ref{q3})
are  not proportional to $\gamma_{CP}$,
they are sensitive  to the Yukawa couplings  
$a,\tilde{a}$ \cite{corr} and should
therefore also be measured in future experiments. The expectation
value $\<Q_3\>$ was computed in terms of 
$a$ and $\tilde{a}$ in \cite{BFH}.
\par 
The amount of numerical work can be drastically simplified if
 no cuts besides the one on $M_{t\bar t}$  are imposed.
In  this case 
 we can derive  a formula that allows for a rather fast computation
of the expectation values $\<Q_{1,2}\>$  for the
dilepton samples. It is  given in the appendix  and  we  also give there
the widths of the distributions of the CP observables. 
We have found that the results for the expectation values and 
statistical sensitivities  obtained in this way differ only slightly
from the ones presented in Tables 2 -- 9. 
\par
The asymmetries $A(Q_{1,2})$ and 
the quantities  ${\cal E}_{1,2}$ and $A({\cal E}_{1,2})$ for the lepton + jets sample
are determined in this case by the general relations given in section 4.
With these formulae we can easily compute the CP asymmetries and the 
statistical 
sensitivities  $|\<Q_1\>|/\delta Q_1$, etc..
 By comparing these numbers we find that the sensitivities of
  $Q_1, Q_2, {\cal E}_{1}$,
and ${\cal E}_{2}$ are larger by about 20\% than those 
for $A(Q_{1}),A(Q_{2}), A({\cal E}_{1})$, and $A({\cal E}_{2})$, respectively.
 On the other hand, as mentioned, 
the asymmetries should be experimentally more robust.  
\par
With the above results  we conclude that for a large
range of $\varphi$ masses values of $|\gamma_{CP}|\gtrsim 0.1$ 
can be traced with the 
observables of section 4.  
As far as we can see it  is, in the foreseeable future,
a rather unique possibility of the LHC to make CP tests for  heavy Higgs 
bosons. In view of this opportunity analyses of hadronization and
detector effects on the sensitivities of these observables, which are beyond the
scope of this paper, would be worthwhile. 
\par 
We close with two remarks. In \cite{BeBra2}  
``experimentally simple"  CP observables involving  momenta in the laboratory
frame were studied. Moreover, no cuts on $M_{t\bar t}$ were made. Therefore
these observables are considerably less sensitive to $\gamma_{CP}$
than those of section 4. In \cite{Zhou} so-called optimal CP observables
-- which are essentially given by the CP-violating terms in the squared
matrix element divided by its CP-invariant part --  
were considered (without  $M_{t\bar t}$ cuts) and found to be sensitive  to
$|\gamma_{CP}|\gtrsim 0.1$.  While the construction and evaluation  of
optimal CP observables is straightforward theoretically \cite{Atopt},
their experimental usefulness in the case at hand remains to be seen: 
apart from depending on many kinematic variables they involve
also  model-dependent parameters, in particular particle masses, which
may be unknown. 
\section{Conclusions}
In several extensions of the SM 
the Higgs sector can, apart
from breaking the electroweak gauge symmetry,  also violate CP.
The  neutral Higgs bosons 
$\varphi$ which are predicted by these models then act 
also as messengers of the latter phenomenon. 
In this paper we have shown that top-quark pair production at the 
 LHC offers a good  possibility to investigate whether or not the interactions
of neutral Higgs bosons are CP-violating.
 The observables and 
asymmetries that we have proposed and investigated were found to
be very sensitive to the product of scalar and pseudoscalar
top-quark Yukawa couplings, namely  $|\gamma_{CP}|\gtrsim 0.1$.
 At least as far as heavy Higgs bosons
are concerned, these would  be rather unique CP tests in the  
foreseeable future.
%                                           
% Acknowledgments                           
%                                           
\subsubsection*{Acknowledgments}
We thank W. Hollik and H. Y. Zhou for discussions. W.B. wishes to thank the 
Theory Division at CERN for the hospitality extended to him.
\newpage
%                                           
% Appendix                                  
%                                           
\section*{Appendix}
In this appendix we give a compact formula for the
expectation values of the observables defined in section 4, which allows for
a rather quick evaluation.
\par
When no cuts  
are applied, the multiple phase-space integrals that appear
in (\ref{calo}) can be performed up to the two-dimensional integrals
over the momentum fractions $x_1, x_2$ of the colliding partons in the 
initial state. 
Here we consider a  neutral $\varphi$ boson with mass
$m_{\varphi} \geq 2m_t$, which is the most interesting case.
Then the CP-violating $\varphi$ contribution to $t\bar t$ production is 
dominated by the $s$-channel $gg \to\varphi\to t\bar t$ diagram, and
the other non-resonant $\varphi$ contributions of order $\gamma_{CP}$
to $gg, q\bar q\to t\bar t$ can be  neglected as  shown in section 5.
\par

\par
For the dilepton channels (\ref{dilepton}) we find for the expectation values
of the observables (\ref{qabs}), (\ref{qdisp}):
\be
\<Q_{1}  \>_{\ell\ell}  = \,\,\,\, \frac{2}{3}\,{\cal F}[B]\,\,,\quad
\<Q_{2}  \>_{\ell\ell} = \,\, \frac{2}{9}\,{\cal F}[C]\,\,,\quad
\label{resll}
\ee
where ${\cal F}[G]$ denotes the ratio of integrals
\bea
{\cal F}[G] &=& 
\frac{\ds\int_0^1\!\!dx_1\!\int_0^1\!\!dx_2\, \Theta(\hat{s}-4m_t^2)\,\, \beta/\hat s 
       \,\,
         \sum_\lambda N_{\lambda}(x_1)N_{\bar\lambda}(x_2)\, G^{(\lambda)}(\hat s)}
     {\ds\int_0^1\!\!dx_1\!\int_0^1\!\!dx_2\, \Theta(\hat{s}-4m_t^2) \,\, \beta/\hat s 
       \,\,
         \sum_\lambda N_{\lambda}(x_1)N_{\bar\lambda}(x_2)\, W^{(\lambda)}(\hat s)}
\, \nn\\
&=& \frac{\ds \int_{r}^1\!\!d\tau\, \beta/\hat{s} \sum\limits_{\lambda}
L^{(\lambda)}(\tau)G^{(\lambda)}(\hat{s})}
{\ds \int_{r}^1\!\!d\tau\, \beta/\hat{s} \sum\limits_{\lambda}
L^{(\lambda)}(\tau)W^{(\lambda)}(\hat{s})} \,\,.
\label{calf}
\eea
Here $N_{\lambda}(x_1)$, $N_{\bar\lambda}(x_2)$ denote the parton distribution
functions, $L^{(\lambda)}$ is the luminosity 
function defined in Eq.\ (\ref{lumi}),
$\tau=\sqrt{\hat{s}/s}$ and $r=2m_t/\sqrt{s}$. The functions
$B^{(\lambda)},C^{(\lambda)}$ and $W^{(\lambda)}$ are given below.
\par
If the expectation values  Eq. (\ref{resll}) 
are evaluated for events
in a given $M_{t\bar{t}}$ bin, $M_{t\bar{t}}=\sqrt{\hat{s}}\in 
[M_{\rm{low}},M_{\rm{high}}]$, the factor 
$\Theta(\sqrt{\hat{s}}-M_{\rm{low}})\Theta(M_{\rm{high}}-\sqrt{\hat{s}})$
has to be inserted in all the integrals of Eq.\ (\ref{calf}).
\par
The quantities ${\cal E}_{1,2}$ defined in Eqs.\ (\ref{cale1}), (\ref{cale2}) 
and the asymmetries defined in Eqs.\ (\ref{asq}) and (\ref{ase}) 
can then also be calculated with the  formulae (\ref{resll}) 
using the relations
(\ref{rele}), (\ref{resa}), and (\ref{rese}) respectively. 
\par
In order to estimate the statistical 
 sensitivity of the observables $Q_{1,2}$ to Higgs sector CP violation,
we need also the widths of the distributions of these observables, i.e.
the expectation values of their squares. We find that
\be
\<Q_{1}^2\>_{\ell\ell} = \frac{2}{3} - \frac{2}{9}\,{\cal F}[D]\,\,,\quad
\<Q_{2}^2\>_{\ell\ell} = \frac{2}{9}\,\, .
\ee
As usual, the widths are given by
\be
 \Delta Q_i =  \sqrt{\<Q_{i}^2\>_{\ell\ell}-\<Q_{i}\>_{\ell\ell}^2} \,\,.
\ee
For  the single lepton channels (\ref{sampleA}),
(\ref{sampleAb}) we find that 
\be
\<O_1^2\>_{\cal A} = \<\bar O_1^2\>_{\bar{\cal A}} = \frac{1}{3} \,\,,\quad
\<O_2^2\>_{\cal A} = \<\bar O_2^2\>_{\bar{\cal A}} = \frac{2}{9} \,\,,
\ee
and the widths are given by
$\Delta O_{i} = \left(\<O_{i}^2\>_{\cal A}-\<O_{i}\>_{\cal A}^2\right)^{1/2}$,
$\Delta {\bar O}_{i} = \left(\<{\bar O}_{i}^2\>_{\bar{\cal A}}
-\<\bar{O}_{i}\>_{\bar{\cal A}}^2\right)^{1/2}$.\\
Note that the above results also hold for individual bins in $M_{t\bar{t}}$.
\par
\noindent The functions $ G^{(\lambda)} \in 
\{B^{(\lambda)},C^{(\lambda)},D^{(\lambda)}\}$ and $W^{(\lambda)}$ read
\bea
W^{(q)} & = &\,\,\frac{2}{9}\, \left[ 1-\frac{1}{3}\beta^2\right]\,\, , \nn \\
W^{(g)} & = &\frac{1}{96}\left[ 31\beta^2-59-(33-18\beta^2+\beta^4) 
                \frac{\ln(\omega)}{\beta}\right] \nn\\
        &   & - \, K \,\,\frac{x^2}{16} \left[\beta^2a^2 {\rm Re}(d) 
                      +{\tilde a}^2 {\rm Re}(\tilde d)\right]
                \frac{\ln(\omega)}{\beta} \nn \\
        &   & + \, K^2 \,\frac{3x^2}{16} (\beta^2a^2+{\tilde a}^2) 
                     \left(a^2|d|^2+{\tilde a}^2|\tilde d|^2 \right) \,\, ,\nn \\
B^{(g)} & = & - \, K \, \frac{x^2}{16} \,\, a{\tilde a}\,\,\, 
                         {\rm Im}(d-\tilde d)\ln(\omega)\,\, , \nn\\
C^{(g)} & = & \,\,\,\,\, K \, \frac{x^2}{16} \,\, a\tilde{a} 
                \left[{\rm Re}(d+\tilde d) 
                \ln(\omega)-6K\beta  \left( a^2|d|^2 
                + {\tilde a}^2|\tilde d|^2 \right)\right] \,\, , \nn \\
D^{(q)} & = & \frac{2}{27}(1+\beta^2) \,\, ,\nn\\
D^{(g)} & = & \frac{1}{96\beta^2}\left[(\beta^6-17\beta^4+33\beta^2-33)
                    \frac{\ln(\omega)}{\beta} - 
                    (31\beta^4-37\beta^2+66) \right] \nn\\
        &   & + \, K \,\,\frac{x^2}{16} \left[\beta^2a^2 {\rm Re}(d) 
                      +{\tilde a}^2 {\rm Re}(\tilde d)\right]
                \frac{\ln(\omega)}{\beta} \nn \\
        &   & - \, K^2 \,\frac{3x^2}{16} (\beta^2a^2+{\tilde a}^2)
                      \left(a^2|d|^2+{\tilde a}^2|\tilde d|^2 \right) \,\, , \nn \\
B^{(q)} & = & C^{(q)} = 0 \,\,.
\eea
where $a$, $\tilde a$ are the reduced top quark
Yukawa couplings of Eq. (\ref{lagr}),
\be
\omega  =  \frac{1-\beta}{1+\beta} \,\, , \quad
\beta   =  \sqrt{1-x^2} \,\, , \quad
x       =  \frac{2m_t}{\sqrt{\hat{s}}} \,\, , \quad
{\hat s} = x_1 x_2 s \, ,
\ee
and we have used the abbreviations
\bea
K         & = & {\sqrt 2} G_F \left(\frac{m_{t}}{4\pi}\right)^2 \,\, , \nn\\
d         & = & (-2+\beta^2\hat{s}C_0)\frac{\hat s}{\hat s-m_\varphi^2
                 +im_\varphi\Gamma_\varphi}  \,\, , \nn \\ 
\tilde{d} & = & \hat sC_0\frac{\hat s}{\hat s-m_\varphi^2
                +im_\varphi\Gamma_\varphi} \,\, ,
\eea
with
\be
C_0 = \frac{1}{2\hat s}\left[
\ln(\omega)+i\pi\right]^2 \,\, ,
\ee
and $\Gamma_{\varphi}$ is the width of $\varphi$. As we work to lowest
order in the Yukawa couplings we have adopted the energy-independent
width approximation for the $\varphi$ propagator. 
The width $\Gamma_{\varphi}$ is the sum of partial widths for
$\varphi\to W^+W^-,ZZ,t\bar{t}$, i.e.,   
$\Gamma_{\varphi}= \Gamma_{W}+\Gamma_{Z}+\Gamma_{t}$ with
\bea
\Gamma_W &=& \Theta(m_{\varphi}-2m_W)\frac{g_{VV}^2\sqrt 2 G_F
m_{\varphi}^3\beta_W}{16\pi}
\left[\beta_W^2+12\frac{m_W^4}{m_{\varphi}^4}\right],\nonumber \\
\Gamma_Z &=& \Theta(m_{\varphi}-2m_Z)\frac{g_{VV}^2\sqrt 2 G_F
m_{\varphi}^3\beta_Z}{32\pi} 
\left[\beta_Z^2+12\frac{m_Z^4}{m_{\varphi}^4}\right],\nonumber \\
\Gamma_t &=& \Theta(m_{\varphi}-2m_t)\frac{3\sqrt 2 G_F
m_{\varphi}m_t^2\beta_t}{8\pi}
\left[\beta_t^2a^2+\tilde{a}^2\right].
\label{width}
\eea  
Here we have used the notation $\beta_{W,Z,t}=
\left(1-4m_{W,Z,t}^2/m_{\varphi}^2\right)^{1/2}$.
\newpage
%                                           
% References                                
%                                           

\newpage
%                                           
% ========================================= 
%                                           
% Figure Captions                           
%                                           
% ========================================= 
%                                           
\section*{Figure Captions}

{\bf Fig.\ 1:}
\parbox[t]{14.3cm}{
Differential expectation value of $Q_{1}$ (see (\ref{diffexp})) 
at $\sqrt{s}=14$ TeV for
$g_{VV}=0$, reduced Yukawa couplings $a=1$, $\tilde{a}=-1$, 
 and Higgs boson masses $m_{\varphi}=320$ GeV (a), $350$ GeV (b),
$400$ GeV (c), and $500$ GeV (d) in the dilepton channel. The dashed 
line represents the resonant and the solid line the sum of the
resonant and non-resonant $\varphi$ contributions to $\exp{Q_1}$.}
\\[6pt]
\noindent{\bf Fig.\ 2:}
\parbox[t]{14.3cm}{Same as Figure 1 for the observable $Q_2$.}
\\[6pt]
\noindent{\bf Fig.\ 3:}
\parbox[t]{14.3cm}{
Differential expectation value of $Q_{1}$ 
at $\sqrt{s}=14$ TeV
for different Higgs boson
masses and couplings $a$, $\tilde{a}$ and $g_{VV}=0$ in the dilepton channel. 
$m_{\varphi}=350$ GeV (solid line),
$m_{\varphi}=370$ GeV (dashed line),
$m_{\varphi}=400$ GeV (dotted line),
$m_{\varphi}=500$ GeV (dash-dotted line).
Figures a, b, c, d correspond to ($a,\, \tilde{a}$) = $(1,-1)$, $(1,-0.3)$, $(0.3,-1)$, 
$(0.3,-0.3)$. Only the resonant $\varphi$ contributions are shown.}
\\[6pt]
\noindent{\bf Fig.\ 4:}
\parbox[t]{14.3cm}{Same as Figure 3, but $g_{VV}=0.4$.}
\\[6pt]
\noindent{\bf Fig.\ 5:}
\parbox[t]{14.3cm}{Same as Figure 3, but observable $Q_2$.}
\\[6pt]
\noindent{\bf Fig.\ 6:}
\parbox[t]{14.3cm}{Same as Figure 5, but $g_{VV}=0.4$.}

\newpage
%                                           
% ========================================= 
%                                           
%  Table Captions                           
%                                           
% ========================================= 
%                                           
\section*{Table Captions}

\noindent{\bf Table\ 1:}
\parbox[t]{14.0cm}{
Selection of the $M_{t\bar{t}}$ interval below and above a Higgs boson
mass $m_{\varphi}$ in units of GeV.}
\\[6pt]
\noindent{\bf Table\ 2:}
\parbox[t]{14.0cm}{
The expectation value of $Q_1$ and its sensitivity  at $\sqrt{s}=14$ TeV
for the dilepton channels. The $M_{t\bar{t}}$ interval is
chosen below $m_{\varphi}$ as given in Table 1. For each pair 
($m_{\varphi}$, $g_{VV}$) the first column is $\<Q_{1}\>$ in percent and the 
second column is the sensitivity in s.d. The rows correspond, in descending 
order, to ($a,\, \tilde{a}$) = $(1,-1)$, $(1,-0.3)$, $(0.3,-1)$, $(0.3,-0.3)$.
Numbers for $m_{\varphi}$ are in GeV. The non-resonant contributions have been
neglected for these values of $m_\varphi$.}
\\[6pt]
\noindent{\bf Table\ 3:}
\parbox[t]{14.0cm}{Same as Table 2, but with $M_{t\bar{t}}$ interval above $m_{\varphi}$
as given in Table 1. For $m_\varphi=320$ GeV and $m_\varphi=350$ GeV the non-resonant
contributions have been included.}
\\[6pt]
\noindent{\bf Table\ 4:}
\parbox[t]{14.0cm}{
Same as Table 2 for the expectation value of $Q_2$.}
\\[6pt]
\noindent{\bf Table\ 5:}
\parbox[t]{14.0cm}{
Same as Table 3 for the expectation value of $Q_2$.}
\\[6pt]
\noindent{\bf Table\ 6:}
\parbox[t]{14.0cm}{
Same as Table 2 for the quantity ${\cal E}_1$ in the lepton + jets channels.}
\\[6pt]
\noindent{\bf Table\ 7:}
\parbox[t]{14.0cm}{
Same as Table 3 for the quantity ${\cal E}_1$in the lepton + jets channels.}
\\[6pt]
\noindent{\bf Table\ 8:}
\parbox[t]{14.0cm}{
Same as Table 2 for the quantity ${\cal E}_2$ in the lepton + jets channels.}
\\[6pt]
\noindent{\bf Table\ 9:}
\parbox[t]{14.0cm}{
Same as Table 3 for the quantity ${\cal E}_2$ in the lepton + jets channels.}

\clearpage
%                                           
% ========================================= 
%                                           
%  Figures                                  
%                                           
% ========================================= 
%                                           
\section*{Figures}
%                        
%  1. figure             
%                        
\begin{figure}[h]
\unitlength1.0cm
\begin{center}
\begin{picture}(13.6,13.)
%\put(1,13.4){\mbox{$m_\varphi=320$ GeV}}
%\put(8,13.4){\mbox{$m_\varphi=400$ GeV}}
%\put(1,6.4){\mbox{$m_\varphi=500$ GeV}}
\put(0,7){\psfig{figure=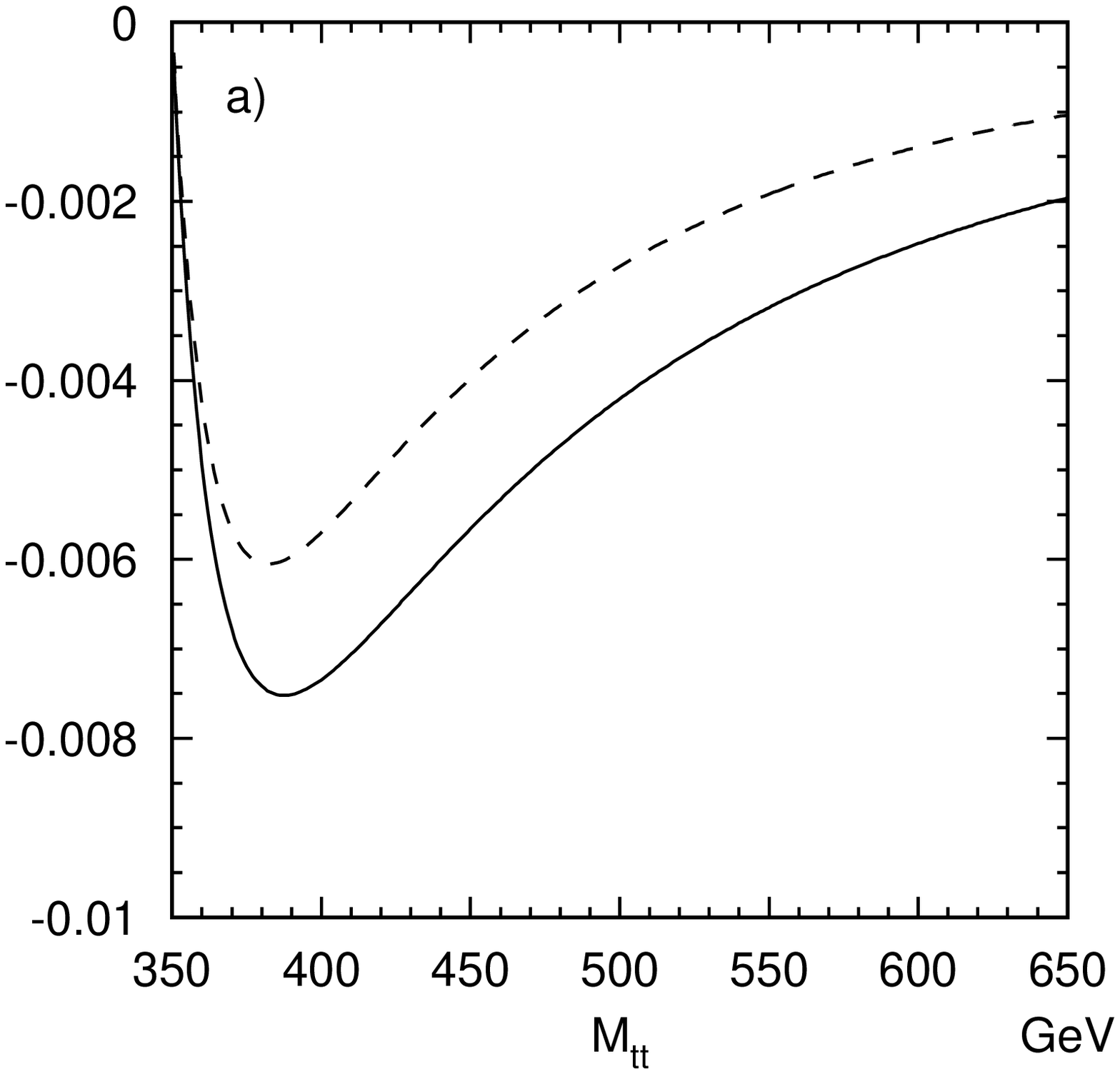,height=7cm,width=7cm}}
\put(7,7){\psfig{figure=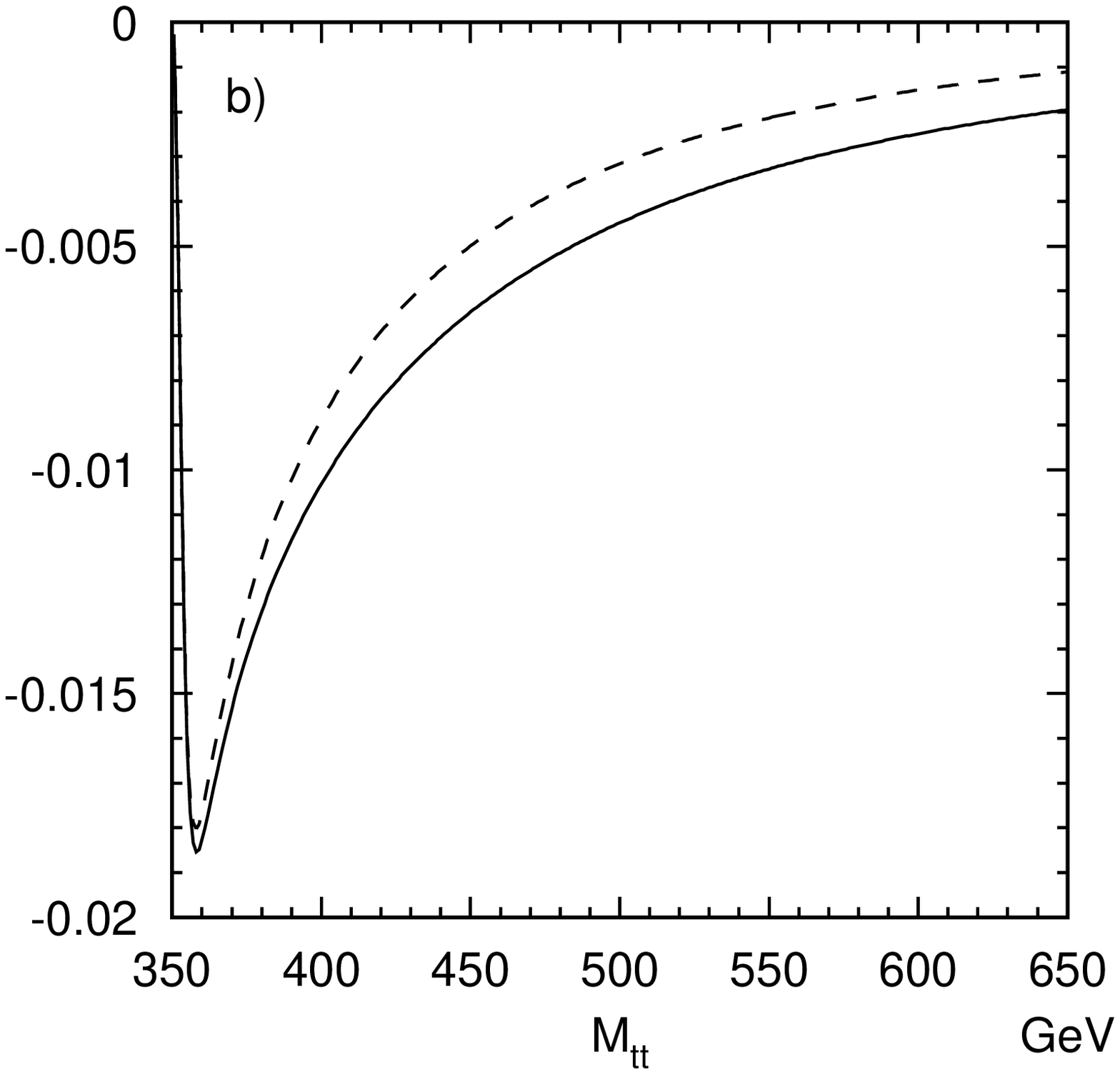,height=7cm,width=7cm}}
\put(0,0){\psfig{figure=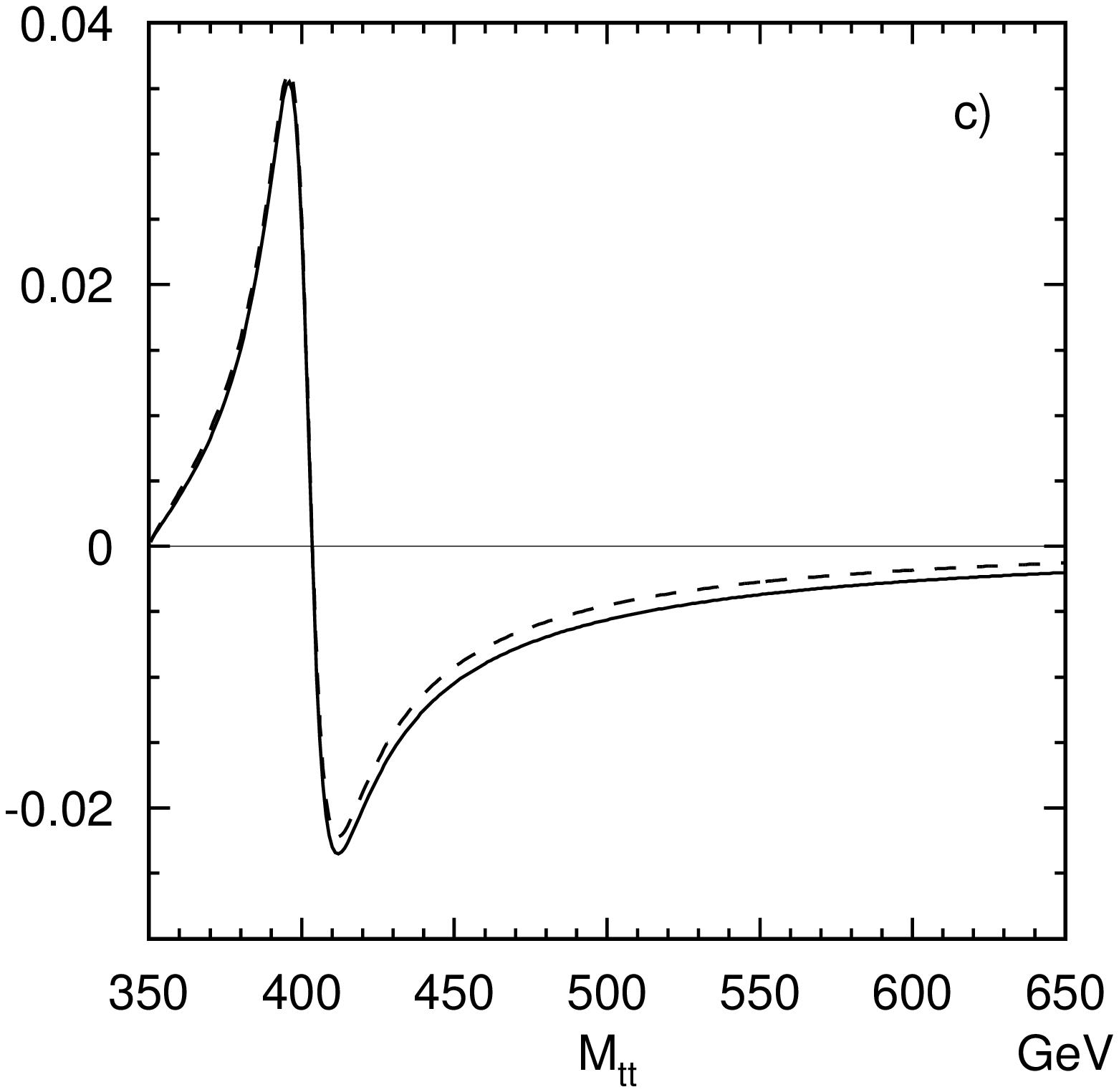,height=7cm,width=7cm}}
\put(7,0){\psfig{figure=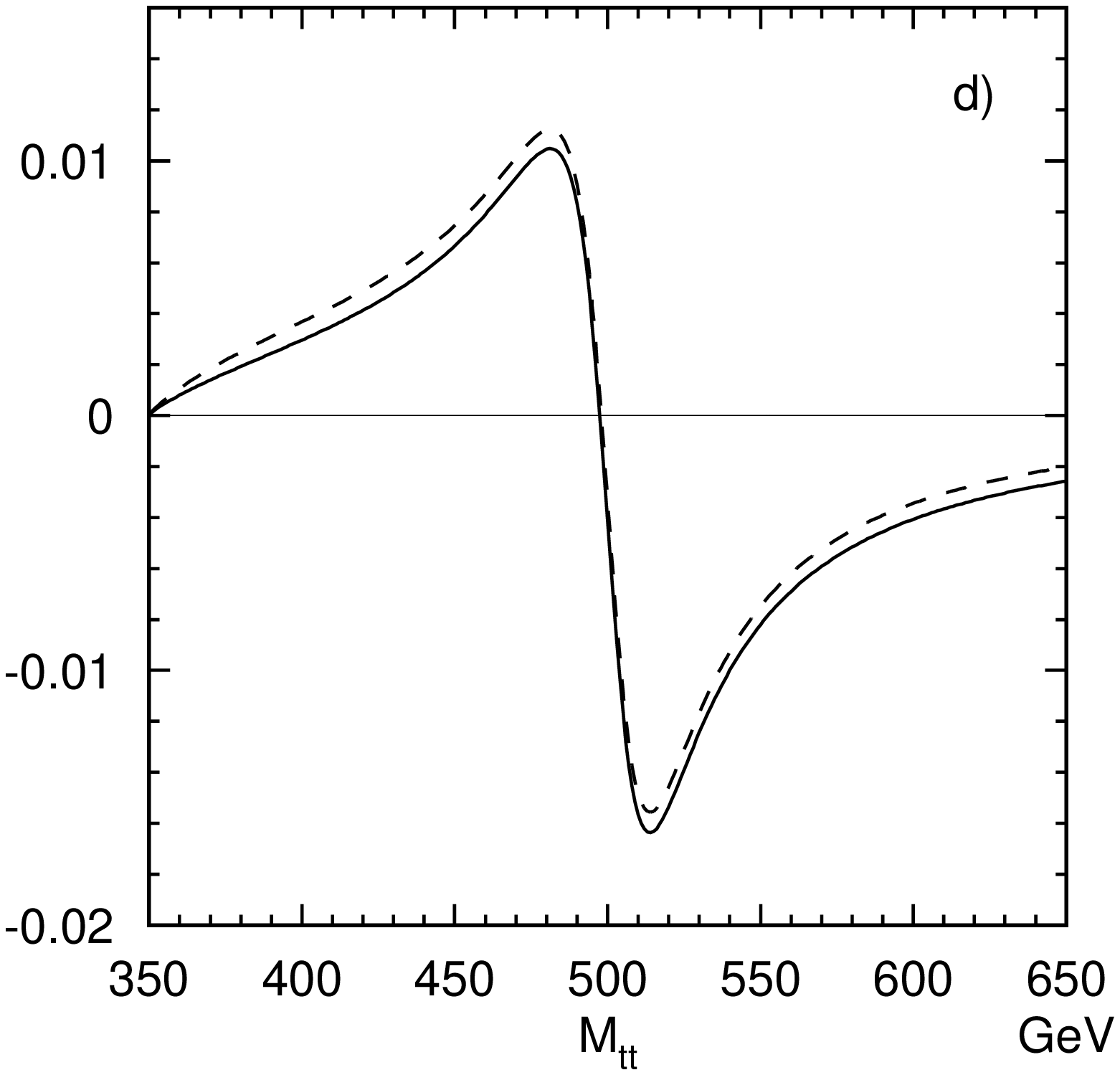,height=7cm,width=7cm}}
\end{picture}
\\[6pt]
Figure 1
\end{center}
\end{figure}
\newpage
%                        
%  2. figure             
%                        
\begin{figure}[h]
\unitlength1.0cm
\begin{center}
\begin{picture}(13.6,13.)
%\put(1,13.4){\mbox{$m_\varphi=320$ GeV}}
%\put(8,13.4){\mbox{$m_\varphi=400$ GeV}}
%\put(1,6.4){\mbox{$m_\varphi=500$ GeV}}
\put(0,7){\psfig{figure=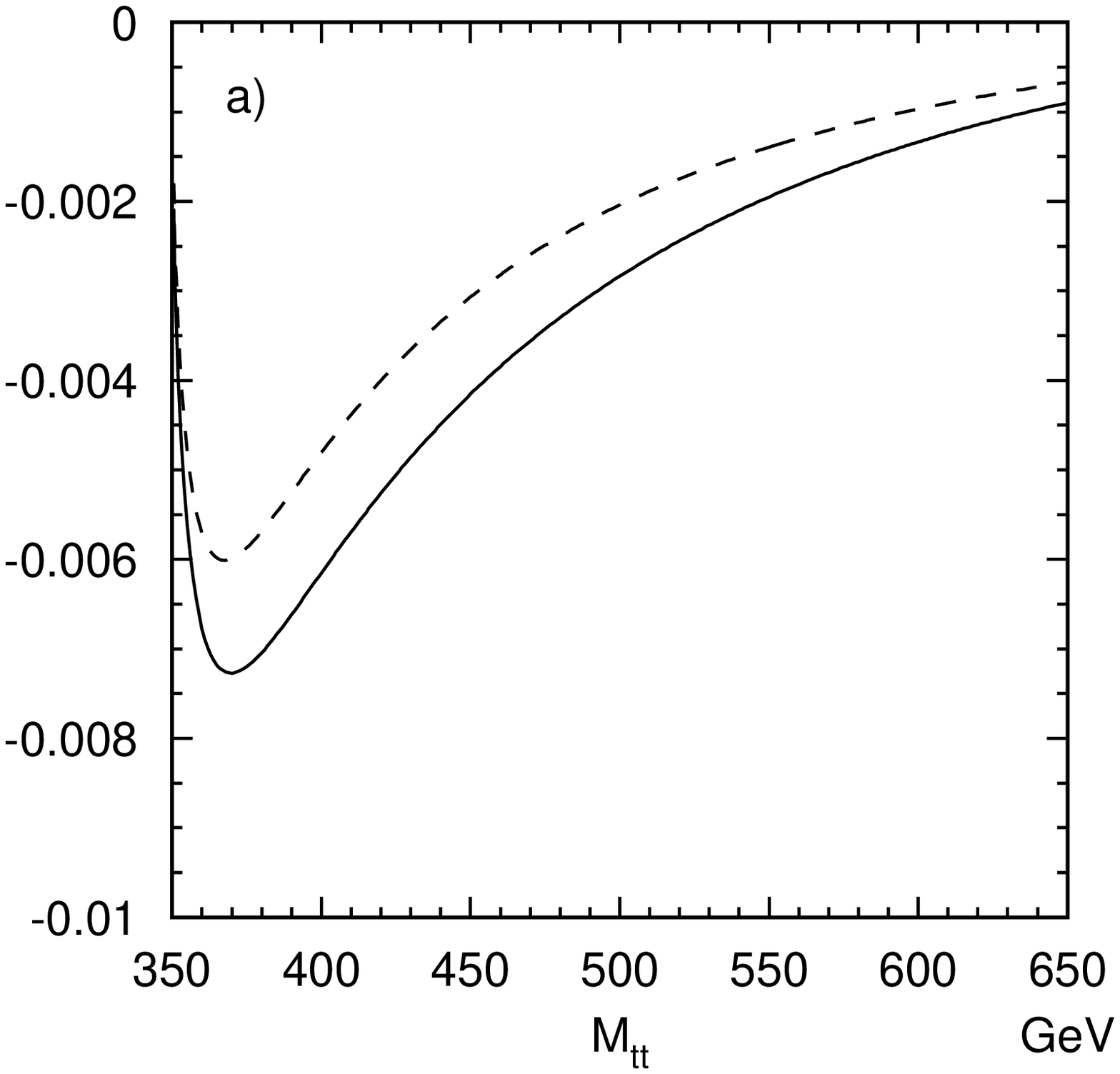,height=7cm,width=7cm}}
\put(7,7){\psfig{figure=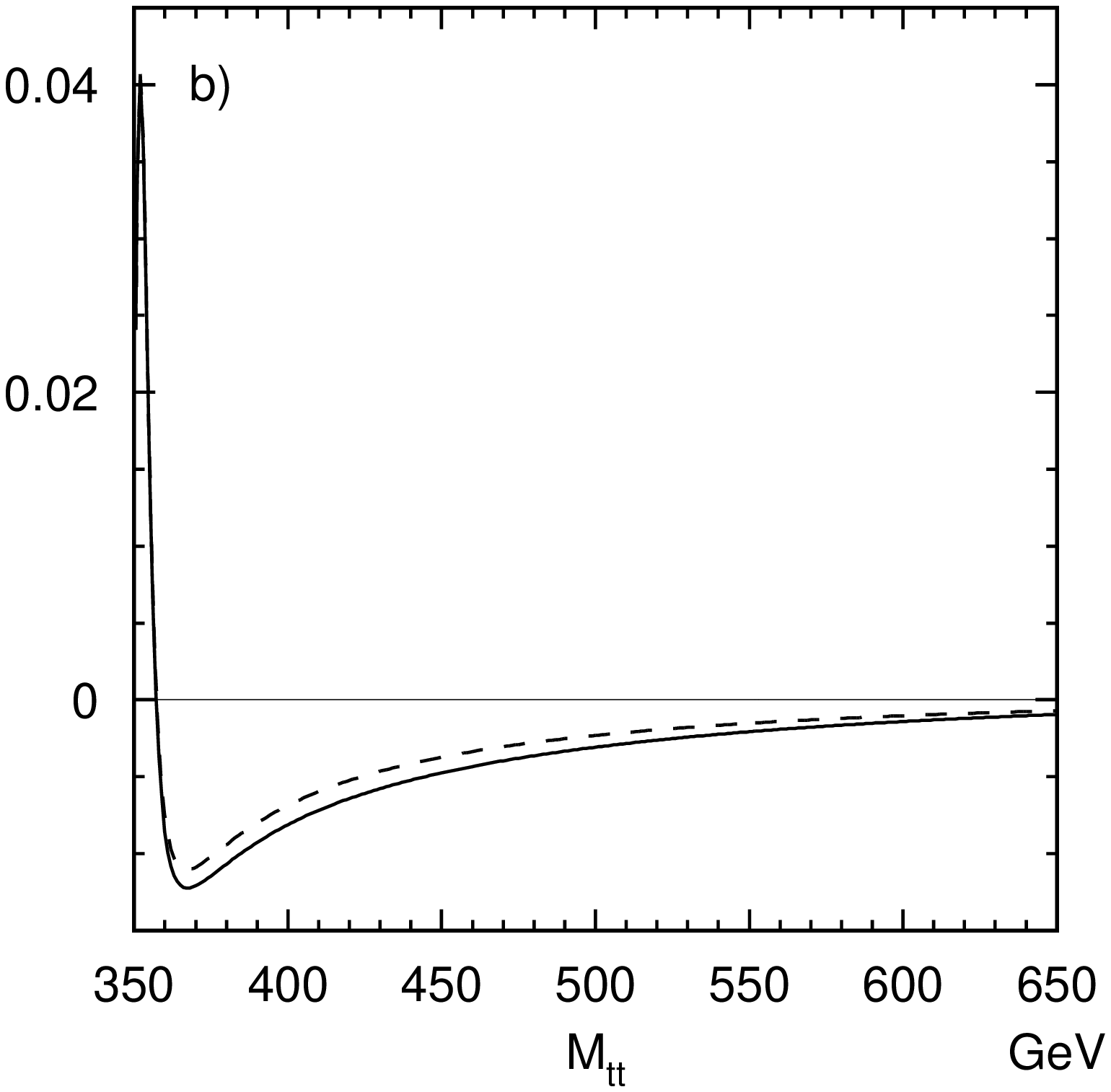,height=7cm,width=7cm}}
\put(0,0){\psfig{figure=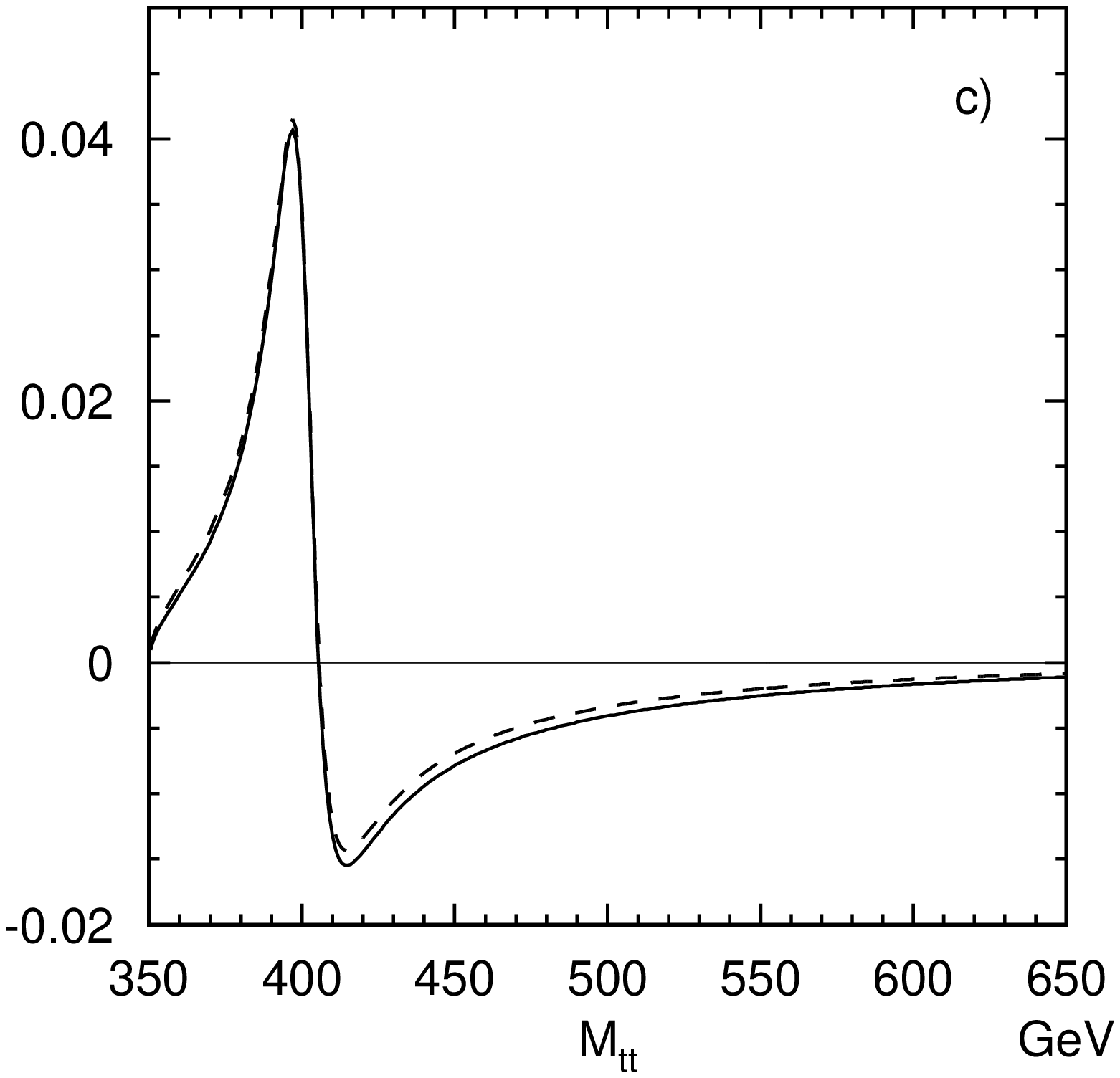,height=7cm,width=7cm}}
\put(7,0){\psfig{figure=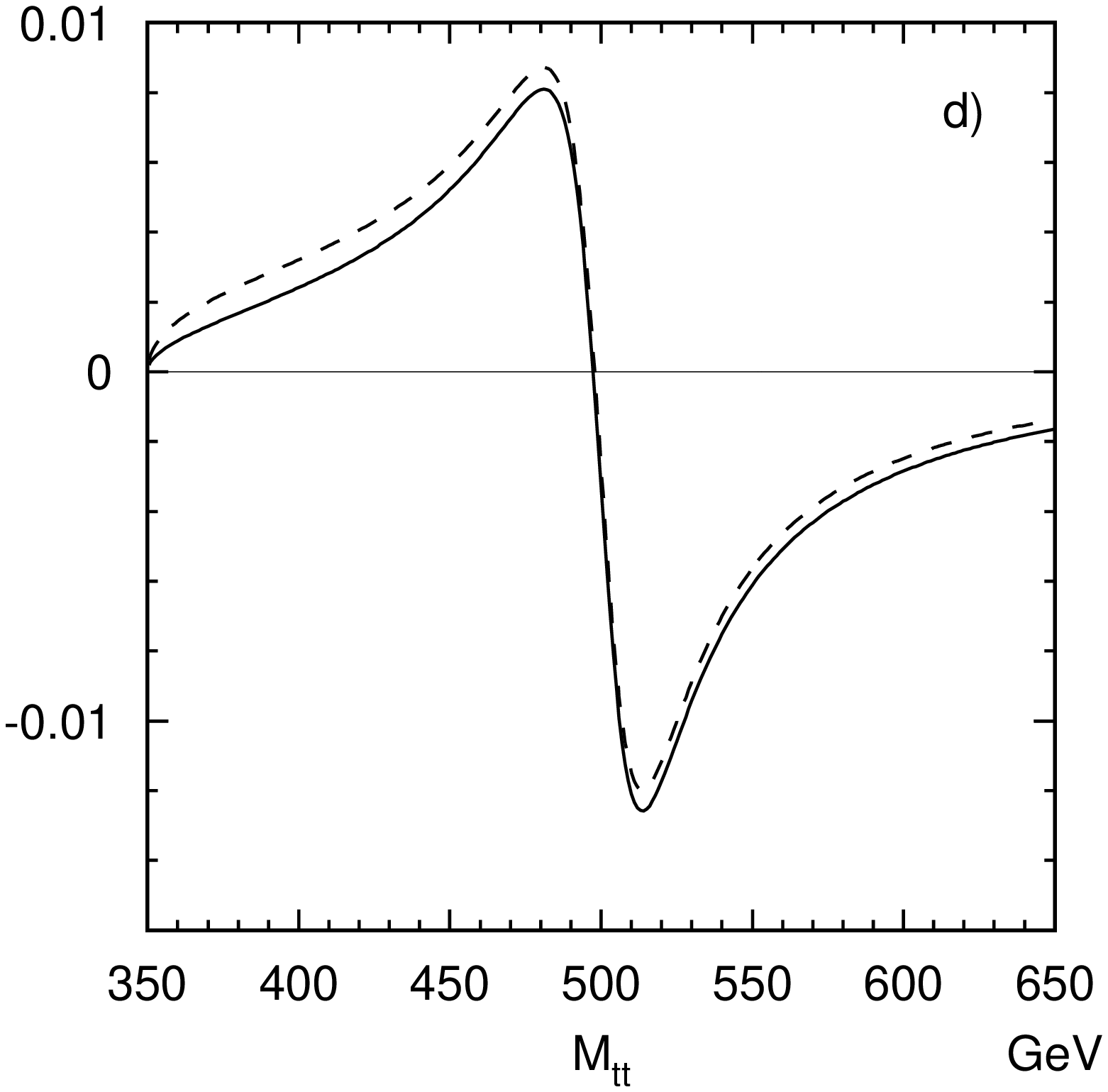,height=7cm,width=7cm}}
\end{picture}
\\[6pt]
Figure 2
\end{center}
\end{figure}
\newpage
%                        
%  3. figure             
%                        
\begin{figure}[h]
\unitlength1.0cm
\begin{center}
\begin{picture}(13.6,13.)
%\put(1,13.2){\mbox{$a=1,\,\,\tilde{a}=-1$}}
%\put(8,13.2){\mbox{$a=1,\,\,\tilde{a}=-0.3$}}
%\put(1,6.2){\mbox{$a=-0.3,\,\,\tilde{a}=-1$}}
%\put(8,6.2){\mbox{$a=-0.3,\,\,\tilde{a}=-0.3$}}
\put(0,7){\psfig{figure=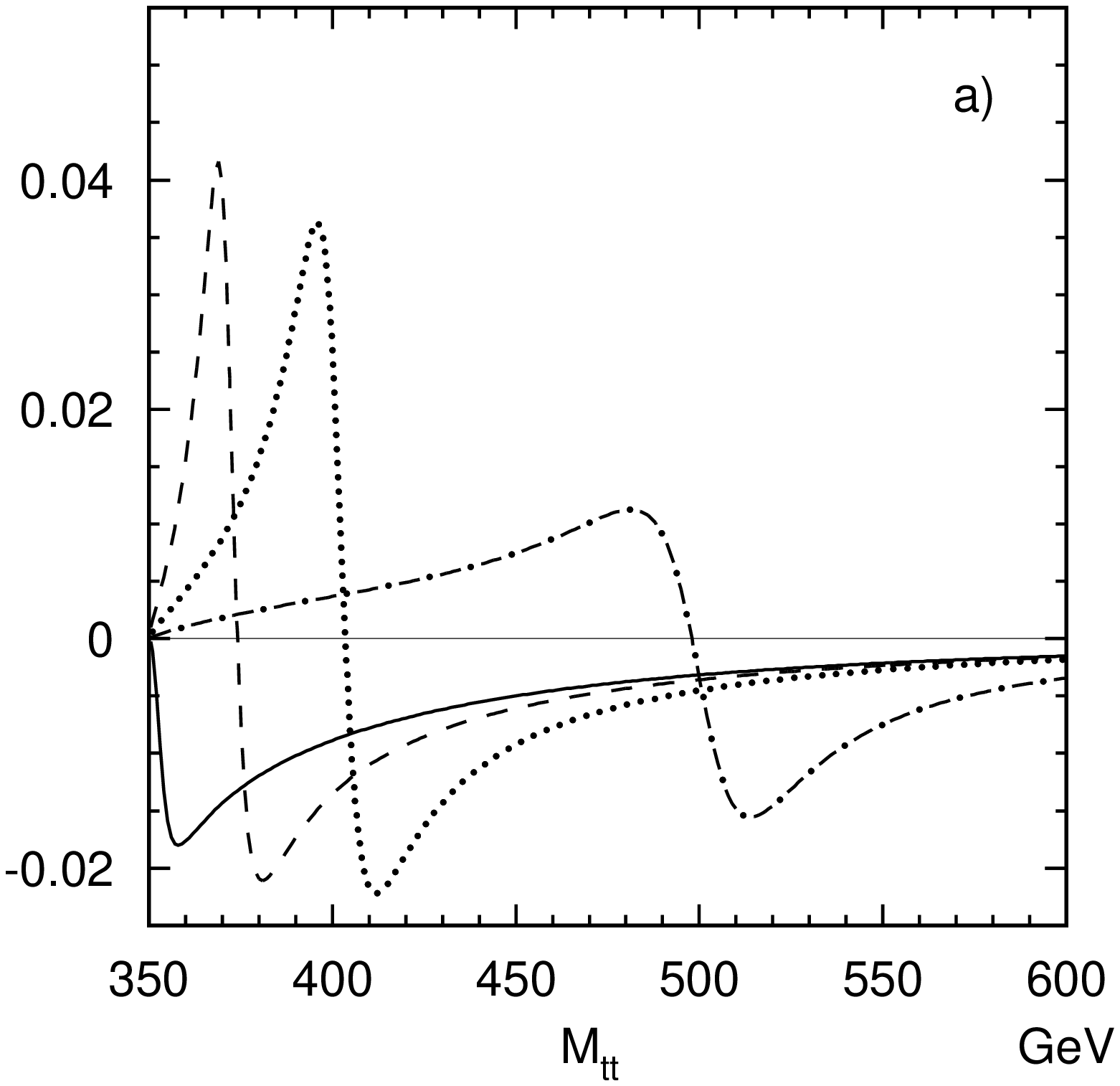,height=7cm,width=7cm}}
\put(7,7){\psfig{figure=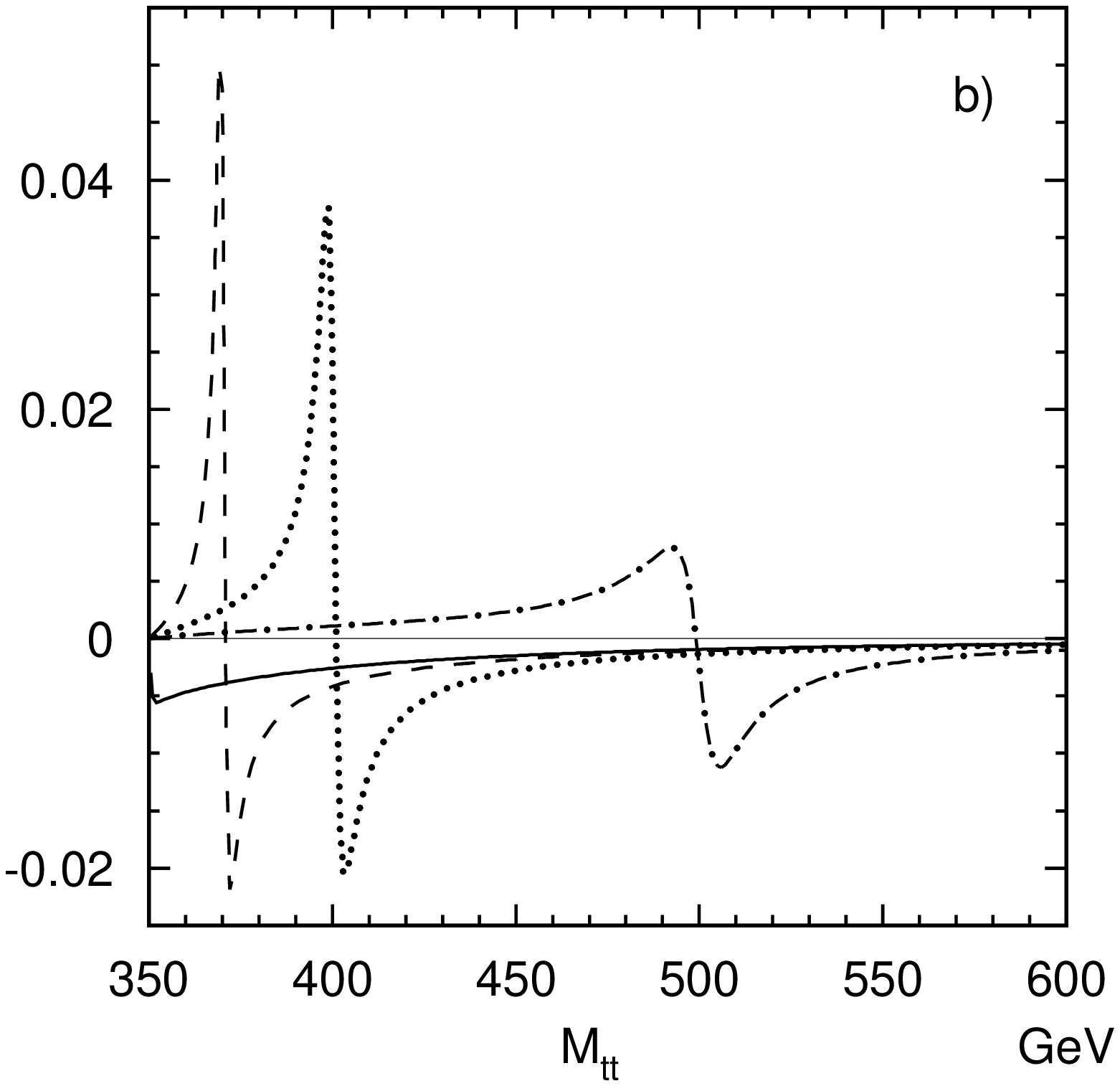,height=7cm,width=7cm}}
\put(0,0){\psfig{figure=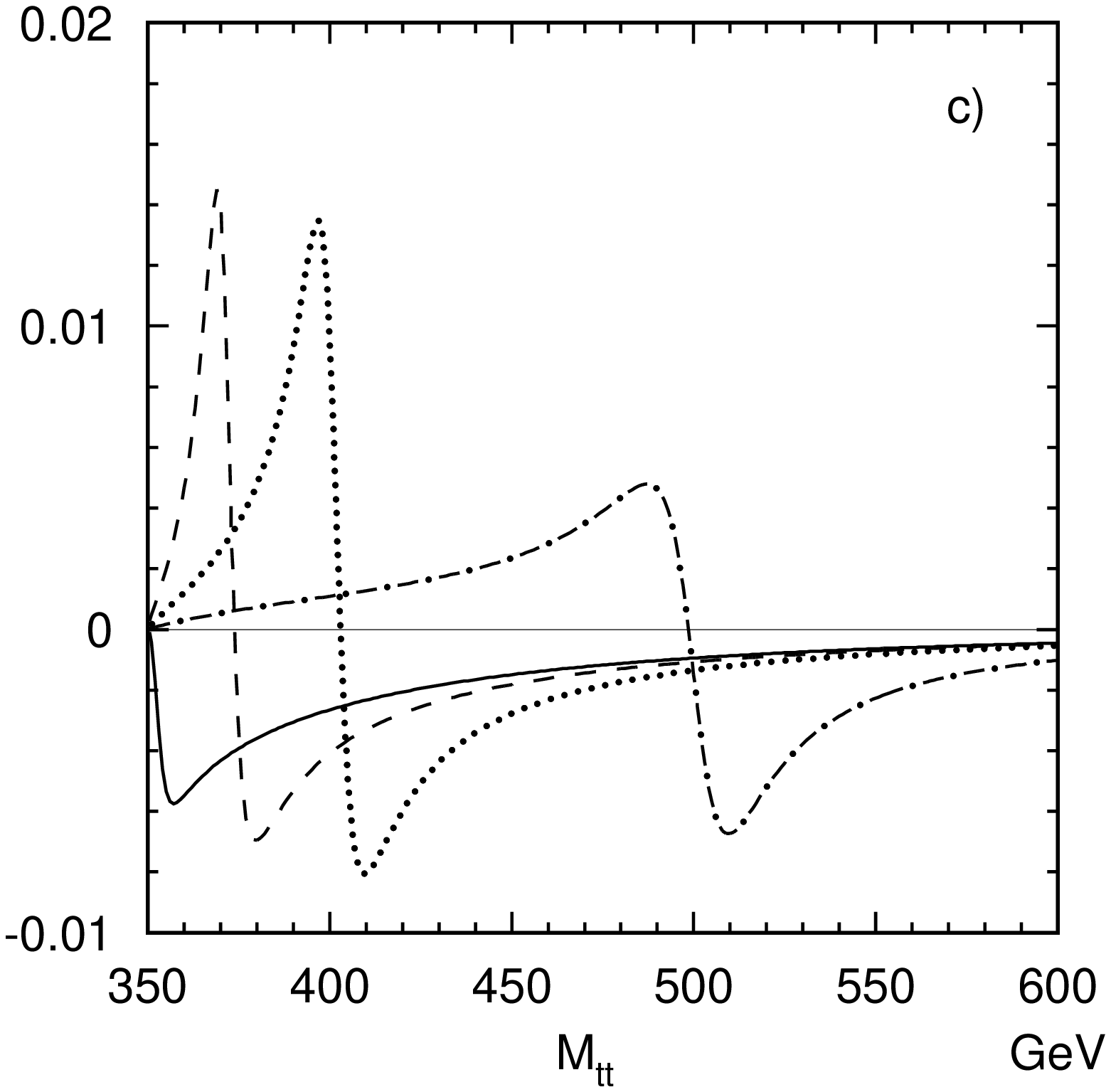,height=7cm,width=7cm}}
\put(7,0){\psfig{figure=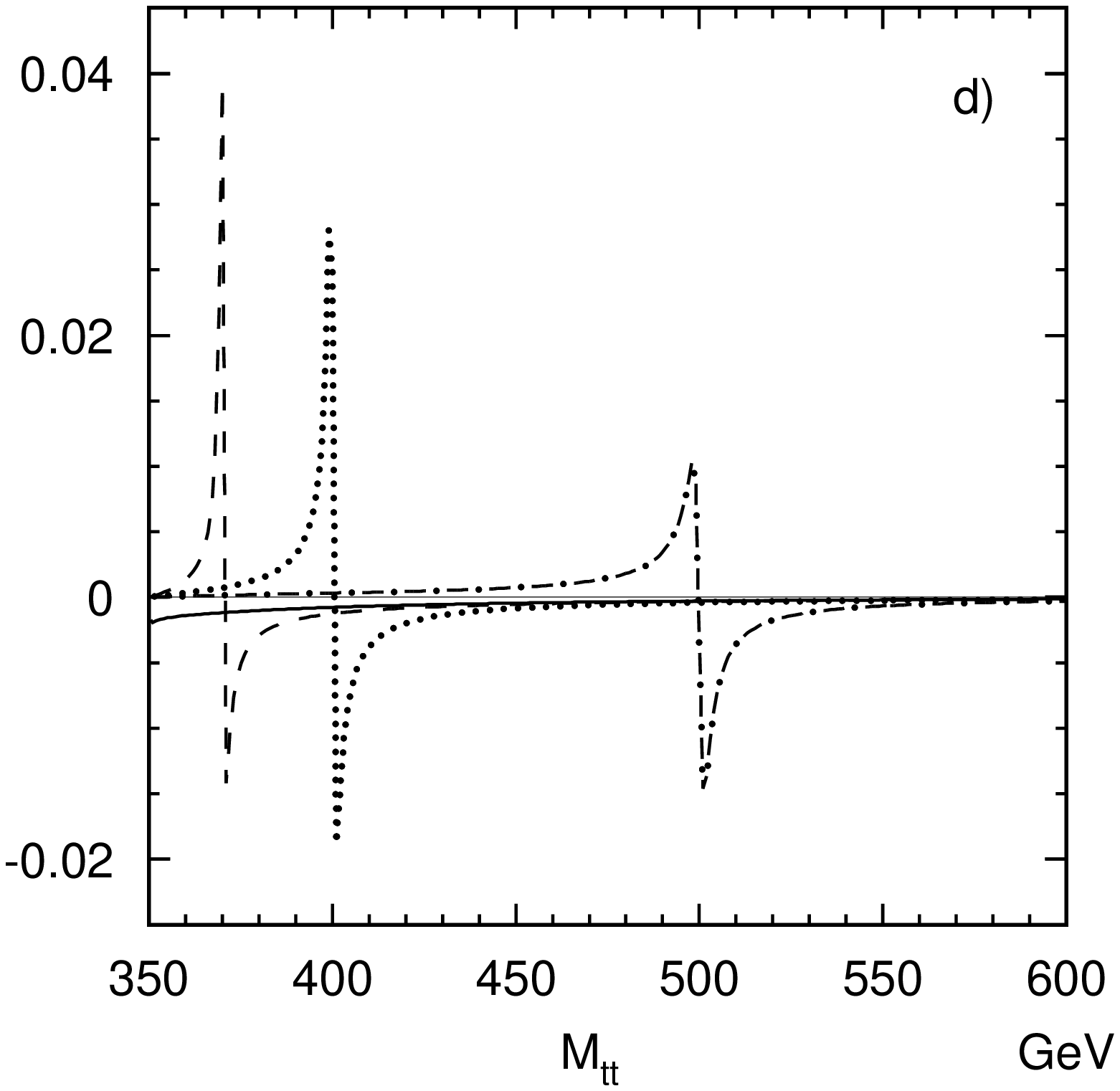,height=7cm,width=7cm}}
\end{picture}
\\[6pt]
Figure 3
\end{center}
\end{figure}
\newpage
%                        
%  4. figure             
%                        
\begin{figure}[h]
\unitlength1.0cm
\begin{center}
\begin{picture}(13.6,13.)
%\put(1,13.2){\mbox{$a=1,\,\,\tilde{a}=-1$}}
%\put(8,13.2){\mbox{$a=1,\,\,\tilde{a}=-0.3$}}
%\put(1,6.2){\mbox{$a=-0.3,\,\,\tilde{a}=-1$}}
%\put(8,6.2){\mbox{$a=-0.3,\,\,\tilde{a}=-0.3$}}
\put(0,7){\psfig{figure=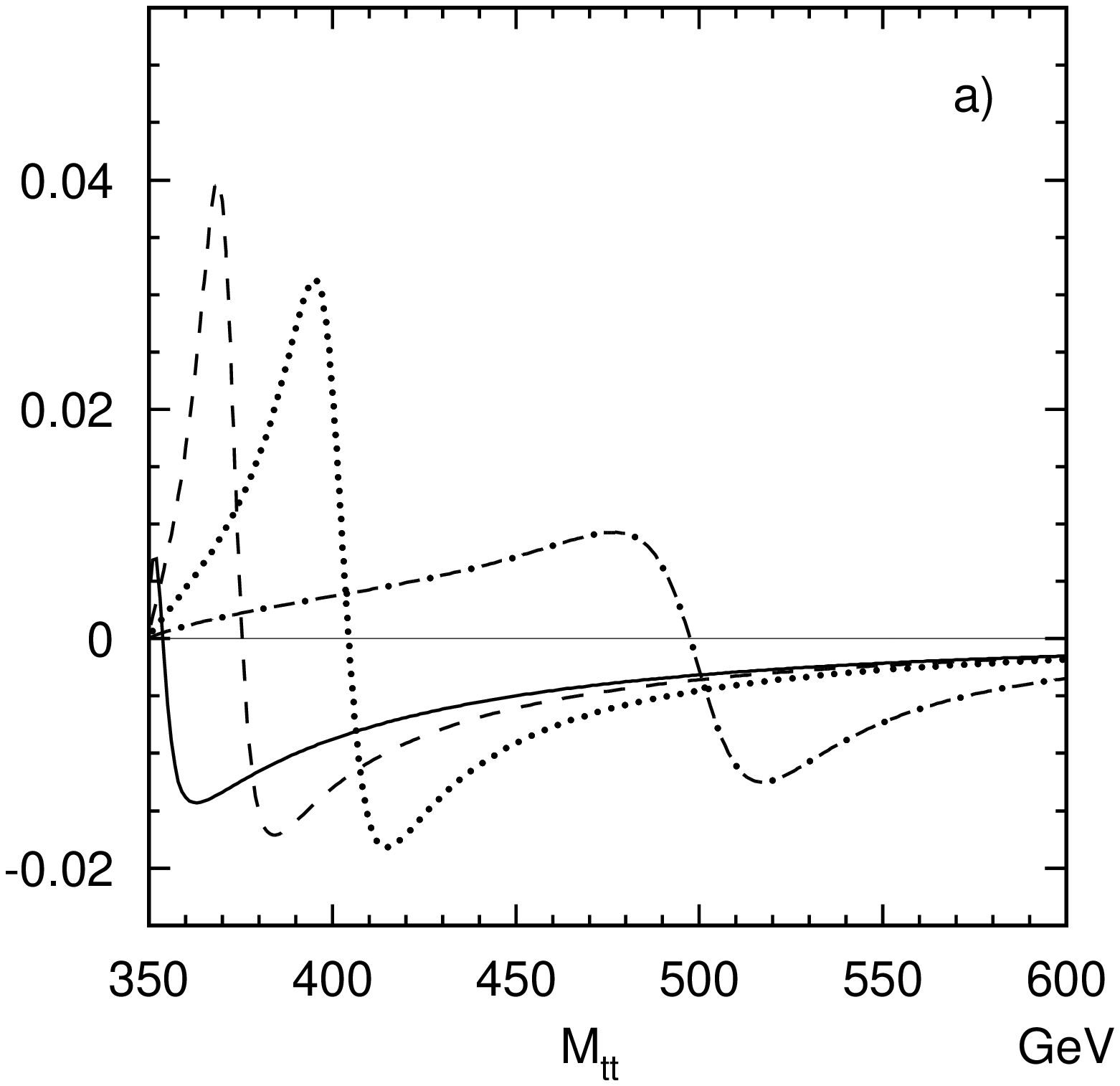,height=7cm,width=7cm}}
\put(7,7){\psfig{figure=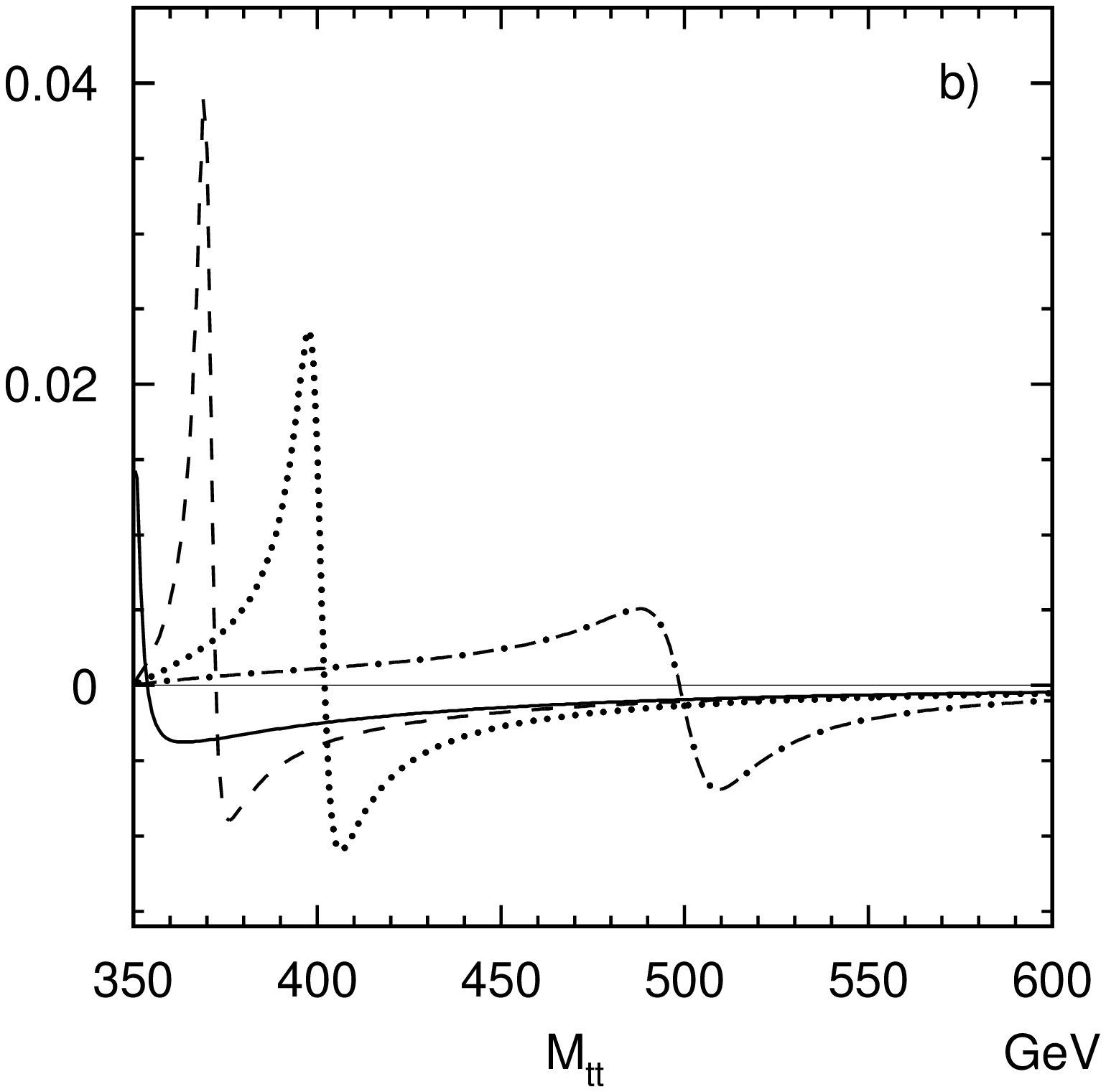,height=7cm,width=7cm}}
\put(0,0){\psfig{figure=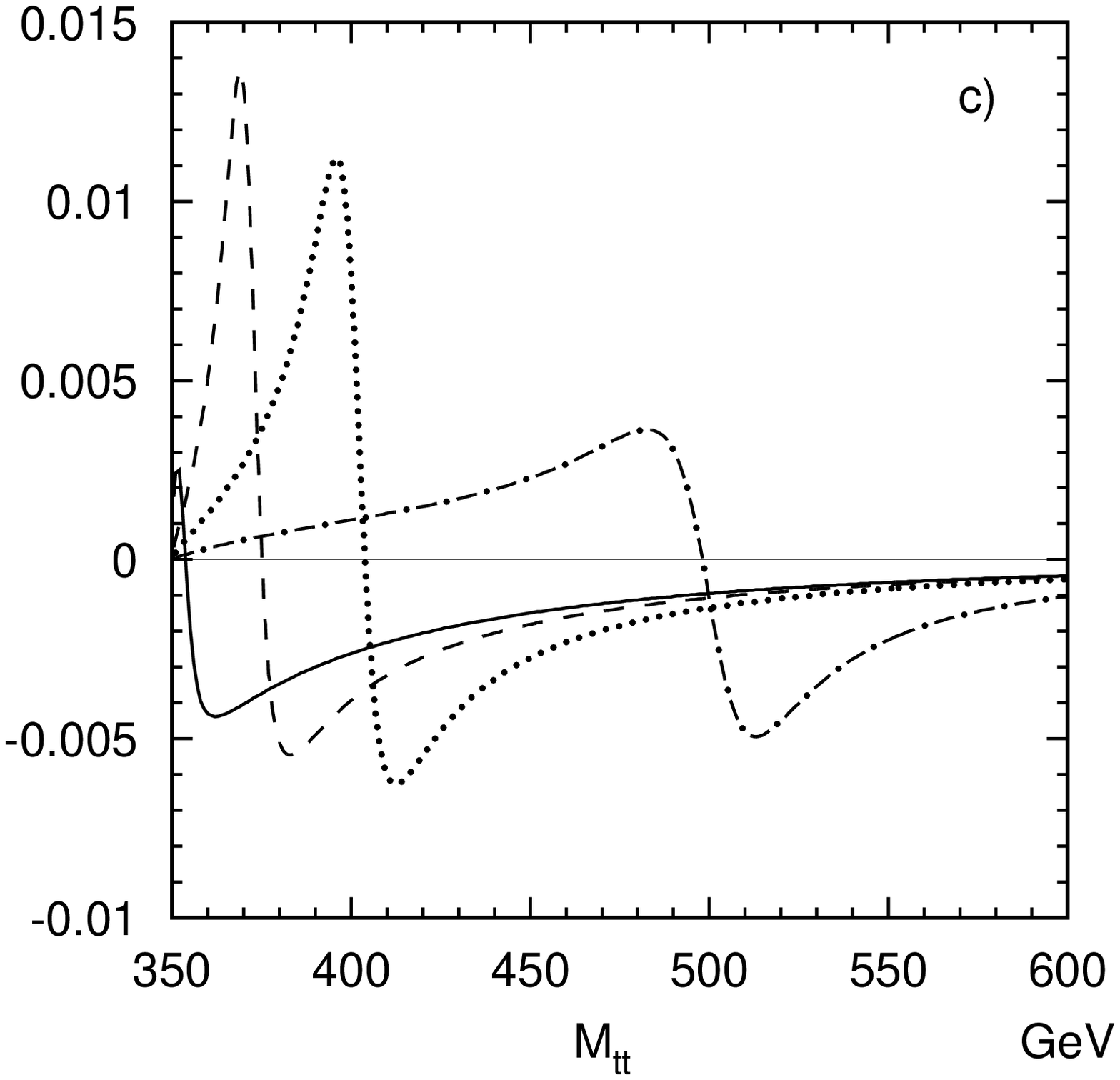,height=7cm,width=7cm}}
\put(7,0){\psfig{figure=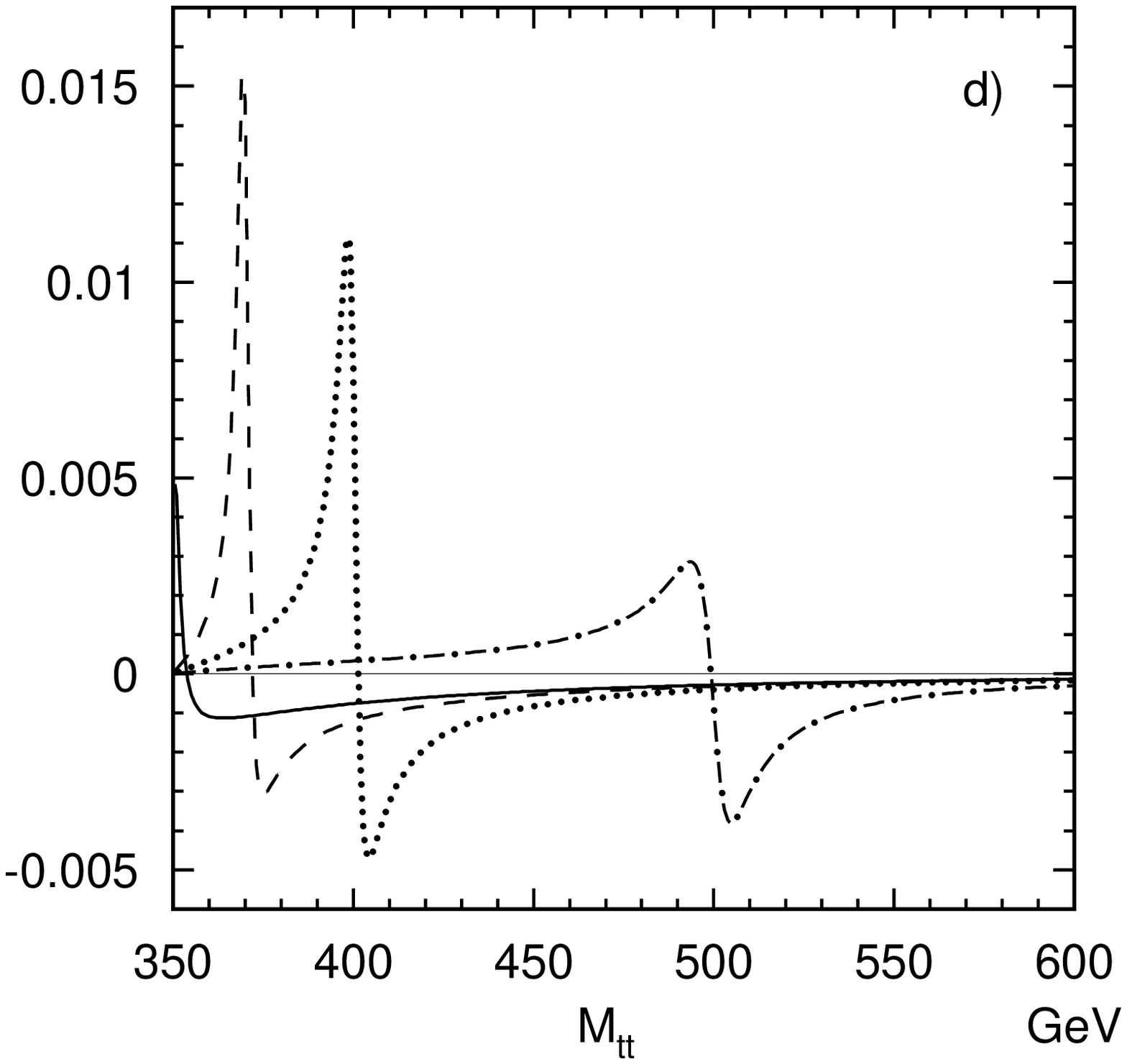,height=7cm,width=7cm}}
\end{picture}
\\[6pt]
Figure 4
\end{center}
\end{figure}
\newpage
%                        
%  5. figure             
%                        
\begin{figure}[h]
\unitlength1.0cm
\begin{center}
\begin{picture}(13.6,13.)
%\put(1,13.2){\mbox{$a=1,\,\,\tilde{a}=-1$}}
%\put(8,13.2){\mbox{$a=1,\,\,\tilde{a}=-0.3$}}
%\put(1,6.2){\mbox{$a=-0.3,\,\,\tilde{a}=-1$}}
%\put(8,6.2){\mbox{$a=-0.3,\,\,\tilde{a}=-0.3$}}
\put(0,7){\psfig{figure=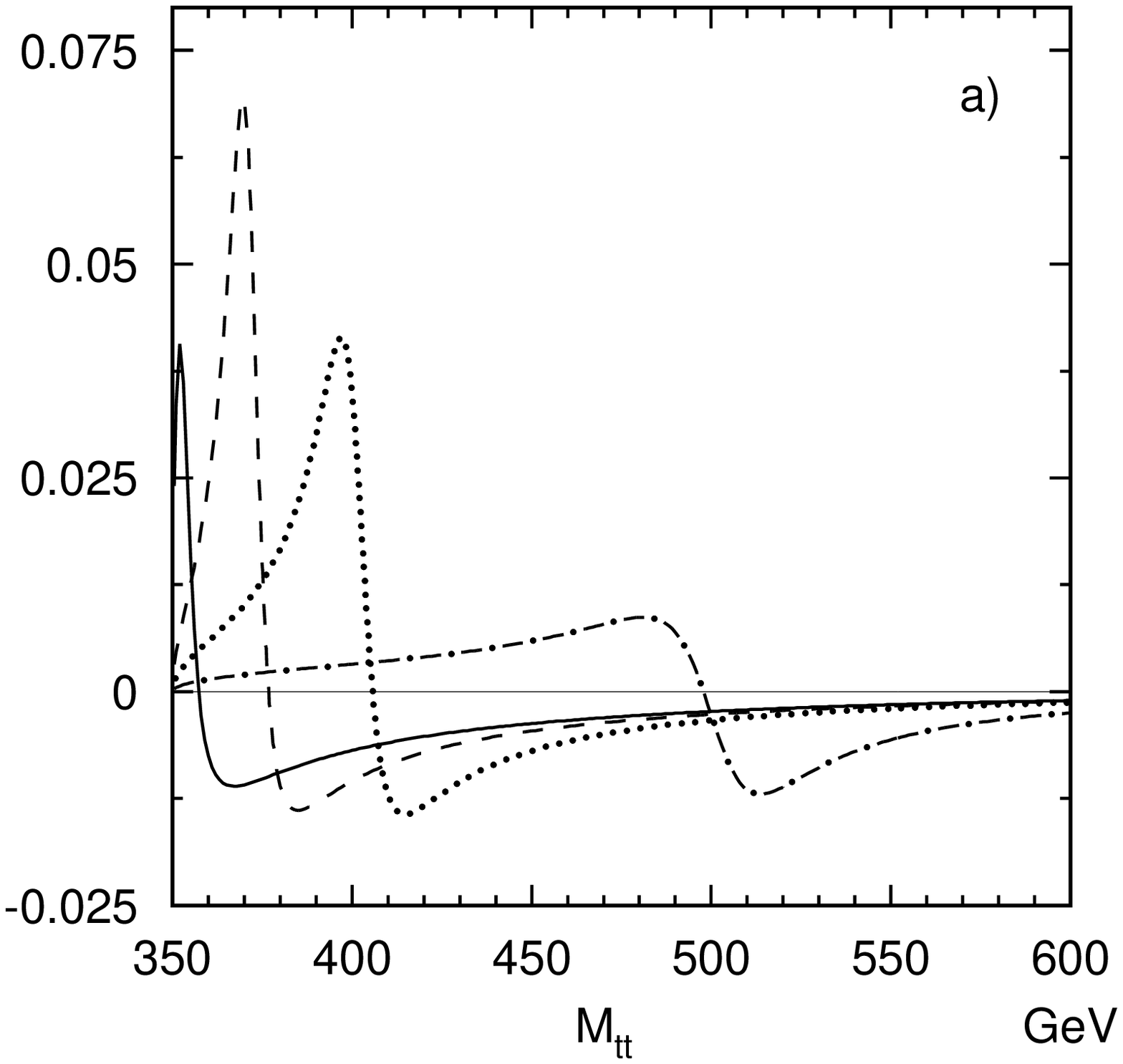,height=7cm,width=7cm}}
\put(7,7){\psfig{figure=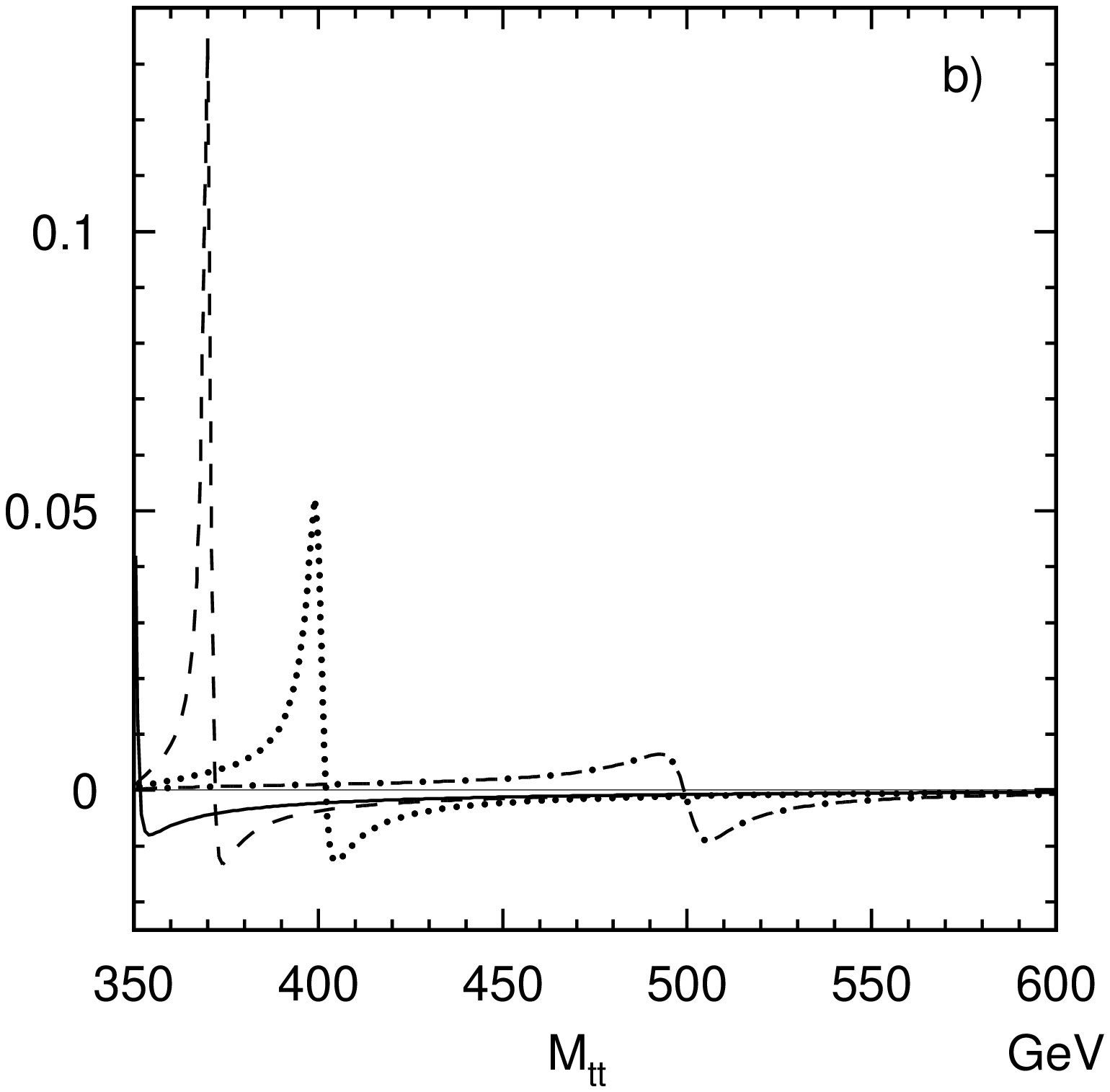,height=7cm,width=7cm}}
\put(0,0){\psfig{figure=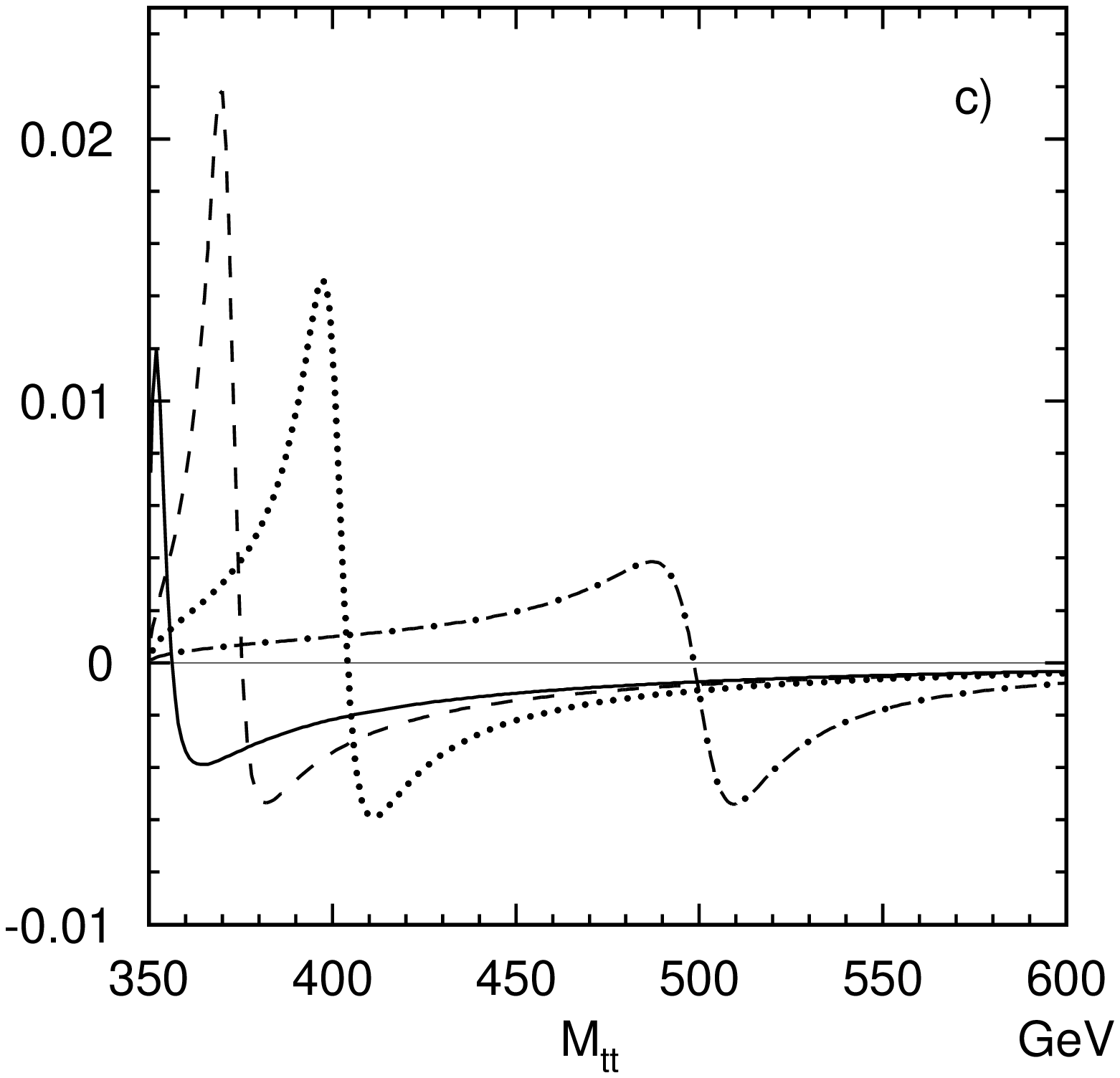,height=7cm,width=7cm}}
\put(7,0){\psfig{figure=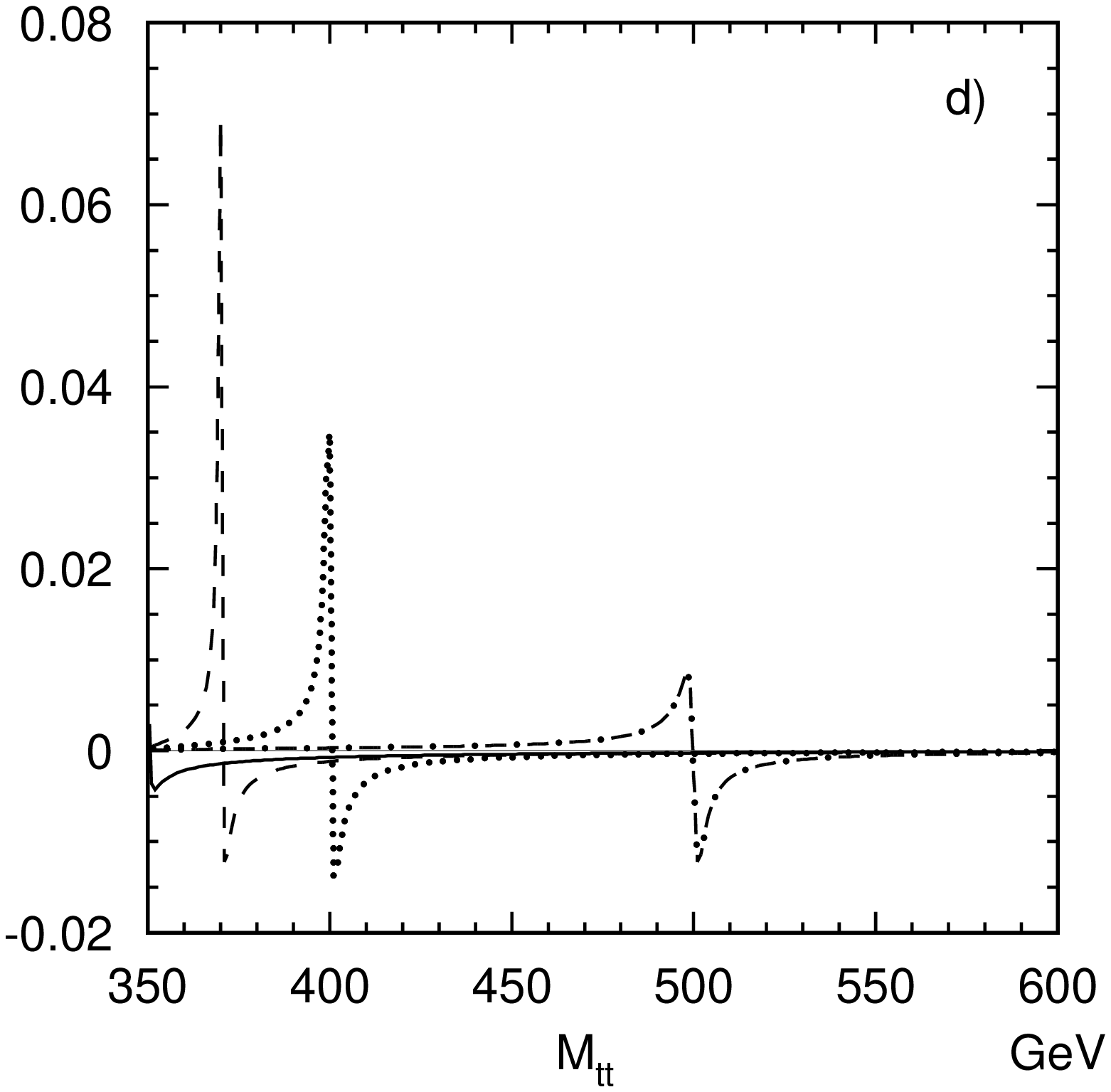,height=7cm,width=7cm}}
\end{picture}
\\[6pt]
Figure 5
\end{center}
\end{figure}
\newpage
%                        
%  6. figure             
%                        
\begin{figure}[h]
\unitlength1.0cm
\begin{center}
\begin{picture}(13.6,13.)
%\put(1,13.2){\mbox{$a=1,\,\,\tilde{a}=-1$}}
%\put(8,13.2){\mbox{$a=1,\,\,\tilde{a}=-0.3$}}
%\put(1,6.2){\mbox{$a=-0.3,\,\,\tilde{a}=-1$}}
%\put(8,6.2){\mbox{$a=-0.3,\,\,\tilde{a}=-0.3$}}
\put(0,7){\psfig{figure=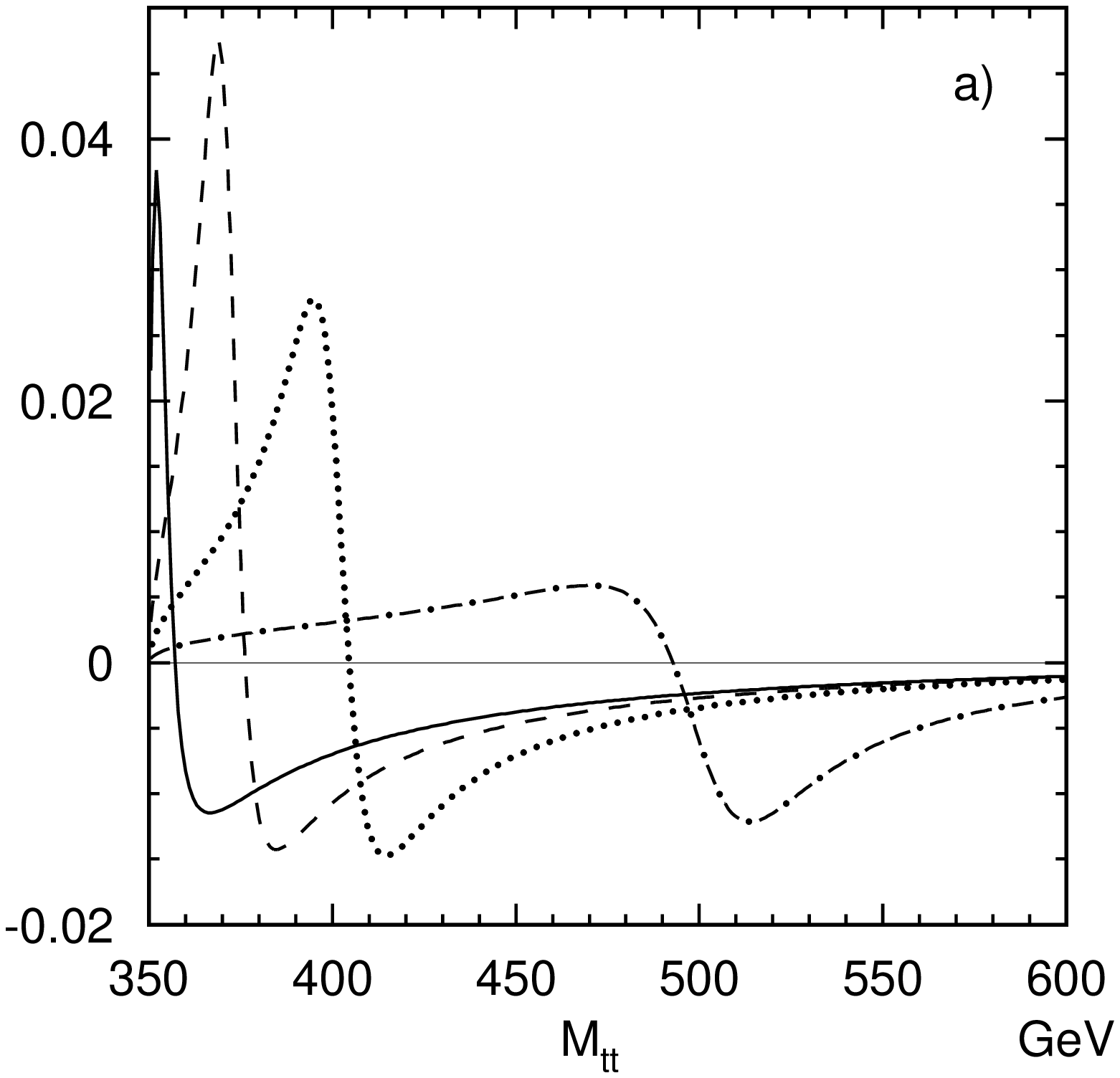,height=7cm,width=7cm}}
\put(7,7){\psfig{figure=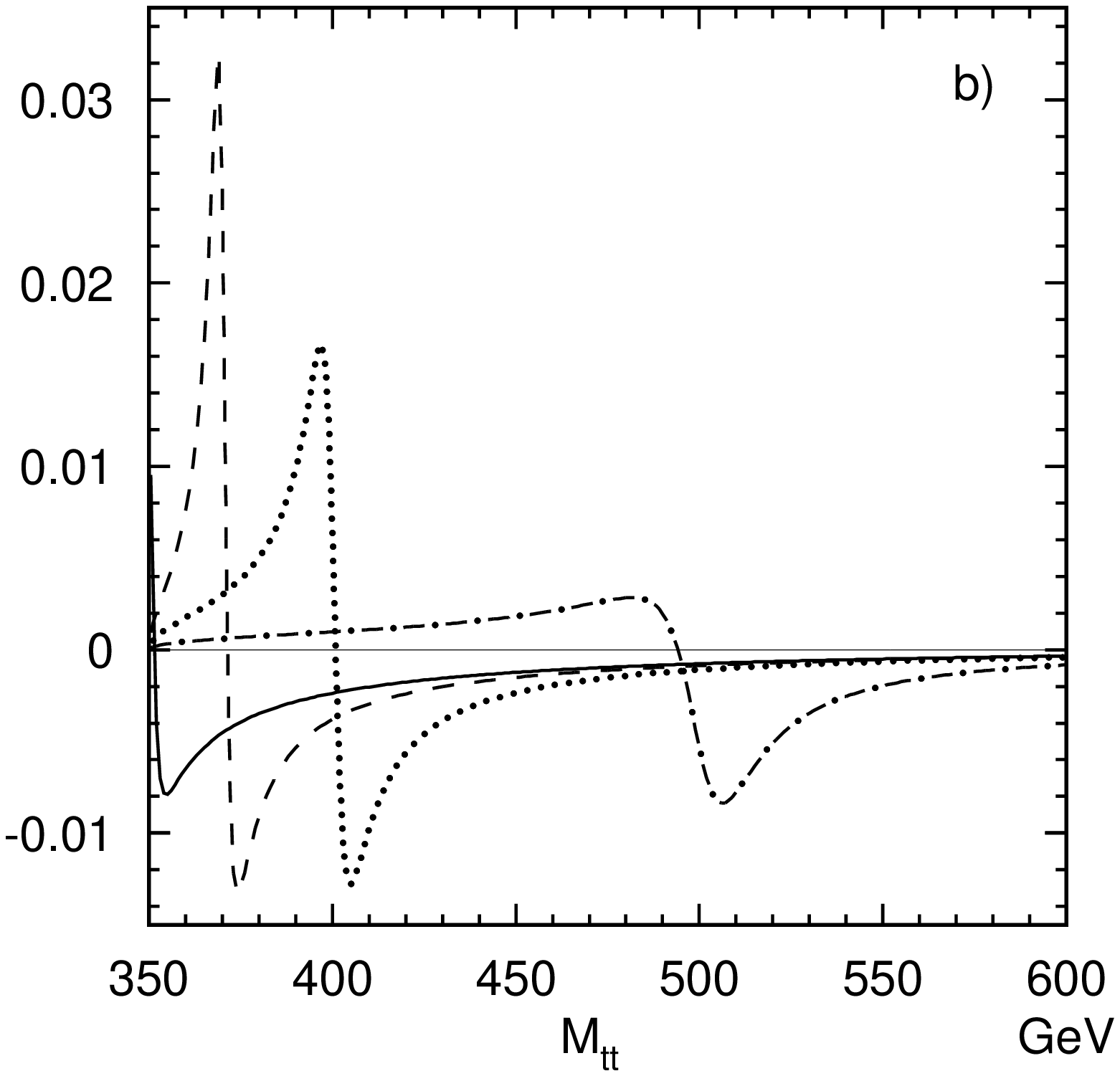,height=7cm,width=7cm}}
\put(0,0){\psfig{figure=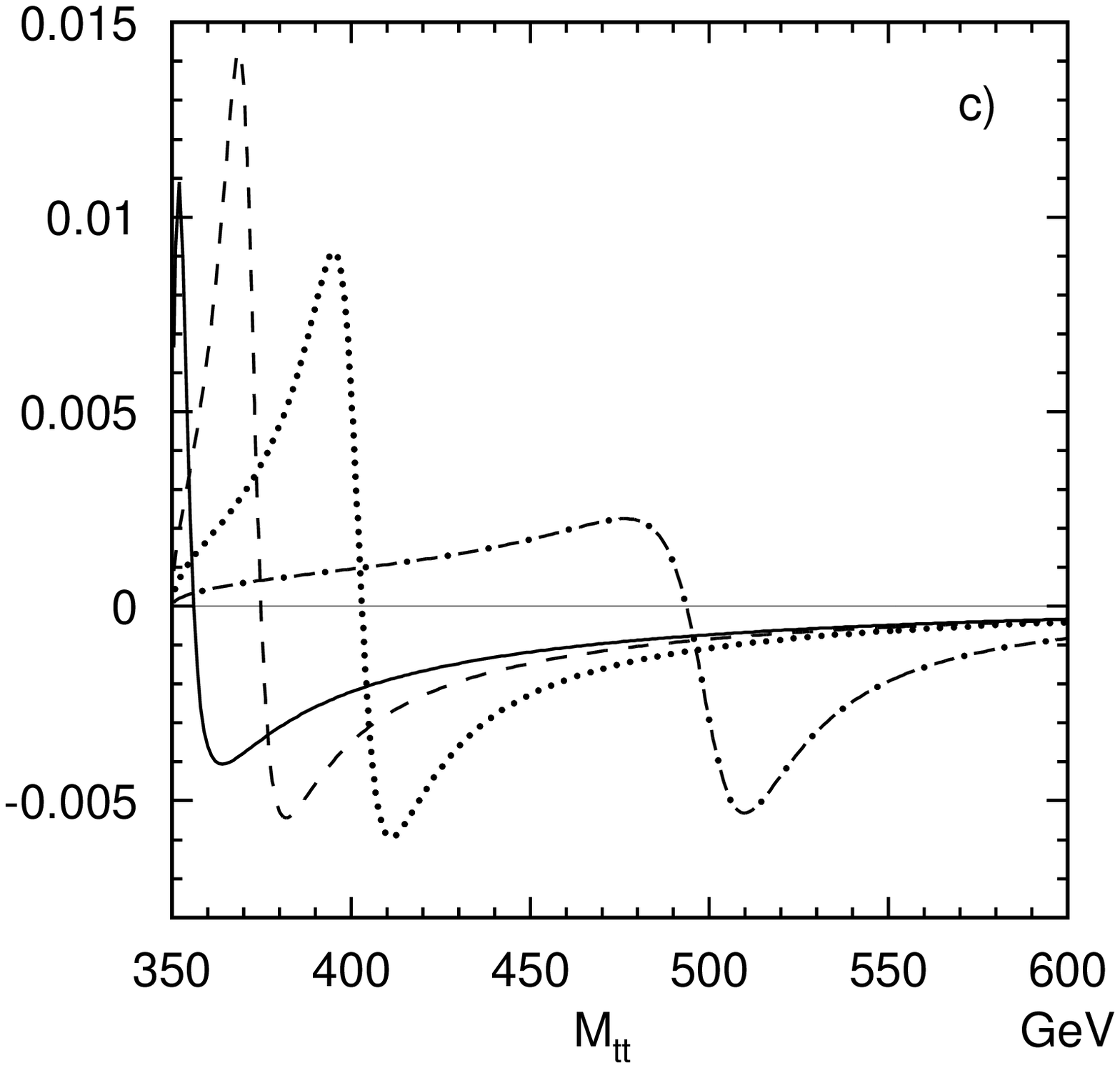,height=7cm,width=7cm}}
\put(7,0){\psfig{figure=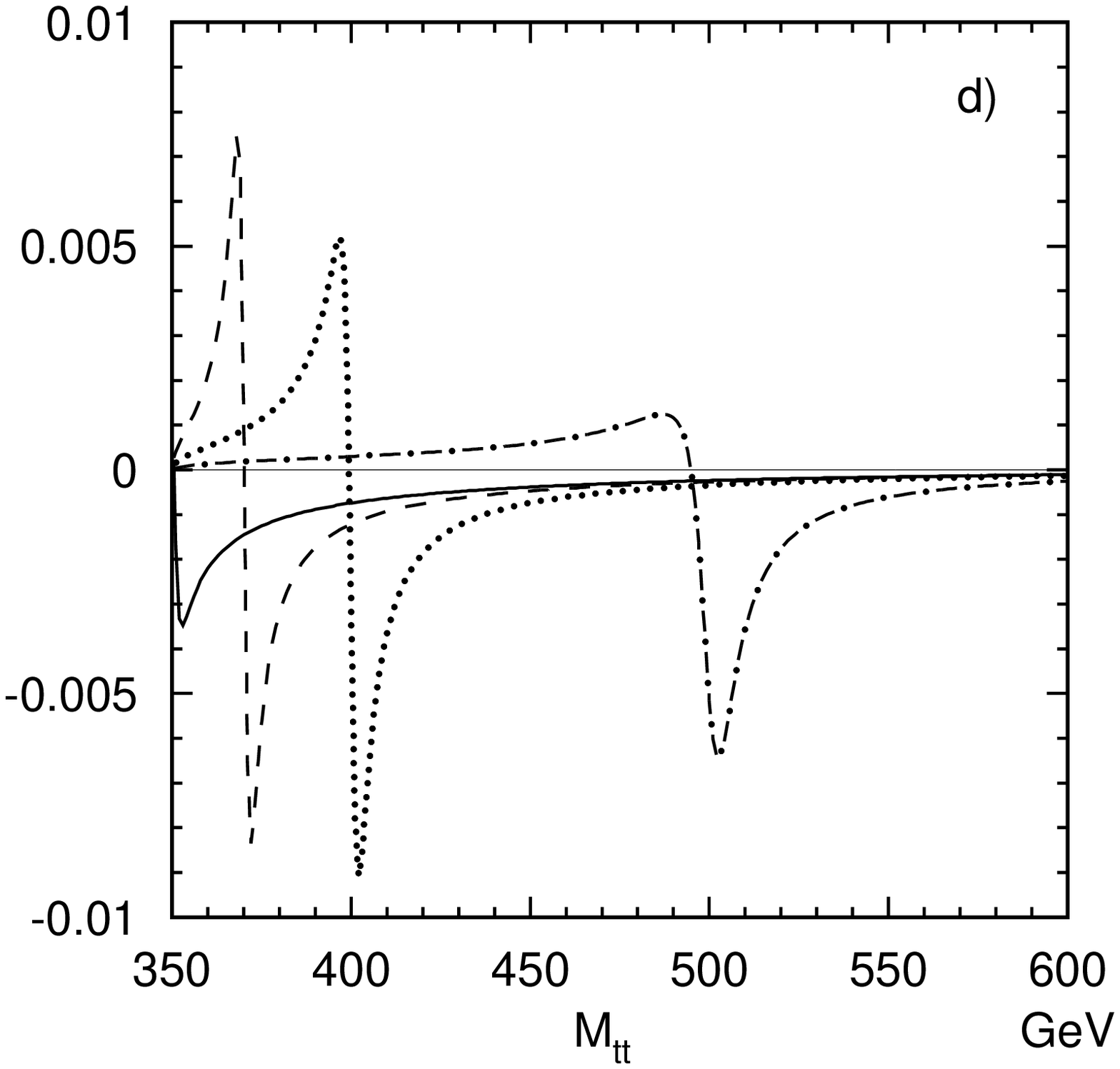,height=7cm,width=7cm}}
\end{picture}
\\[6pt]
Figure 6
\end{center}
\end{figure}

\newpage
%                                           
% ========================================= 
%                                           
%  Tables                                   
%                                           
% ========================================= 
%                                           
\section*{Tables}
%                                    
% position of cuts                   
%                                    
\begin{table}[ht]
\begin{center}
\begin{tabular}{r||c|c||c|c}
\hline
\hline
 & \multicolumn{2}{|c||}{below $m_\varphi$} & \multicolumn{2}{|c}{above $m_\varphi$} \\
\hline
 $m_{\varphi}$ & $M_{t\bar{t}}$ & $\Delta M_{t\bar{t}}$ & $M_{t\bar{t}}$ 
 & $\Delta M_{t\bar{t}}$ \\
\hline
\hline
$ 320 $ & $    -    $ & $  - $ & $ 350-450 $ & $ 100 $ \\ 
$ 350 $ & $    -    $ & $  - $ & $ 360-400 $ & $  40 $ \\ 
$ 370 $ & $ 355-370 $ & $ 15 $ & $ 375-420 $ & $  45 $ \\ 
$ 400 $ & $ 360-400 $ & $ 40 $ & $ 405-450 $ & $  45 $ \\ 
$ 500 $ & $ 420-500 $ & $ 80 $ & $ 500-560 $ & $  60 $ \\ 
\hline
\hline
\end{tabular}
\\[6pt]
Table 1
\end{center}
\end{table}
%                                
%                                
%  Q1 mit Schnitten unterhalb    
%                                
%                                
\begin{table}[ht]
\begin{center}
\begin{tabular}{|r|c|c|c|c|c|c|}
\hline
 & \multicolumn{6}{c|}{$g_{VV}$}  \\
 $m_\varphi$ & \multicolumn{2}{c|}{0.0} &  \multicolumn{2}{c|}{0.2}&\multicolumn
{2}{c|}{0.4}\\
\hline
\raisebox{-4.6ex}[0pt][0pt]{370} 
  & $ 2.7    $ & $ 9.4    $ & $ 2.7    $ & $ 9.3    $ & $ 2.6    $ & $ 8.6 $   \\ 
  & $ 1.9    $ & $ 5.9    $ & $ 1.9    $ & $ 5.7    $ & $ 1.6    $ & $ 4.6 $   \\ 
  & $ 0.86   $ & $ 3.0    $ & $ 0.87   $ & $ 3.0    $ & $ 0.83   $ & $ 2.7 $   \\ 
  & $ 0.73   $ & $ 2.2    $ & $ 0.65   $ & $ 1.9    $ & $ 0.53   $ & $ 1.5 $   \\ 
\hline
\raisebox{-4.6ex}[0pt][0pt]{400} 
  & $ 1.9    $ & $ 11.1   $ & $ 1.9    $ & $ 10.8   $ & $ 1.8    $ & $ 10.1 $  \\ 
  & $ 1.0    $ & $ 5.7    $ & $ 0.95   $ & $ 5.3    $ & $ 0.83   $ & $ 4.5  $  \\ 
  & $ 0.63   $ & $ 3.6    $ & $ 0.62   $ & $ 3.5    $ & $ 0.58   $ & $ 3.3  $  \\ 
  & $ 0.41   $ & $ 2.3    $ & $ 0.36   $ & $ 2.0    $ & $ 0.29   $ & $ 1.6  $  \\ 
\hline
\raisebox{-4.6ex}[0pt][0pt]{500} 
  & $ 0.69   $ & $ 5.4    $ & $ 0.67   $ & $ 5.3    $ & $ 0.61   $ & $ 4.8  $  \\ 
  & $ 0.31   $ & $ 2.4    $ & $ 0.29   $ & $ 2.3    $ & $ 0.26   $ & $ 2.0  $  \\ 
  & $ 0.25   $ & $ 1.9    $ & $ 0.24   $ & $ 1.9    $ & $ 0.21   $ & $ 1.7  $  \\ 
  & $ 0.14   $ & $ 1.1    $ & $ 0.12   $ & $ 0.97   $ & $ 0.10   $ & $ 0.76 $  \\ 
\hline
\end{tabular}
\\[6pt]
Table 2
\end{center}
\end{table}
%                                
%                                
%  Q1 mit Schnitten oberhalb     
%                                
%                                
\begin{table}[ht]
\begin{center}
\begin{tabular}{|r|c|c|c|c|c|c|}
\hline
 & \multicolumn{6}{c|}{$g_{VV}$}  \\
 $m_\varphi$ & \multicolumn{2}{c|}{0.0} &  \multicolumn{2}{c|}{0.2}&\multicolumn
{2}{c|}{0.4}\\
\hline
\raisebox{-4.6ex}[0pt][0pt]{320} 
  & $ -0.63  $ & $ 5.3    $ & $ -0.63  $ & $ 5.3    $ & $ -0.62  $ & $ 5.3  $  \\ 
  & $ -0.18  $ & $ 1.6    $ & $ -0.18  $ & $ 1.6    $ & $ -0.18  $ & $ 1.6  $  \\ 
  & $ -0.19  $ & $ 1.6    $ & $ -0.19  $ & $ 1.6    $ & $ -0.19  $ & $ 1.6  $  \\ 
  & $ -0.06  $ & $ 0.47   $ & $ -0.06  $ & $ 0.47   $ & $ -0.05  $ & $ 0.47 $  \\ 
\hline
\raisebox{-4.6ex}[0pt][0pt]{350} 
  & $ -1.3   $ & $ 6.7    $ & $ -1.3   $ & $ 6.6    $ & $ -1.2   $ & $ 6.3  $  \\ 
  & $ -0.36  $ & $ 1.9    $ & $ -0.36  $ & $ 1.9    $ & $ -0.34  $ & $ 1.8  $  \\ 
  & $ -0.38  $ & $ 2.0    $ & $ -0.38  $ & $ 2.0    $ & $ -0.37  $ & $ 1.9  $  \\ 
  & $ -0.11  $ & $ 0.57   $ & $ -0.11  $ & $ 0.57   $ & $ -0.10  $ & $ 0.55 $  \\ 
\hline
\raisebox{-4.6ex}[0pt][0pt]{370} 
  & $ -1.4   $ & $ 8.1    $ & $ -1.3   $ & $ 7.8    $ & $ -1.2   $ & $ 7.1  $  \\ 
  & $ -0.51  $ & $ 3.1    $ & $ -0.50  $ & $ 3.0    $ & $ -0.46  $ & $ 2.8  $  \\ 
  & $ -0.43  $ & $ 2.5    $ & $ -0.41  $ & $ 2.4    $ & $ -0.38  $ & $ 2.2  $  \\ 
  & $ -0.16  $ & $ 0.95   $ & $ -0.15  $ & $ 0.92   $ & $ -0.14  $ & $ 0.85 $  \\ 
\hline
\raisebox{-4.6ex}[0pt][0pt]{400} 
  & $ -1.4   $ & $ 8.6    $ & $ -1.4   $ & $ 8.3    $ & $ -1.3   $ & $ 7.5  $  \\ 
  & $ -0.60  $ & $ 3.6    $ & $ -0.58  $ & $ 3.5    $ & $ -0.53  $ & $ 3.2  $  \\ 
  & $ -0.47  $ & $ 2.8    $ & $ -0.45  $ & $ 2.7    $ & $ -0.41  $ & $ 2.5  $  \\ 
  & $ -0.19  $ & $ 1.2    $ & $ -0.18  $ & $ 1.1    $ & $ -0.17  $ & $ 1.0  $  \\ 
\hline
\raisebox{-4.6ex}[0pt][0pt]{500} 
  & $ -0.96  $ & $ 5.4    $ & $ -0.93  $ & $ 5.2    $ & $ -0.83  $ & $ 4.6  $  \\ 
  & $ -0.46  $ & $ 2.6    $ & $ -0.43  $ & $ 2.4    $ & $ -0.36  $ & $ 2.0  $  \\ 
  & $ -0.36  $ & $ 2.0    $ & $ -0.34  $ & $ 1.9    $ & $ -0.30  $ & $ 1.7  $  \\ 
  & $ -0.23  $ & $ 1.3    $ & $ -0.19  $ & $ 1.1    $ & $ -0.15  $ & $ 0.82 $  \\ 
\hline
\end{tabular}
\\[6pt]
Table 3
\end{center}
\end{table}
%                                
%                                
%  Q2 mit Schnitten unterhalb    
%                                
%                                
\begin{table}[ht]
\begin{center}
\begin{tabular}{|r|c|c|c|c|c|c|}
\hline
 & \multicolumn{6}{c|}{$g_{VV}$}  \\
 $m_\varphi$ & \multicolumn{2}{c|}{0.0} &  \multicolumn{2}{c|}{0.2}&\multicolumn
{2}{c|}{0.4}\\
\hline
\raisebox{-4.6ex}[0pt][0pt]{370} 
  & $ 4.4    $ & $ 29.8   $ & $ 4.1    $ & $ 27.4   $ & $ 3.3    $ & $ 20.9 $  \\ 
  & $ 3.9    $ & $ 23.4   $ & $ 2.9    $ & $ 16.7   $ & $ 1.6    $ & $ 9.0  $  \\ 
  & $ 1.3    $ & $ 8.7    $ & $ 1.2    $ & $ 8.0    $ & $ 0.98   $ & $ 6.1  $  \\ 
  & $ 1.2    $ & $ 6.6    $ & $ 0.75   $ & $ 4.1    $ & $ 0.39   $ & $ 2.1  $  \\ 
\hline
\raisebox{-4.6ex}[0pt][0pt]{400} 
  & $ 2.3    $ & $ 24.4   $ & $ 2.1    $ & $ 22.8   $ & $ 1.8    $ & $ 18.7 $  \\ 
  & $ 1.3    $ & $ 13.4   $ & $ 1.1    $ & $ 11.1   $ & $ 0.75   $ & $ 7.5  $  \\ 
  & $ 0.73   $ & $ 7.8    $ & $ 0.67   $ & $ 7.1    $ & $ 0.55   $ & $ 5.8  $  \\ 
  & $ 0.49   $ & $ 4.9    $ & $ 0.35   $ & $ 3.5    $ & $ 0.21   $ & $ 2.1  $  \\ 
\hline
\raisebox{-4.6ex}[0pt][0pt]{500} 
  & $ 0.65   $ & $ 8.6    $ & $ 0.59   $ & $ 7.9    $ & $ 0.46   $ & $ 6.0  $  \\ 
  & $ 0.31   $ & $ 4.1    $ & $ 0.26   $ & $ 3.5    $ & $ 0.18   $ & $ 2.4  $  \\ 
  & $ 0.24   $ & $ 3.2    $ & $ 0.22   $ & $ 2.9    $ & $ 0.16   $ & $ 2.1  $  \\ 
  & $ 0.14   $ & $ 1.9    $ & $ 0.10   $ & $ 1.4    $ & $ 0.06   $ & $ 0.77 $  \\ 
\hline
\end{tabular}
\\[6pt]
Table 4
\end{center}
\end{table}
%                                
%                                
%  Q2 mit Schnitten oberhalb     
%                                
%                                
\begin{table}[ht]
\begin{center}
\begin{tabular}{|r|c|c|c|c|c|c|}
\hline
 & \multicolumn{6}{c|}{$g_{VV}$}  \\
 $m_\varphi$ & \multicolumn{2}{c|}{0.0} &  \multicolumn{2}{c|}{0.2}&\multicolumn
{2}{c|}{0.4}\\
\hline
\raisebox{-4.6ex}[0pt][0pt]{320} 
  & $ -0.61  $ & $ 9.2    $ & $ -0.61  $ & $ 9.2    $ & $ -0.61  $ & $ 9.2  $  \\ 
  & $ -0.19  $ & $ 3.0    $ & $ -0.19  $ & $ 3.0    $ & $ -0.19  $ & $ 3.0  $  \\ 
  & $ -0.19  $ & $ 2.8    $ & $ -0.19  $ & $ 2.8    $ & $ -0.19  $ & $ 2.8  $  \\ 
  & $ -0.06  $ & $ 0.91   $ & $ -0.06  $ & $ 0.91   $ & $ -0.06  $ & $ 0.91 $  \\ 
\hline
\raisebox{-4.6ex}[0pt][0pt]{350} 
  & $ -1.1   $ & $ 10.0   $ & $ -1.1   $ & $ 10.1   $ & $ -1.1   $ & $ 10.2 $  \\ 
  & $ -0.40  $ & $ 4.0    $ & $ -0.40  $ & $ 4.0    $ & $ -0.41  $ & $ 4.0  $  \\ 
  & $ -0.34  $ & $ 3.3    $ & $ -0.35  $ & $ 3.3    $ & $ -0.35  $ & $ 3.3  $  \\ 
  & $ -0.13  $ & $ 1.3    $ & $ -0.13  $ & $ 1.3    $ & $ -0.13  $ & $ 1.3  $  \\ 
\hline
\raisebox{-4.6ex}[0pt][0pt]{370} 
  & $ -1.0   $ & $ 10.6   $ & $ -1.0   $ & $ 10.7   $ & $ -1.1   $ & $ 11.1 $  \\ 
  & $ -0.50  $ & $ 5.4    $ & $ -0.51  $ & $ 5.5    $ & $ -0.52  $ & $ 5.6  $  \\ 
  & $ -0.37  $ & $ 3.9    $ & $ -0.37  $ & $ 3.9    $ & $ -0.38  $ & $ 4.0  $  \\ 
  & $ -0.17  $ & $ 1.9    $ & $ -0.17  $ & $ 1.9    $ & $ -0.17  $ & $ 1.9  $  \\ 
\hline
\raisebox{-4.6ex}[0pt][0pt]{400} 
  & $ -1.1   $ & $ 11.3   $ & $ -1.1   $ & $ 11.4   $ & $ -1.2   $ & $ 11.9 $  \\ 
  & $ -0.55  $ & $ 5.8    $ & $ -0.56  $ & $ 5.8    $ & $ -0.57  $ & $ 6.0  $  \\ 
  & $ -0.41  $ & $ 4.2    $ & $ -0.42  $ & $ 4.3    $ & $ -0.42  $ & $ 4.3  $  \\ 
  & $ -0.19  $ & $ 2.0    $ & $ -0.20  $ & $ 2.1    $ & $ -0.20  $ & $ 2.1  $  \\ 
\hline
\raisebox{-4.6ex}[0pt][0pt]{500} 
  & $ -0.92  $ & $ 8.2    $ & $ -0.94  $ & $ 8.4    $ & $ -0.98  $ & $ 8.8  $  \\ 
  & $ -0.46  $ & $ 4.1    $ & $ -0.48  $ & $ 4.3    $ & $ -0.49  $ & $ 4.4  $  \\ 
  & $ -0.35  $ & $ 3.2    $ & $ -0.36  $ & $ 3.3    $ & $ -0.38  $ & $ 3.4  $  \\ 
  & $ -0.24  $ & $ 2.1    $ & $ -0.25  $ & $ 2.2    $ & $ -0.23  $ & $ 2.1  $  \\ 
\hline
\end{tabular}
\\[6pt]
Table 5
\end{center}
\end{table}
%                                
%                                
%  E1 mit Schnitten unterhalb    
%                                
%                                
\begin{table}[ht]
\begin{center}
\begin{tabular}{|r|c|c|c|c|c|c|}
\hline
 & \multicolumn{6}{c|}{$g_{VV}$}  \\
 $m_\varphi$ & \multicolumn{2}{c|}{0.0} &  \multicolumn{2}{c|}{0.2}&\multicolumn
{2}{c|}{0.4}\\
\hline
\raisebox{-4.6ex}[0pt][0pt]{370} 
  & $ 2.7    $ & $ 16.6   $ & $ 2.7    $ & $ 16.1   $ & $ 2.6    $ & $ 14.9  $ \\ 
  & $ 1.8    $ & $ 9.8    $ & $ 1.9    $ & $ 9.6    $ & $ 1.6    $ & $ 7.8   $ \\ 
  & $ 0.84   $ & $ 5.1    $ & $ 0.87   $ & $ 5.2    $ & $ 0.84   $ & $ 4.7   $ \\ 
  & $ 0.76   $ & $ 3.9    $ & $ 0.66   $ & $ 3.3    $ & $ 0.53   $ & $ 2.6   $ \\ 
\hline
\raisebox{-4.6ex}[0pt][0pt]{400} 
  & $ 1.9    $ & $ 19.0   $ & $ 1.9    $ & $ 18.4   $ & $ 1.8    $ & $ 17.2  $ \\ 
  & $ 1.0    $ & $ 9.5    $ & $ 0.96   $ & $ 9.0    $ & $ 0.84   $ & $ 7.8   $ \\ 
  & $ 0.64   $ & $ 6.2    $ & $ 0.62   $ & $ 6.1    $ & $ 0.59   $ & $ 5.6   $ \\ 
  & $ 0.42   $ & $ 3.8    $ & $ 0.36   $ & $ 3.4    $ & $ 0.29   $ & $ 2.7   $ \\ 
\hline
\raisebox{-4.6ex}[0pt][0pt]{500} 
  & $ 0.70   $ & $ 8.9    $ & $ 0.68   $ & $ 8.7    $ & $ 0.63   $ & $ 8.0   $ \\ 
  & $ 0.31   $ & $ 4.0    $ & $ 0.30   $ & $ 3.8    $ & $ 0.26   $ & $ 3.3   $ \\ 
  & $ 0.25   $ & $ 3.2    $ & $ 0.24   $ & $ 3.1    $ & $ 0.22   $ & $ 2.8   $ \\ 
  & $ 0.14   $ & $ 1.8    $ & $ 0.13   $ & $ 1.6    $ & $ 0.10   $ & $ 1.3   $ \\ 
\hline
\end{tabular}
\\[6pt]
Table 6
\end{center}
\end{table}
%                                
%                                
%  E1 mit Schnitten oberhalb     
%                                
%                                
\begin{table}[ht]
\begin{center}
\begin{tabular}{|r|c|c|c|c|c|c|}
\hline
 & \multicolumn{6}{c|}{$g_{VV}$}  \\
 $m_\varphi$ & \multicolumn{2}{c|}{0.0} &  \multicolumn{2}{c|}{0.2}&\multicolumn
{2}{c|}{0.4}\\
\hline
\raisebox{-4.6ex}[0pt][0pt]{320} 
  & $ -0.64  $ & $ 9.1    $ & $ -0.64  $ & $ 9.0    $ & $ -0.64  $ & $ 9.0   $ \\ 
  & $ -0.19  $ & $ 2.7    $ & $ -0.19  $ & $ 2.7    $ & $ -0.19  $ & $ 2.7   $ \\ 
  & $ -0.19  $ & $ 2.7    $ & $ -0.19  $ & $ 2.7    $ & $ -0.19  $ & $ 2.7   $ \\ 
  & $ -0.06  $ & $ 0.80   $ & $ -0.06  $ & $ 0.80   $ & $ -0.06  $ & $ 0.80  $ \\ 
\hline
\raisebox{-4.6ex}[0pt][0pt]{350} 
  & $ -1.3   $ & $ 11.3   $ & $ -1.3   $ & $ 11.2   $ & $ -1.2   $ & $ 10.8  $ \\ 
  & $ -0.36  $ & $ 3.3    $ & $ -0.36  $ & $ 3.2    $ & $ -0.35  $ & $ 3.1   $ \\ 
  & $ -0.40  $ & $ 3.4    $ & $ -0.39  $ & $ 3.4    $ & $ -0.37  $ & $ 3.2   $ \\ 
  & $ -0.11  $ & $ 1.0    $ & $ -0.11  $ & $ 1.0    $ & $ -0.10  $ & $ 0.9   $ \\ 
\hline
\raisebox{-4.6ex}[0pt][0pt]{370} 
  & $ -1.4   $ & $ 13.5   $ & $ -1.3   $ & $ 13.1   $ & $ -1.2   $ & $ 11.9  $ \\ 
  & $ -0.52  $ & $ 5.2    $ & $ -0.51  $ & $ 5.1    $ & $ -0.47  $ & $ 4.7   $ \\ 
  & $ -0.44  $ & $ 4.2    $ & $ -0.42  $ & $ 4.1    $ & $ -0.38  $ & $ 3.7   $ \\ 
  & $ -0.16  $ & $ 1.6    $ & $ -0.15  $ & $ 1.6    $ & $ -0.14  $ & $ 1.4   $ \\ 
\hline
\raisebox{-4.6ex}[0pt][0pt]{400} 
  & $ -1.5   $ & $ 14.3   $ & $ -1.4   $ & $ 13.7   $ & $ -1.3   $ & $ 12.5  $ \\ 
  & $ -0.61  $ & $ 6.0    $ & $ -0.59  $ & $ 5.9    $ & $ -0.54  $ & $ 5.4   $ \\ 
  & $ -0.48  $ & $ 4.7    $ & $ -0.46  $ & $ 4.5    $ & $ -0.42  $ & $ 4.1   $ \\ 
  & $ -0.19  $ & $ 1.9    $ & $ -0.19  $ & $ 1.9    $ & $ -0.17  $ & $ 1.7   $ \\ 
\hline
\raisebox{-4.6ex}[0pt][0pt]{500} 
  & $ -0.99  $ & $ 8.7    $ & $ -0.95  $ & $ 8.3    $ & $ -0.85  $ & $ 7.4   $ \\ 
  & $ -0.47  $ & $ 4.2    $ & $ -0.44  $ & $ 3.9    $ & $ -0.37  $ & $ 3.3   $ \\ 
  & $ -0.37  $ & $ 3.2    $ & $ -0.35  $ & $ 3.1    $ & $ -0.30  $ & $ 2.7   $ \\ 
  & $ -0.23  $ & $ 2.1    $ & $ -0.20  $ & $ 1.8    $ & $ -0.15  $ & $ 1.3   $ \\ 
\hline
\end{tabular}
\\[6pt]
Table 7
\end{center}
\end{table}
%                                
%                                
%  E2 mit Schnitten unterhalb    
%                                
%                                
\begin{table}[ht]
\begin{center}
\begin{tabular}{|r|c|c|c|c|c|c|}
\hline
 & \multicolumn{6}{c|}{$g_{VV}$}  \\
 $m_\varphi$ & \multicolumn{2}{c|}{0.0} &  \multicolumn{2}{c|}{0.2}&\multicolumn
{2}{c|}{0.4}\\
\hline
\raisebox{-4.6ex}[0pt][0pt]{370} 
  & $ 3.5    $ & $ 26.4   $ & $ 3.3    $ & $ 24.3   $ & $ 2.7    $ & $ 18.7  $ \\ 
  & $ 3.1    $ & $ 20.4   $ & $ 2.4    $ & $ 14.8   $ & $ 1.3    $ & $ 7.9   $ \\ 
  & $ 1.1    $ & $ 7.7    $ & $ 0.98   $ & $ 7.0    $ & $ 0.80   $ & $ 5.5   $ \\ 
  & $ 0.93   $ & $ 5.8    $ & $ 0.61   $ & $ 3.7    $ & $ 0.32   $ & $ 1.9   $ \\ 
\hline
\raisebox{-4.6ex}[0pt][0pt]{400} 
  & $ 1.8    $ & $ 21.4   $ & $ 1.7    $ & $ 20.2   $ & $ 1.4    $ & $ 16.6  $ \\ 
  & $ 1.0    $ & $ 11.8   $ & $ 0.88   $ & $ 9.9    $ & $ 0.60   $ & $ 6.7   $ \\ 
  & $ 0.59   $ & $ 6.9    $ & $ 0.55   $ & $ 6.3    $ & $ 0.45   $ & $ 5.1   $ \\ 
  & $ 0.39   $ & $ 4.3    $ & $ 0.28   $ & $ 3.1    $ & $ 0.17   $ & $ 1.9   $ \\ 
\hline
\raisebox{-4.6ex}[0pt][0pt]{500} 
  & $ 0.52   $ & $ 7.6    $ & $ 0.47   $ & $ 7.0    $ & $ 0.37   $ & $ 5.3   $ \\ 
  & $ 0.24   $ & $ 3.6    $ & $ 0.21   $ & $ 3.1    $ & $ 0.14   $ & $ 2.1   $ \\ 
  & $ 0.19   $ & $ 2.8    $ & $ 0.17   $ & $ 2.5    $ & $ 0.13   $ & $ 1.9   $ \\ 
  & $ 0.12   $ & $ 1.7    $ & $ 0.08   $ & $ 1.2    $ & $ 0.05   $ & $ 0.68  $ \\ 
\hline
\end{tabular}
\\[6pt]
Table 8
\end{center}
\end{table}
%                                
%                                
%  E2 mit Schnitten oberhalb     
%                                
%                                
\begin{table}[ht]
\begin{center}
\begin{tabular}{|r|c|c|c|c|c|c|}
\hline
 & \multicolumn{6}{c|}{$g_{VV}$}  \\
 $m_\varphi$ & \multicolumn{2}{c|}{0.0} &  \multicolumn{2}{c|}{0.2}&\multicolumn
{2}{c|}{0.4}\\
\hline
\raisebox{-4.6ex}[0pt][0pt]{320} 
  & $ -0.48  $ & $ 8.1    $ & $ -0.48  $ & $ 8.1    $ & $ -0.48  $ & $ 8.1   $ \\ 
  & $ -0.15  $ & $ 2.6    $ & $ -0.15  $ & $ 2.6    $ & $ -0.15  $ & $ 2.6   $ \\ 
  & $ -0.15  $ & $ 2.5    $ & $ -0.15  $ & $ 2.5    $ & $ -0.15  $ & $ 2.5   $ \\ 
  & $ -0.05  $ & $ 0.80   $ & $ -0.05  $ & $ 0.80   $ & $ -0.05  $ & $ 0.80  $ \\ 
\hline
\raisebox{-4.6ex}[0pt][0pt]{350} 
  & $ -0.84  $ & $ 8.9    $ & $ -0.85  $ & $ 8.9    $ & $ -0.86  $ & $ 9.0   $ \\ 
  & $ -0.32  $ & $ 3.5    $ & $ -0.32  $ & $ 3.5    $ & $ -0.33  $ & $ 3.5   $ \\ 
  & $ -0.28  $ & $ 2.9    $ & $ -0.28  $ & $ 2.9    $ & $ -0.28  $ & $ 3.0   $ \\ 
  & $ -0.10  $ & $ 1.1    $ & $ -0.10  $ & $ 1.1    $ & $ -0.10  $ & $ 1.1   $ \\ 
\hline
\raisebox{-4.6ex}[0pt][0pt]{370} 
  & $ -0.81  $ & $ 9.4    $ & $ -0.82  $ & $ 9.5    $ & $ -0.86  $ & $ 9.9   $ \\ 
  & $ -0.41  $ & $ 4.8    $ & $ -0.41  $ & $ 4.9    $ & $ -0.42  $ & $ 4.9   $ \\ 
  & $ -0.30  $ & $ 3.4    $ & $ -0.30  $ & $ 3.5    $ & $ -0.31  $ & $ 3.5   $ \\ 
  & $ -0.14  $ & $ 1.6    $ & $ -0.14  $ & $ 1.7    $ & $ -0.14  $ & $ 1.7   $ \\ 
\hline
\raisebox{-4.6ex}[0pt][0pt]{400} 
  & $ -0.88  $ & $ 9.9    $ & $ -0.89  $ & $ 10.1   $ & $ -0.93  $ & $ 10.5  $ \\ 
  & $ -0.44  $ & $ 5.1    $ & $ -0.45  $ & $ 5.2    $ & $ -0.46  $ & $ 5.3   $ \\ 
  & $ -0.33  $ & $ 3.7    $ & $ -0.33  $ & $ 3.8    $ & $ -0.34  $ & $ 3.8   $ \\ 
  & $ -0.16  $ & $ 1.8    $ & $ -0.16  $ & $ 1.8    $ & $ -0.16  $ & $ 1.9   $ \\ 
\hline
\raisebox{-4.6ex}[0pt][0pt]{500} 
  & $ -0.73  $ & $ 7.2    $ & $ -0.75  $ & $ 7.4    $ & $ -0.78  $ & $ 7.8   $ \\ 
  & $ -0.37  $ & $ 3.7    $ & $ -0.38  $ & $ 3.8    $ & $ -0.39  $ & $ 3.9   $ \\ 
  & $ -0.28  $ & $ 2.8    $ & $ -0.29  $ & $ 2.9    $ & $ -0.30  $ & $ 3.0   $ \\ 
  & $ -0.19  $ & $ 1.9    $ & $ -0.20  $ & $ 2.0    $ & $ -0.18  $ & $ 1.8   $ \\ 
\hline
\end{tabular}
\\[6pt]
Table 9
\end{center}
\end{table}


\begin{thebibliography}{99}
\baselineskip=12pt
\bibitem{Lee}
T. D. Lee, Phys. Rev. D8 (1973) 1226; Phys. Rep. C9 (1974) 143.
\bibitem{DepMa}
N. G. Deshpande and E. Ma, Phys. Rev. D16 (1977) 1583.
\bibitem{Branco}
G. C. Branco and M. N. Rebelo, Phys. Lett. B160 (1985) 117.
\bibitem{Wolf}
J. Liu and L. Wolfenstein, Nucl. Phys. B289 (1987) 1.
\bibitem{Wein}
S. Weinberg, Phys. Rev. D42 (1990) 860.
\bibitem{Matsuda}
M. Matsuda and M. Tanimoto, Phys. Rev. D52 (1995) 3100;\\
N. Haba, Prog. Theor. Phys. 97 (1997) 301.
\bibitem{Babu}
K. S. Babu, C. Kolda, J. March-Russell and F. Wilczek, hep-ph/9804355.
\bibitem{Pil}
A. Pilaftsis, Phys. Lett. B435 (1998) 88.
\bibitem{BeBra1}
W. Bernreuther and A. Brandenburg, Phys. Lett.  B314  (1993) 104.
\bibitem{BeBra2}
W. Bernreuther and A. Brandenburg,
Phys. Rev. D49 (1994) 4481.
\bibitem{Chang}
D. Chang, W. Y. Keung and I. Phillips, Phys. Rev. D48 (1993) 3225.
\bibitem{corr}
W. Bernreuther, A. Brandenburg and M. Flesch,
Phys. Rev. D56  (1997) 90.
\bibitem{Osland}
A. Skjold and P. Osland, Phys. Lett. B329 (1994) 305.
\bibitem{Peskin}
C.R. Schmidt and  M.E. Peskin, Phys. Rev. Lett. 69  (1992) 410.
\bibitem{Zhou}
H. Y. Zhou, Phys. Rev. D58 (1998) 114002.
\bibitem{He}
X. Zhang et al., Phys. Rev. D50 (1994) 7042;\\
S. Bar-Shalom et al., Phys. Rev. D53 (1996) 1162;\\
J. F. Gunion and X. G. He, Phys. Rev. Lett. 76 (1996) 4468.
\bibitem{KLY}
G. L. Kane, G. A. Ladinsky and C. P. Yuan, Phys. Rev. D45 (1992) 124.
\bibitem{Aeppli}
D. Atwood, A. Aeppli and A. Soni, Phys. Rev. Lett. 69 (1992) 2756.
\bibitem{BM}
 A. Brandenburg and J. P. Ma, Phys. Lett. B298 (1993) 211.
\bibitem{Haberl}
P. Haberl, O. Nachtmann and A. Wilch,  Phys. Rev. D53 (1996) 4875.
\bibitem{Lampe}
B. Grzadkowski, B. Lampe and K. J. Abraham, Phys. Lett. B415 (1997) 193.
\bibitem{Schmidt}
C. R. Schmidt, Phys. Lett. B293 (1992) 111.
\bibitem{Atwood}
D. Atwood et al., Phys. Rev. D54 (1996) 5412.
\bibitem{Bar}
S. Bar-Shalom, D. Atwood and A. Soni, Phys. Rev. D57 (1998) 1495.
\bibitem{KM}
M. Kobayashi and T. Maskawa, Prog. Theor. Phys. 49 (1973) 652.
\bibitem{BSP}
W. Bernreuther, T. Schr\"oder and T.N. Pham, Phys. Lett. B279 (1992) 389.
\bibitem{BorzG}
F. M. Borzumati and C. Greub, Phys. Rev. D58 (1998) 074004;
hep-ph/9809438.
\bibitem{Buras}
A. J. Buras, P. Krawczyk, M. E. Lautenbacher and C. Salazar,
Nucl. Phys. B337 (1990) 284.
\bibitem{neutr}
K. F. Smith et al., Phys. Lett. B234 (1990) 191;\\
I. S. Altarev et al., Phys. Lett. B276 (1992) 242.
\bibitem{elect}
E. D. Commins, S. B. Ross, D. DeMille and B. C. Regan,
Phys. Rev. A50 (1994) 2960.
\bibitem{Hayashi}
T. Hayashi et al., Phys. Lett. B348 (1995) 489.
\bibitem{Chemtob}
M. Chemtob, Phys. Rev. D45 (1992) 1649; \\
J. Ellis and R. Flores, Phys. Lett. B377 (1996) 83;\\ 
I. B. Khriplovich, Phys. Lett. B382 (1996) 145.
\bibitem{CPrev}
W. Bernreuther, hep-ph/9808453.
\bibitem{BFH}
W. Bernreuther, M. Flesch and P. Haberl, Phys. Rev. D58 (1998) 114031.
\bibitem{SSH}
S. Dawson, A. Djouadi and M. Spira, Phys. Rev. Lett. 77  (1996) 16.
\bibitem{SH}
A. Djouadi, M. Spira and P. Zerwas, Phys. Lett. B264 
(1991) 440; \\ D. Graudenz,  M. Spira and P. Zerwas, Phys. Rev. Lett.  70
  (1993) 1372; \\ S. Dawson, Nucl. Phys.  B359  (1991) 283.
\bibitem{SH2}
M. Spira, A. Djouadi, D. Graudenz  and P. Zerwas,
Nucl. Phys.   B453  (1995) 17.
\bibitem{SH3}
M. Kr\"amer, E. Laenen and M. Spira, Nucl. Phys. B511 (1998) 523.
\bibitem{been}
W. Beenakker, A. Denner, W. Hollik, R. Mertig,
T. Sack and D. Wackeroth, Nucl. Phys. B411 (1994) 343.
\bibitem{Gaemers}
K. Gaemers and F. Hoogeveen, Phys. Lett. B146 (1984) 347. 
\bibitem{Dicus}
D. Dicus, A. Stange and S. Willenbrock,
Phys. Lett. B333 (1994) 126.
\bibitem{BO}
W. Bernreuther and P. Overmann, Z. Phys. C61 (1994) 599.
\bibitem{BBU}
W. Bernreuther, A. Brandenburg and P. Uwer, Phys. Lett. B368 (1996) 153.
\bibitem{Goldstein}
W.G.D. Dharmaratna and G. R. Goldstein, Phys. Rev. D53 (1996) 1073.
\bibitem{BNOS}
W. Bernreuther, O. Nachtmann, P. Overmann and T. Schr\"oder,
Nucl. Phys. B388 (1992) 53; B406 (1993)  516 (E).
\bibitem{MB}
J. P. Ma and A. Brandenburg,
Z. Phys. C56 (1992) 97.
\bibitem{Kuhn}
A. Czarnecki, M. Jezabek and J. H. K\"uhn, Nucl. Phys. B351 (1991) 70. 
\bibitem{GG}
B. Grzadkowski and J. F. Gunion, Phys. Lett. B287 (1992) 237;\\
T. Hasuike, T. Hattori and S. Wakaizumi, Phys. Rev. D58 (1998) 095008.
\bibitem{Sola}
J.A. Coarasa, J. Guasch, J. Sola and  W. Hollik,  hep-ph/980827.
\bibitem{CDF}
F. Abe et al. (CDF Collab.), Phys. Rev. Lett. 79 (1997) 357.
\bibitem{Lad}
G. A. Ladinsky, Phys. Rev. D46  (1992) 3789; D47  (1993) 3086 (E).
\bibitem{GRV}
M. Gl\"uck, E. Reya and A. Vogt, Z. Phys. C67  (1995) 433.
\bibitem{MRSCTEQ}
A. D. Martin, W. J. Stirling and R. G. Roberts, Phys. Lett.
B354  (1995) 155; \\ H. L. Lai {\it et al.} (CTEQ Collab.), 
Phys. Rev. D51  (1995) 4763.
\bibitem{Atopt}
D. Atwood and A. Soni, Phys. Rev. D45 (1992) 2405;\\
M. Diehl and O. Nachtmann, Z. Phys. C62 (1994) 397.

\end{thebibliography}
\end{document}